\documentclass{PoS}

\title{Introduction to Early Universe Cosmology}

\ShortTitle{Cosmology}

\author{\speaker{Robert H. Brandenberger}\thanks{An earlier version of these lectures was 
delivered at the Observatorio Nacional in November 2009 \cite{ON}.}\\
        Physics Department, McGill University, Montreal, Quebec, H3A 2T8, Canada\\
        E-mail: \email{rhb@physics.mcgill.ca}}

\abstract{Observational cosmology is in its ``golden age" with a vast amount of
recent data on the distribution of matter and light in the universe. This
data can be used to probe theories of the very early universe. It is
small amplitude cosmological fluctuations which encode the information
about the very early universe and relate it to current data. Hence,
a central topic in these lectures is the ``theory of cosmological perturbations",
the theory which describes the generation of inhomogeneities in the very
early universe and their evolution until the current time. I will apply this
theory to three classes of models of the very early universe. The first
is ``Inflationary Cosmology", the current paradigm for understanding the
early evolution of the universe. I will review the successes of inflationary
cosmology, but will also focus on some conceptual challenges which
inflationary cosmology is facing, challenges which motivate the search
for possible alternatives. I will introduce two alternative scenarios,
the ``Matter Bounce" model and ``String Gas Cosmology", and I will
discuss how cosmological fluctuations which can explain the current
data are generated in those models.}

\FullConference{4th International Conference on Fundamental Interactions -ICFI2010,\\
		August 1-7, 2010\\
		Viçosa, Brazil}

\def\be{\begin{equation}} 
\def\ee{\end{equation}} 
\def\bea{\begin{eqnarray}} 
\def\eea{\end{eqnarray}} 

\begin{document}
\input epsf.tex

\section{Introduction}

\subsection{Overview}

Observational cosmology is currently in its ``Golden Age". 
A wealth of new observational results are being uncovered.
The cosmic microwave background (CMB) has been measured
to high precision, the distribution of visible and of dark matter
is being mapped out to greater and greater depths. New
observational windows to probe the structure of the universe
are opening up. To explain these observational results it is
necessary to consider processes which happened in the
very early universe.

The inflationary scenario \cite{Guth} 
(see also \cite{Brout, Starob1, Sato}) is the current paradigm for
the evolution of the very early universe. Inflation can explain
some of the puzzles which the previous paradigm - Standard
Big Bang cosmology - could not address. More importantly, however,
inflationary cosmology gave rise to the first explanation for
the origin of inhomogeneities in the universe based on causal
physics \cite{Mukh} (see also \cite{Press,Sato,Starob2})). The
theory of structure formation in inflationary cosmology predicted
the detailed shape of the angular power spectrum of CMB
anisotropies, a prediction which was verified many years
later observationally \cite{WMAP}.

In spite of this phenomenological success, inflationary cosmology
is not without its conceptual problems. These problems
motivate the search for alternative proposals for the evolution
of the early universe and for the generation of structure. These
alternatives must be consistent with the current observations,
and they must make predictions with which they can be
observationally distinguished from inflationary cosmology.
In these lectures I will discuss two alternative scenarios, the
``Matter Bounce" (see e.g. \cite{CosPA08}) and ``String
Gas Cosmology" \cite{RHBSGrev}.

I begin these lectures with a brief survey of the Standard Big Bang (SBB)
model and its problems. Possibly the most important problem
is the absence of a structure formation scenario based on causal
physics. I will then introduce inflationary cosmology and the two
alternatives and show how they address the conceptual problems of
the SBB model, in particular how structure is formed in these
models.

Since the information about the early universe is encoded in the spectrum
of cosmological fluctuations about the expanding background, it is
these fluctuations which provide a link between the physics in the very
early universe and current cosmological observations. 
Section 2 is devoted to an overview of the theory of linearized cosmological
perturbations, the main technical tool in modern cosmology. The analysis
of this section is applicable to all early universe theories. In Section 3
I give an overview of inflationary cosmology, focusing on the basic
principles and emphasizing recent progress and conceptual problems.

The conceptual problems of inflationary cosmology motivate the
search for alternatives. In Section 4 the ``Matter Bounce" alternative
is discussed, in Section 5 the ``String Gas" alternative.

\subsection{Standard Big Bang Model}

Standard Big Bang cosmology is based on three principles. The first is the
assumption that space is homogeneous and isotropic on large scales. The second
is the assumption that the dynamics of space-time is described by the 
Einstein field equations. The third basis of the theory is that matter can
be described as a superposition of two classical perfect fluids, a radiation 
fluid with relativistic equation of state and pressure-less (cold) matter.
 
The first principle of Standard Cosmology implies that the metric
takes the following most general form for a homogeneous and isotropic
four-dimensional universe
\be \label{metric}
ds^2 \, = \, dt^2 - a(t)^2 d{\bf x}^2 \, ,
\ee
where $t$ is physical time, ${\bf x}$ denote the three co-moving spatial
coordinates (points at rest in an expanding space have constant co-moving
coordinates), and the scale factor $a(t)$ is proportional to the radiius of
space. For simplicity we have assumed that the universe is
spatially flat. 

The expansion rate $H(t)$ of the universe is given by
\be
H(t) \, = \, \frac{{\dot a}}{a} \, ,
\ee
where the overdot represents the derivative with respect to time. The
cosmological redshift $z(t)$ at time $t$ yields the amount of expansion
which space has undergone between the time $t$ and the present time $t_0$:
\be
z(t) + 1 \, = \, \frac{a(t_0)}{a(t)} \, .
\ee

In the case of a homogeneous and isotropic metric (\ref{metric}), the
Einstein equations which describe how matter induces curvature
of space-time reduce to a set of ordinary differential equations, the
Friedmann equations. Written for simplicity in the case of no cosmological 
constant and no spatial curvature they are
\be
H^2 \, = \, \frac{8 \pi G}{3} \rho \, ,
\ee
where $\rho$ is the energy density of matter and $G$ is Newton's
gravitational constant, and
\be
\frac{\ddot{a}}{a} \, = \, - \frac{4 \pi G}{3} \bigl( \rho + 3 p \bigr) \,
\ee
or equivalently
\be
{\dot{\rho}} \, = \, - 3 H \bigl( \rho + p \bigr) \, ,
\ee
where $p$ is the pressure density of matter.

The third principle of Standard Cosmology says that the energy
density and pressure are the sums of the contributions from cold
matter (symbols with subscript ``m") and radiation (subscripts ``r"):
\bea
\rho \, &=& \, \rho_m + \rho_r \, \\
p \, &=& \, p_m + p_r \, \nonumber
\eea
with $p_m = 0$ and $p_r = 1/3 \rho_r$.

The first principle of Standard Cosmology, the homogeneity and isotropy of
space on large scales, was initially introduced soleley to simplify the mathematics. 
However, it is now very well confirmed observationally by the near isotropy
of the cosmic microwave background and by the convergence to homogeneity
in the distribution of galaxies as the length scale on which the universe
is probed increases.

Standard Cosmology rests on three observational pillars. Firstly, it
explains Hubble's redshift-distance relationship. Secondly, and
most importantly for the theme of these lectures, it predicted the
existence and black body nature of the cosmic microwave background (CMB).
The argument is as follows. If we consider regular matter and go back
in time, the temperature of matter increases. Eventually, it exceeds
the ionization temperature of atoms. Before that time, matter was
a plasma, and space was permeated by a thermal gas of photons.
A gas of photons will thus remain at the current time. The photons
last scatter at a time $t_{rec}$, the ``time of recombination"
which occurs at a redshift of about $10^3$. After that, the gas of photons
remains in a thermal distribution with a temperature $T$ which
redshifts as the universe expands, i.e. $T \sim a^{-1}$. The CMB
is this remnant gas of photons from the early universe. The
precision measurement of the black body nature of the CMB
\cite{CMBblack} can be viewed as the beginning of the ``Golden Age"
of observational cosmology. The third observational pillar of
Standard Cosmology is the good agreement between the 
predicted abundances of light elements and the observed ones.
This agreement tells us that Standard Cosmology is a very good
description of how the universe evolved back to the time of
nucleosynthesis. Any modifications to the time evolution of
Standard Cosmology must take place before then. 

\begin{figure}
\includegraphics[height=6cm]{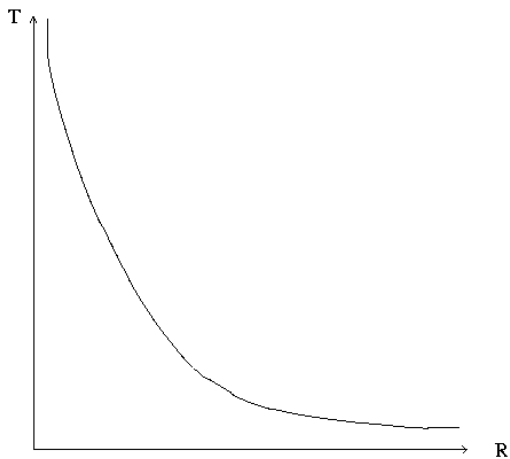}
\caption{A sketch of the temperature (vertical axis) - time (horizontal axis)
relation in Standard Cosmology. As the beginning of time is approached,
the temperature diverges.}
\label{fig1}
\end{figure}

At the present time $t_0$ matter is dominated by the cold pressureless
component. Thus, as we go back in time, the universe is (except
close to the present time when it appears that a residual cosmological
constant is beginning to dominate the cosmological dynamics)
initially in a matter-dominated phase during which $a(t) \sim t^{2/3}$.
Since the energy density in cold matter scales as $a^{-3}$ whereas
that of relativistic radiation scales as $a^{-4}$, there will be
a time before which the energy density in radiation was larger than
that of cold matter. The time when the two energy densities are
equal is $t_{eq}$ and corresponds to a redshift of around $z = 10^4$.
For $t < t_{eq}$ the universe is radiation-dominated and $a(t) \sim t^{1/2}$.
As we go back into the past the density and temperature increase
without bound and a singularity is reached at which energy density,
temperature and curvature are all infinite. This is the ``Big Bang"
(see Fig. \ref{fig1} for a sketch of the temperature/time relationship in
Standard Cosmology). 

\subsection{Problems of the Standard Big Bang}

Obviously, the assumptions on which Standard Cosmology is based
break down long before the singularity is reached. This is the
``singularity problem" of the model, a problem which also
arises in inflationary cosmology. It is wrong to say that Standard
Cosmology predicts a Big Bang. Instead, one should say that
Standard Cosmology is incomplete and one does not understand
the earliest moments of the evolution of the universe.

\begin{figure}
  \includegraphics[height=6cm]{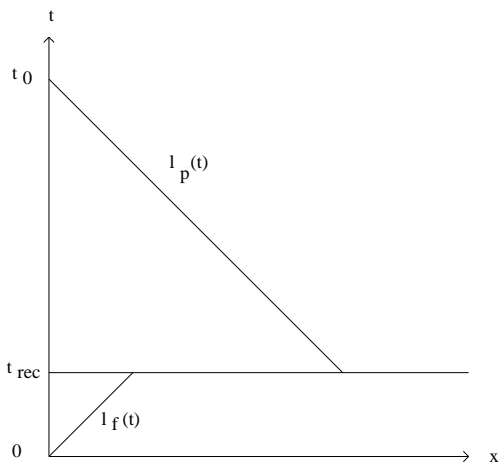}
  \caption{Sketch illustrating the horizon problem of Standard
  Cosmology: our past horizon at $t_{rec}$ is larger than the
  causal horizon (forward light cone) at that time, thus making
  it impossible to causally explain the observed isotropy of
  the CMB.}
  \label{dfig1}
\end{figure}

There are, however, more ``practical" ``problems" of the
Standard Big Bang model - problems in the sense that the
model is unable to explain certain key features of the
observed universe. The first such problem is the ``horizon problem"
\cite{Guth}: within Standard Cosmology there is no possible
explanation for the observed homogeneity and isotropy of
the CMB. Let us consider photons reaching us from
opposite angles in the sky. As sketched in Figure \ref{dfig1},
the source points for these photons on the last scattering
surface are separated by a distance greater than the horizon
at that time. Hence, there is no possible causal mechanism
which can relate the temperatures at the two points.

A related problem is the ``size problem": if the spatial sections
of the universe are finite, then the only length scale available at
the Planck time $t_{pl}$ is the Planck length. However, it we
extrapolate the size of our currently observed horizon back to
when the temperature was equal to the Planck temperature, then the
corresponding wavelength is larger than $1 {\rm{\mu m}}$, many
orders of magnitude larger than the Planck length. Standard
Cosmology offers no explanation for these initial conditions.
  
The third mystery of Standard Cosmology concerns the observed
degree of spatial flatness of the universe. At the current time
the observed energy density equals the ``critical" energy
density $\rho_c$ of a spatially flat universe to within $10\%$ or
better. However, in Standard Cosmology $\rho = \rho_c$ is
an unstable critical point in an expanding universe. As the
universe expands, the relative difference between $\rho$ and
$\rho_c$ increases. This can be seen by taking the
Friedmann equation in the presence of spatial curvature
\be \label{FRW11}
H^2 + \epsilon T^2 \, = \, \frac{8 \pi G}{3} \rho \,
\ee
where
\be
\epsilon \, = \, \frac{k}{(a T)^2} \,
\ee
$T$ being the temperature and $k$ the curvature constant which
is $k = \pm 1$ or $k = 0$ for closed, open or flat spatial sections,
and comparing  (\ref{FRW11}) with the corresponding equation
in a spatially flat universe ($k = 0$ and $\rho_c$ replacing $\rho$.
If entropy is conserved (as it is in Standard Cosmology)
then $\epsilon$ is constant and we obtain
\be \label{flateq}
\frac{\rho - \rho_c}{\rho_c} \, = \, \frac{3}{8 \pi G} \frac{\epsilon T^2}{\rho_c}
\, \sim \, T^{-2} \, .
\ee
Hence, to explain the currently observed degree of spatial flatness, the
initial spatial curvature had to have been tuned to a very high 
accuracy. This is the ``flatness problem".

\begin{figure}
  \includegraphics[height=6cm]{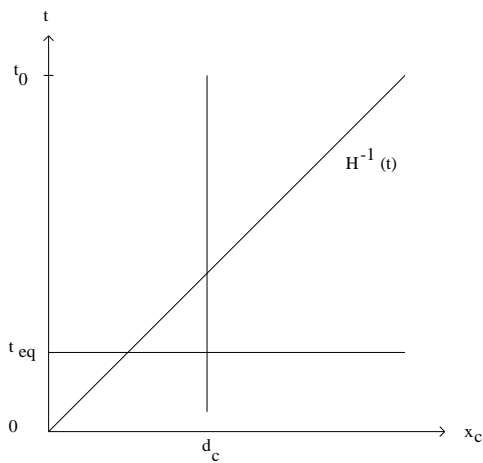}
  \caption{Sketch illustrating the formation of structure problem
  of Standard Cosmology: the physical wavelength of fluctuations
  on fixed comoving scales which correspond to the large-scale
  structures observed in the universe is larger than the horizon at
  the time $t_{eq}$ of equal matter and radiation, the time when
  matter fluctuations begin to grow. Hence, it is impossible to
  explain the origin of non-trivial correlations of the seeds for
  the fluctuations which had to have been present at that time.}
  \label{dfig2}
\end{figure}

We observe highly non-random correlations in the distribution of
galaxies in the universe. The only force which can act on
the relevant distances is gravity. Gravity is a weak force, and therefore
the seed fluctuations which develop into the observed structures
had to have been non-randomly distributed at the time $t_{eq}$,
the time when gravitational clustering begins (see Section 2). 
However, as illustrated in Figure \ref{dfig2}, the physical length
corresponding to the fluctuations which we observe today on the
largest scales (they have constant comoving scale) is larger than
the horizon at that time.  Hence, there can be no causal generation
mechanism for these perturbations \footnote{This is the usual
textbook argument. The student is invited to find (at least) two flaws
in this argument.}. This is the ``fluctuation problem" of Standard
Cosmology.
  
There are also more conceptual problems: Standard Cosmology
is based on treating matter as a set of perfect fluids. However,
we know that at high energies and temperatures a classical
description of matter breaks down. Thus, Standard Cosmology
must break down at sufficiently high energies. It does not
contain  the adequate matter physics to describe the very
early universe. Similarly, the singularity of space-time which
Standard Cosmology contains corresponds to a breakdown
of the assumptions on which General Relativity is based.
This is the ``singularity problem" of Standard Cosmology.

All of the early universe scenarios which I will discuss in the
following provide solutions to the formation of structure problem.
They can successfully explain the wealth of observational
data on the distribution of matter in the universe and on the
CMB anisotropies. The rest of these lectures will focus on
this point. However, the reader should also ask under which
circumstances the scenarios to be discussed below address
the other problems mentioned above.

\subsection{Inflation as a Solution}

\begin{figure}
\includegraphics[height=2.5cm]{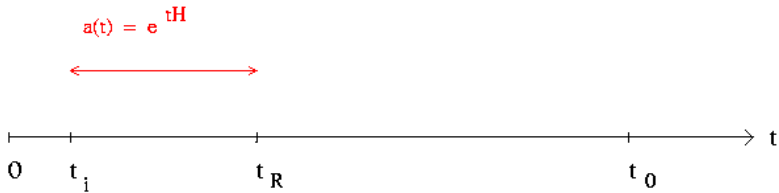}
\caption{A sketch showing the time line of inflationary cosmology.
The period of accelerated expansion begins at time $t_i$ and
end at $t_R$. The time evolution after $t_R$ corresponds to
what happens in Standard Cosmology.}
\label{timeline}
\end{figure}

The idea of inflationary cosmology is to add a period to the evolution of
the very early universe during which the scale factor undergoes
accelerated expansion - most often nearly exponential growth.
The time line of inflationary cosmology is sketched in Figure \ref{timeline}.
The time $t_i$ is the beginning of the inflationary period, and
$t_R$ is its end (the meaning of the subscript $R$ will become
clear later). Although inflation is usually associated with physics
at very high energy scales, e.g. $E \sim 10^{16} {\rm Gev}$, all that
is required from the initial basic considerations is that inflation
ends before the time of nucleosynthesis.

During the period of inflation, the density of any pre-existing particles is
red-shifted. Hence, if inflation is to be viable, it must contain a
mechanism to heat the universe at $t_R$, a ``reheating"
mechanism - hence the subscript $R$ on $t_R$. This mechanism must
involve dramatic entropy generation. It is this non-adiabatic evolution
which leads to a solution of the flatness problem, as the reader can
verify by inspecting equation (\ref{flateq}) and allowing for entropy
generation at the time $t_R$.

\begin{figure} 
\includegraphics[height=9cm]{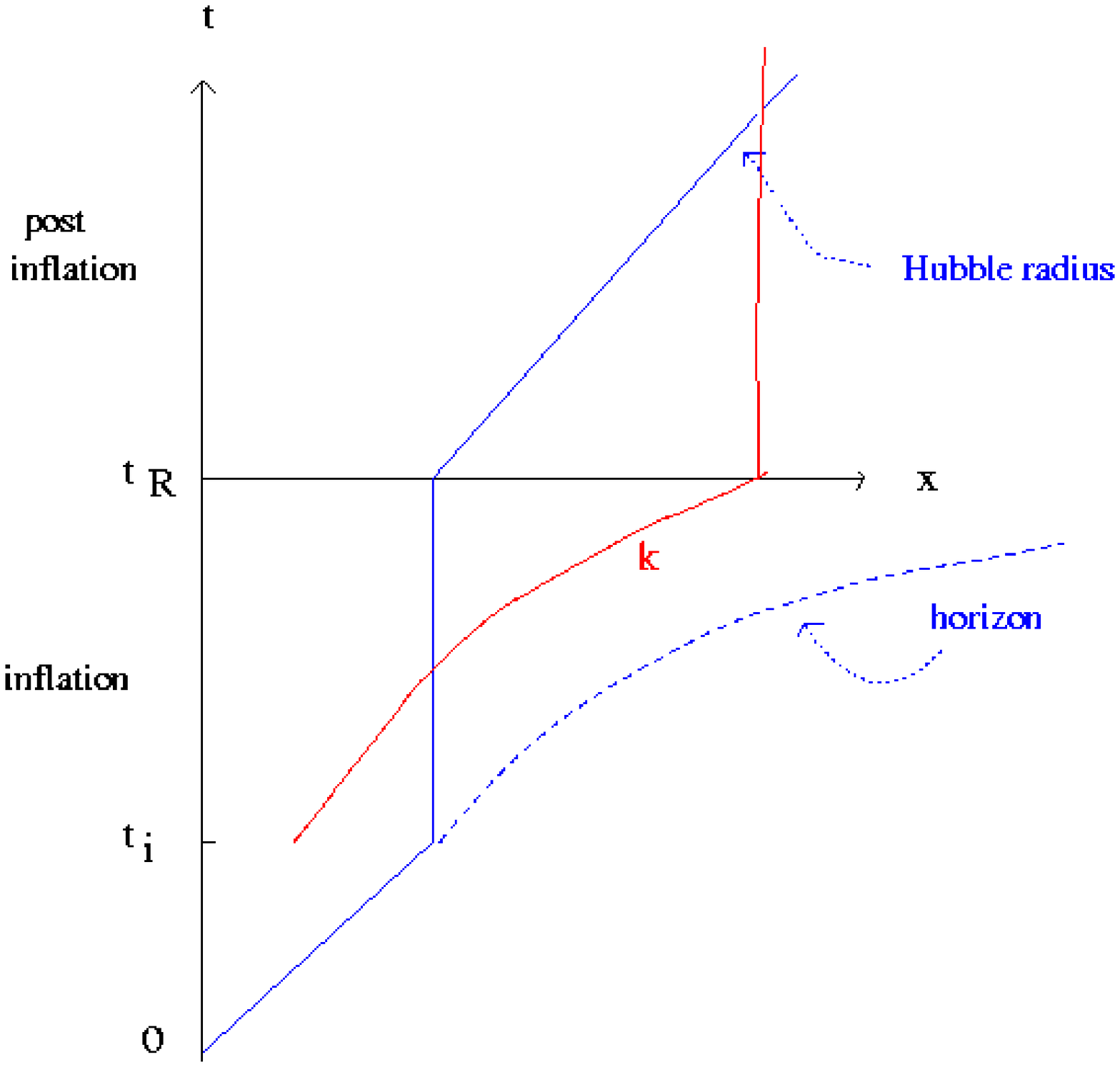}
\caption{Space-time sketch of inflationary cosmology.
The vertical axis is time, the horizontal axis corresponds
to physical distance. The solid line labelled $k$ is the
physical length of a fixed comoving fluctuation scale. The
role of the Hubble radius and the horizon are discussed
in the text.}
\label{infl1}
\end{figure}

A space-time sketch of inflationary cosmology is given in
Figure \ref{infl1} . The vertical axis is time, the horizontal
axis corresponds to physical distance.  Three different
distance scales are shown. The solid line labelled by $k$
is the physical length corresponding to a fixed comoving
perturbation. The second solid line (blue) is the Hubble
radius 
\be
l_H(t) \, \equiv \, H^{-1}(t) \, .
\ee
The Hubble radius separates scales where microphysics
dominates over gravity (sub-Hubble scales) from ones 
on which the effects of microphysics are negligible 
(super-Hubble scales) \footnote{This statement will be
demonstrated later in these lectures.}. Hence, a
necessary requirement for a causal theory of structure
formation is that scales we observe today originate
at sub-Hubble lengths in the early universe. The third
length is the ``horizon", the forward light cone of our position
at the Big Bang. The horizon is the causality limit. Note
that because of the exponential expansion of space during
inflation, the horizon is exponentially larger than the
Hubble radius. It is important not to confuse these two
scales. Hubble radius and horizon are the same in
Standard Cosmology, but in all three early universe
scenarios which will be discussed in the following they
are different.

{F}rom Fig. \ref{infl1} it is clear that provided that the period
of inflation is sufficiently long, all scales which are currently
observed originate as sub-Hubble scales at the beginning
of the inflationary phase. Thus, in inflationary cosmology 
it is possible to have a causal generation mechanism of
fluctuations \cite{Mukh,Press,Sato}. Since matter pre-existing
at $t_i$ is redshifted away, we are left with a matter vacuum.
The inflationary universe scenario of structure formation is
based on the hypothesis that all current structure originated
as quantum vacuum fluctuations. From Figure \ref{infl1} it
is also clear that the horizon problem of standard cosmology
can be solved provided that the period of inflation lasts
sufficiently long. The reader should convince him/herself
that the required period of inflation is about $50 H^{-1}$ if
inflation takes place at an enegy scale of about $10^{16} {\rm GeV}$.
Inflation thus solve both the horizon and the structure
formation problems.

To obtain exponential expansion of space in the context
of Einstein gravity the energy density must be constant.
Thus, during inflation the total energy and size of the
universe both increase exponentially. In this way,
inflation can solve the size and entropy problems
of Standard Cosmology.

To summarize the main point concerning the generation
of cosmological fluctuations (the main theme of these
lectures) in inflationary cosmology: the first crucial
criterium which must be satisfied in order to have a successful
theory of structure formation is that fluctuation scales
originate inside the Hubble radius. In inflationary cosmology
it is the accelerated expansion of space during the inflationary
phase which provides this success. In the following we
will emphasize what is responsible for the corresponding
success in the two alternative scenarios which we
will discuss.

\subsection{Matter Bounce as a Solution}

The first alternative to cosmological
inflation as a theory of structure formation is the
 ``matter bounce" , an alternative which is not
yet well appreciated (for an overview the reader is
referred to \cite{CosPA08}). The scenario is
based on a cosmological background in which the
scale factor $a(t)$ bounces in a non-singular manner.

\begin{figure}[htbp] 
\includegraphics[height=9cm]{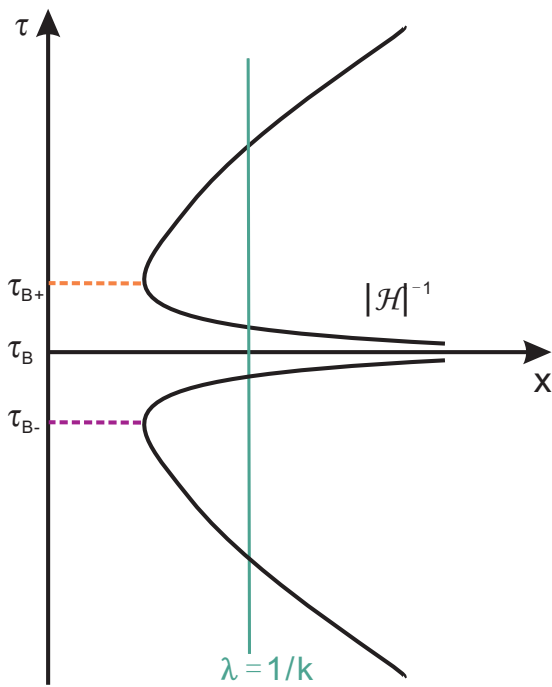}
\caption{Space-time sketch in the matter bounce scenario. The vertical axis
is conformal time $\eta$, the horizontal axis denotes a co-moving space coordinate.
Also, ${\cal H}^{-1}$ denotes the co-moving Hubble radius.}
\label{bounce}
\end{figure}

{F}igure \ref{bounce} shows a space-time sketch
of such a bouncing cosmology. Without loss of generality
we can adjust the time axis such that the bounce point
(minimal value of the scale factor) occurs at $t = 0$.
There are three phases in such a non-singular bounce: the initial
contracting phase during which the Hubble radius is
decreasing linearly in $|t|$, a bounce phase during which
a transition from contraction to expansion takes place,
and thirdly the usual expanding phase of Standard
Cosmology. There is no prolonged inflationary phase 
after the bounce, nor is there a time-symmetric deflationary 
contracting period before the bounce point. As is obvious
from the Figure, scales which we observe today started
out early in the contracting phase at sub-Hubble lengths.
The matter bounce scenario assumes that the contracting
phase is matter-dominated at the times when scales
we observe today exit the Hubble radius. A model in
which the contracting phase is the time reverse of our
current expanding phase would obey this condition.
The assumption of an initial matter-dominated phase
will be seen later in these lectures to be important if
we want to obtain a scale-invariant spectrum of
cosmological perturbations from initial vacuum
fluctuations \cite{Wands,FB2,Wands2}.

Let us make a first comparison with inflation. A non-deflationary
contracting phase replaces the accelerated expanding phase
as a mechanism to bring fixed comoving scales within the Hubble
radius as we go back in time, allowing us to consider the possibility
of a causal generation mechanism of fluctuations. Starting with
vacuum fluctuations, a matter-dominated contracting phase is
required in order to obtain a scale-invariant spectrum. This corresponds
to the requirement in inflationary cosmology that the accelerated expansion be
nearly exponential.

With Einstein gravity and matter satisfying the usual energy conditions
it is not possible to obtain a non-singular bounce. However, as
mentioned before, it is unreasonable to expect that Einstein gravity
will provide a good description of the physics near the bounce.
There are a large number of ways to obtain a non-singular
bouncing cosmology. To mention but a few, it has been
shown that a bouncing cosmology results naturally from the
special ghost-free higher derivative gravity Lagrangian 
introduced in \cite{Biswas1}. Bounces also arise in the higher-derivative
non-singular universe construction of \cite{MBS}, in ``mirage cosmology"
(see e.g. \cite{BFS}) which is the cosmology on a brane moving through
a curved higher-dimensional bulk space-time (the time dependence for
a brane observer is induced by the motion through the bulk), or - within
the context of Einstein gravity - by making use of ``quintom matter" (matter
consisting of two components, one with regular kinetic term in the action,
the other one with opposite sign kinetic action) \cite{Cai1}. For an
in-depth review of ways of obtaining bouncing cosmologies see
\cite{Novello}. Very recently, it has been shown \cite{HLbounce} 
that a bouncing cosmology
can easily emerge from Horava-Lifshitz gravity \cite{Horava}, a
new approach to quantizing gravity.

In the matter bounce scenario the universe begins cold and therefore 
large. Thus, the size problem of Standard Cosmology does not arise.
As is obvious from Figure \ref{bounce}, there is no horizon problem
for the matter bounce scenario as long as the contracting period is long
(to be specific, of similar duration as the post-bounce expanding
phase until the present time). By the same argument, it is possible
to have a causal mechanism for generating the primordial
cosmological perturbations which evolve into the structures we observe
today. Specifically, as will be discussed in the section of the matter
bounce scenario, if the fluctuations originate as vacuum perturbations
on sub-Hubble scales in the contracting phase, then the
resulting spectrum at late times for scales exiting the Hubble
radius in the matter-dominated phase of contraction is
scale-invariant \cite{Wands,FB2,Wands2}. The propagation
of infrared (IR) fluctuations through the non-singular bounce
was analyzed in the case of the higher derivative gravity model
of \cite{Biswas1} in \cite{ABB}, in mirage cosmology in \cite{BFS},
in the case of the quintom bounce in \cite{Cai2,LWbounce} and for a Horava-Lifshitz
bounce in \cite{HLbounce2}. The result of these studies is that the
scale-invariance of the spectrum before the bounce goes persists
after the bounce as long as the time period of the bounce phase is
short compared to the wavelengths of the modes being considered.
 Note that if the fluctuations have a thermal origin, then the condition
 on the background cosmology to yield scale-invariance of the
 spectrum of fluctuations is different \cite{Thermalflucts}.
 
\subsection{String Gas Cosmology as a Solution}

\begin{figure}
\includegraphics[height=6cm]{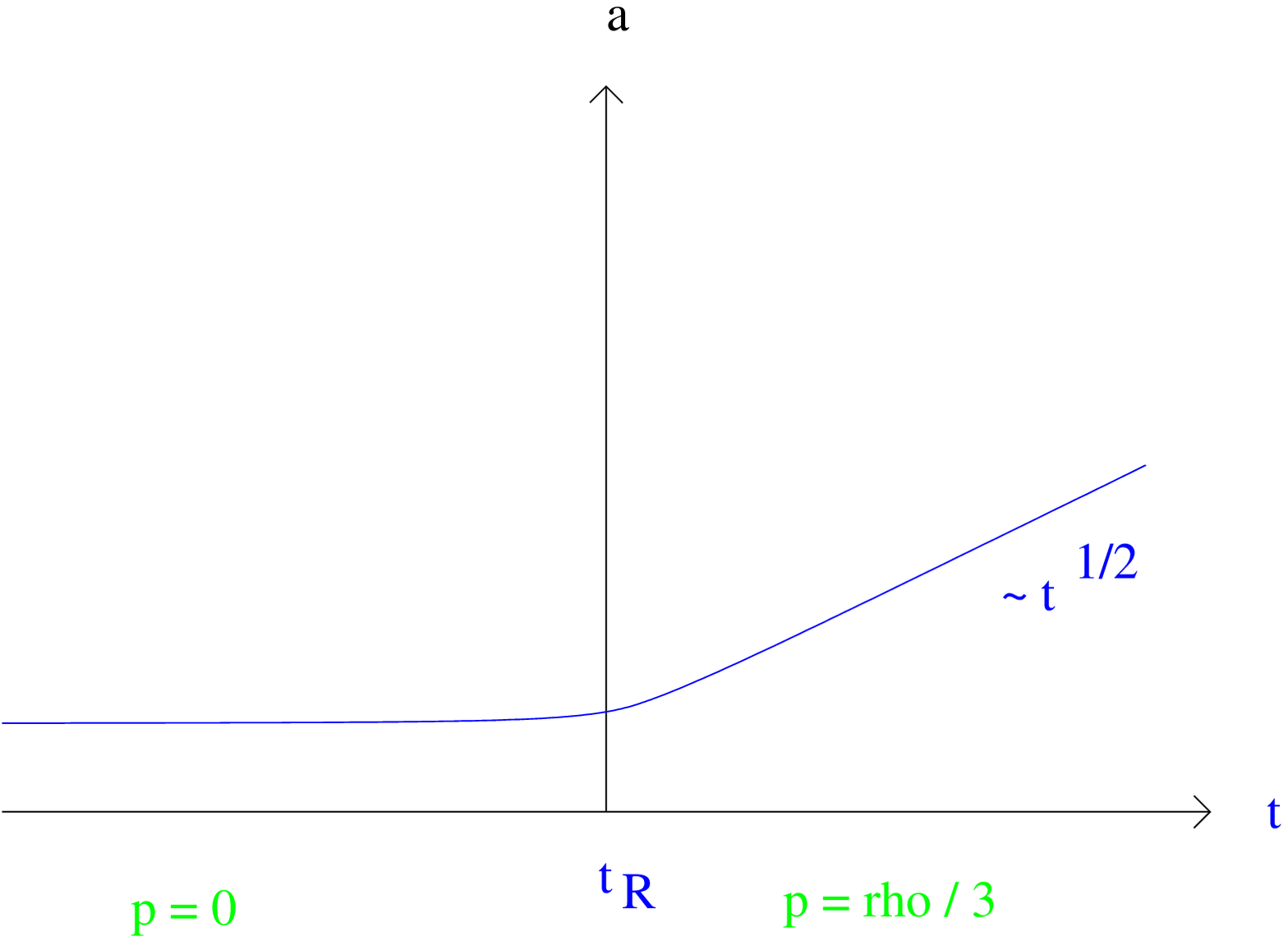}
\caption{The dynamics of string gas cosmology. The vertical axis
represents the scale factor of the universe, the horizontal axis is
time. Along the horizontal axis, the approximate equation of state
is also indicated. During the Hagedorn phase the pressure is negligible
due to the cancellation between the positive pressure of the momentum
modes and the negative pressure of the winding modes, after time $t_R$
the equation of state is that of a radiation-dominated universe.}
 \label{timeevol}
\end{figure}

String gas cosmology \cite{BV} (see also \cite{Perlt}, and see 
\cite{RHBSGrev}  for a comprehensive review) is
a toy model of cosmology which results from coupling a
gas of fundamental strings to a background space-time metric.
It is assumed that the spatial sections are compact.
It is argued that the universe starts in a quasi-static
phase during which the temperature of the string gas
hovers at the Hagedorn value \cite{Hagedorn}, the
maximal temperature of a gas of closed strings in
thermal equilibrium. The string gas in this early
phase is dominated by strings winding the compact
spatial sections. The annihilation
of winding strings will produce string loops and lead
to a transition from the early quasi-static phase to the
radiation phase of Standard Cosmology. Fig. \ref{timeevol}
shows a sketch of the evolution of the scale factor in
string gas cosmology. 

\begin{figure} 
\includegraphics[height=10cm]{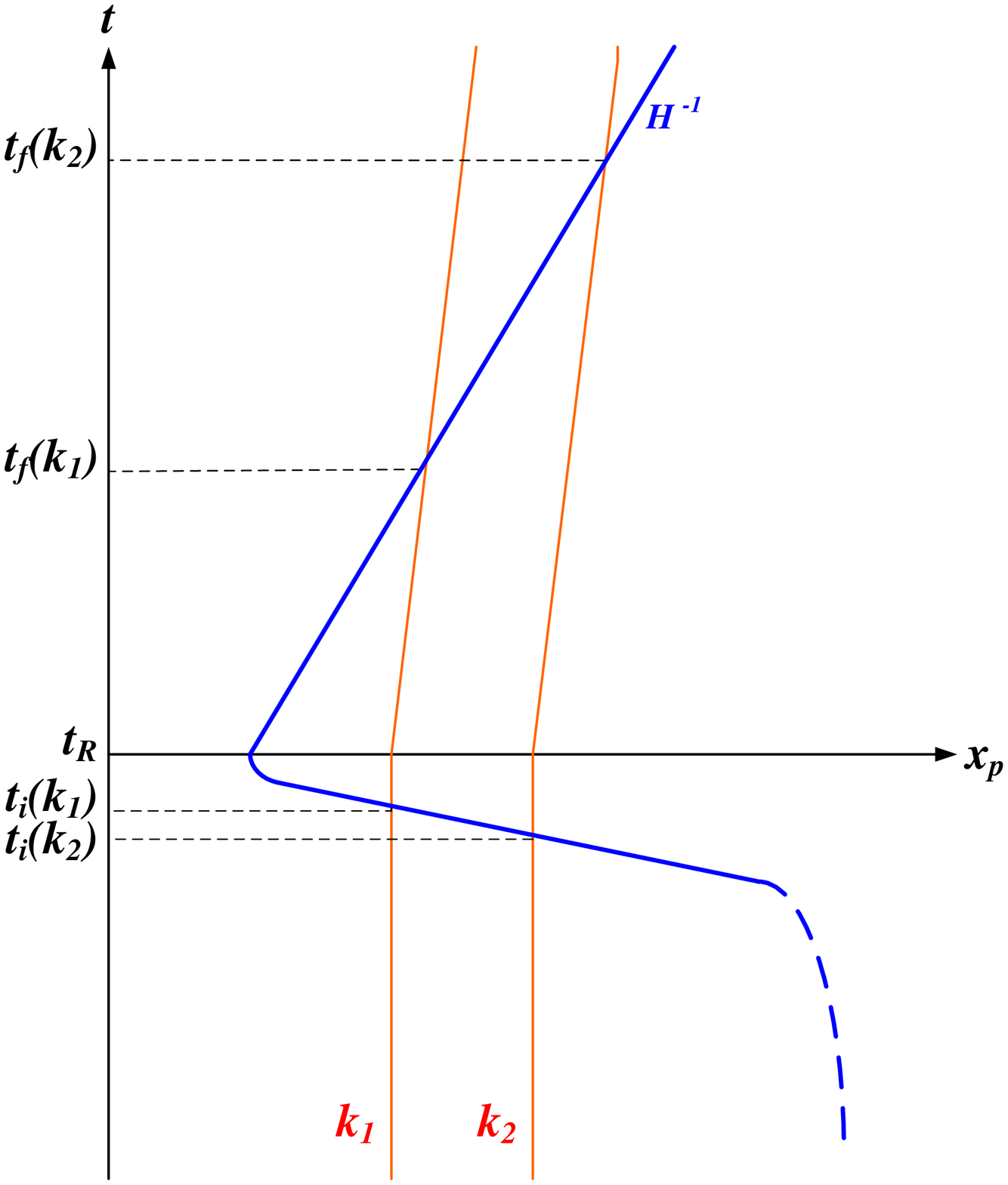}
\caption{Space-time diagram (sketch) showing the evolution of fixed 
co-moving scales in string gas cosmology. The vertical axis is time, 
the horizontal axis is physical distance.  
The solid curve represents the Einstein frame Hubble radius 
$H^{-1}$ which shrinks abruptly to a micro-physical scale at $t_R$ and then 
increases linearly in time for $t > t_R$. Fixed co-moving scales (the 
dotted lines labeled by $k_1$ and $k_2$) which are currently probed 
in cosmological observations have wavelengths which are smaller than 
the Hubble radius before $t_R$. They exit the Hubble 
radius at times $t_i(k)$ just prior to $t_R$, and propagate with a 
wavelength larger than the Hubble radius until they reenter the 
Hubble radius at times $t_f(k)$.}
\label{spacetimenew}
\end{figure}

In Figure \ref{spacetimenew} we sketch the space-time diagram
in string gas cosmology. Since the early Hagedorn phase is
quasi-static, the Hubble radius is infinite. For the same reason,
the physical wavelength of fluctuations remains constant in
this phase. At the end of the Hagedorn phase, the Hubble radius
decreases to a microscopic value and makes a transition to
its evolution in Standard Cosmology.

Once again, we see that fluctuations originate on sub-Hubble scales.
In string gas cosmology, it is the existence of a quasi-static phase
which leads to this result. As will be discussed in the section on
string gas cosmology, the source of perturbations in string gas
cosmology is thermal: string thermodynamical fluctuations in
a compact space with stable winding modes in fact leads to
a scale-invariant spectrum \cite{NBV}.

\section{Theory of Cosmological Perturbations}

The key tool which is used in modern cosmology to connect
theories of the very early universe with cosmological
observations today is the theory of cosmological perturbations.
In the following, we will give an overview of this theory
(similar to the overview in \cite{RHBrev1} which in turn is
based on the comprehensive review \cite{MFB}). We begin
with the analysis of perturbations in Newtonian cosmology,
a useful exercise to develop intuition and notations.

\subsection{Newtonian Theory}

The growth of density fluctuations is a consequence of the
purely attractive nature of the gravitational force. Imagine (first
in a non-expanding background)
a density excess $\delta \rho$ localized about some point ${\bf x}$ in space.
This fluctuation produces an attractive force which pulls
the surrounding matter towards ${\bf x}$. The magnitude of this
force is proportional to $\delta \rho$. Hence, by Newton's
second law
\begin{equation} \label{rhbeq1}
\ddot{\delta \rho} \, \sim \, G \delta \rho \, ,
\end{equation}
where $G$ is Newton's gravitational constant. Hence, there is
an exponential instability of flat space-time to the development
of fluctuations. 

Obviously, in General Relativity it is inconsistent to consider
density fluctuations in a non-expanding background. If we
consider density fluctuations in an expanding background,
then the expansion of space leads to a friction term in (\ref{rhbeq1}).
Hence, instead of an exponential instability to the development of
fluctuations, fluctuations in an expanding Universe
will grow as a power of time. We will now determine what this power
is and how it depends both on the background cosmological expansion rate
and on the length scale of the fluctuations.

We first consider the evolution of hydrodynamical matter fluctuations in a fixed
non-expanding background.  
In this context, matter is described
by a perfect fluid, and gravity by the Newtonian gravitational
potential $\varphi$. The fluid variables are the energy density
$\rho$, the pressure $p$, the fluid velocity ${\bf v}$, and
the entropy density $s$. The basic hydrodynamical equations
are
\begin{eqnarray} \label{rhbeq2}
\dot{\rho} + \nabla \cdot (\rho {\bf v}) & = & 0 \nonumber \\
\dot{{\bf v}} + ({\bf v} \cdot \nabla) {\bf v} + {1 \over {\rho}} \nabla p
+ \nabla \varphi & = & 0 \nonumber \\
\nabla^2 \varphi & = & 4 \pi G \rho \\
\dot{s} + ({\bf v} \cdot \nabla) s & = & 0 \nonumber \\
p & = & p(\rho, s) \, . \nonumber
\end{eqnarray}
The first equation is the continuity equation, the second is the Euler
(force) equation, the third is the Poisson equation of Newtonian gravity,
the fourth expresses entropy conservation, and the last describes
the equation of state of matter. The derivative with respect to time
is denoted by an over-dot.

The background is given by the energy density $\rho_o$, 
the pressure $p_0$, vanishing velocity, constant gravitational potential
$\varphi_0$ and constant entropy density $s_0$. As mentioned above, it
does not satisfy the background Poisson equation.

The equations for cosmological perturbations are obtained by perturbing
the fluid variables about the background,
\begin{eqnarray} \label{rhbeq3}
\rho & = & \rho_0 + \delta \rho \nonumber \\
{\bf v} & = & \delta {\bf v} \nonumber \\
p & = & p_0 + \delta p \\
\varphi & = & \varphi_0 + \delta \varphi \nonumber \\
s & = & s_0 + \delta s \, , \nonumber
\end{eqnarray}
where the fluctuating fields $\delta \rho, \delta {\bf v}, \delta p,
\delta \varphi$ and $\delta s$ are functions of space and time, by
inserting these expressions into the basic hydrodynamical equations
(\ref{rhbeq2}), by linearizing, and by combining the resulting equations
which are of first order in time. We get the following
differential equations for the energy density fluctuation $\delta \rho$
and the entropy perturbation $\delta s$
\begin{eqnarray} \label{rhbeq4}
\ddot{\delta \rho} - c_s^2 \nabla^2 \delta \rho - 4 \pi G \rho_0 \delta \rho
& = & \sigma \nabla^2 \delta s \\
\dot \delta s \ & = & 0 \, , \nonumber
\end{eqnarray}
where the variables $c_s^2$ and $\sigma$ describe the equation of state
\begin{equation} \label{pressurepert}
\delta p \, = \, c_s^2 \delta \rho + \sigma \delta S
\end{equation}
with
\begin{equation}
c_s^2 \, = \, \bigl({{\delta p} \over {\delta \rho}}\bigr)_{|_S}
\end{equation}
denoting the square of the speed of sound.

The fluctuations can be classified as follows: If 
$\delta s$ vanishes, we have {\bf adiabatic} fluctuations. If
$\delta s$ is non-vanishing but 
$\dot{\delta \rho} = 0$, we speak of an {\bf entropy} fluctuation.

The first conclusions we can draw from the basic perturbation
equations (\ref{rhbeq4}) are that \\
1) entropy fluctuations do not grow, \\
2) adiabatic fluctuations are time-dependent, and \\
3) entropy fluctuations seed  adiabatic ones.\\
All of these conclusions will remain valid in the relativistic theory.

Since the equations are linear, we can work in Fourier space. Each
Fourier component $\delta \rho_k(t)$ of the fluctuation field 
$\delta \rho({\bf x}, t)$ evolves independently. 
In the case of adiabatic fluctuations, the cosmological 
perturbations are described by a single field which obeys a
second order differential equation and hence has two
fundamental solutions. We will see that this conclusion
remains true in the relativistic theory.

Taking a closer look at the equation of motion for
$\delta \rho$, we see that the third term on the left hand side
represents the force due to gravity, a purely attractive force 
yielding an instability of flat space-time to the development of
density fluctuations (as discussed earlier, see (\ref{rhbeq1})).
The second term on the left hand side of (\ref{rhbeq4}) represents
a force due to the fluid pressure which tends to set up pressure waves.
In the absence of entropy fluctuations, the evolution of $\delta \rho$
is governed by the combined action of both pressure and gravitational
forces.

Restricting our attention to adiabatic fluctuations, we see from
(\ref{rhbeq4}) that there is a critical wavelength, the Jeans length,
whose wavenumber $k_J$ is given by
\begin{equation} \label{Jeans}
k_J \, = \, \bigl({{4 \pi G \rho_0} \over {c_s^2}}\bigr)^{1/2} \, .
\end{equation}
Fluctuations with wavelength longer than the Jeans length ($k \ll k_J$)
grow exponentially
\begin{equation} \label{expgrowth}
\delta \rho_k(t) \, \sim \, e^{\omega_k t} \,\, {\rm with} \,\,
\omega_k \sim 4 (\pi G \rho_0)^{1/2}
\end{equation}
whereas short wavelength modes ($k \gg k_J$) oscillate with 
frequency $\omega_k \sim c_s k$. Note that the value of the
Jeans length depends on the equation of state of the background.
For a background dominated by relativistic radiation, the Jeans
length is large (of the order of the Hubble radius $H^{-1}(t)$), 
whereas for pressure-less matter it vanishes.

Let us now improve on the previous analysis and study Newtonian
cosmological fluctuations about an expanding background. In this
case, the background equations are consistent (the non-vanishing
average energy density leads to cosmological expansion). However,
we are still neglecting general relativistic effects (the 
fluctuations of the metric). Such effects
turn out to be dominant on length scales larger than the Hubble
radius $H^{-1}(t)$, and thus the analysis of this section is
applicable only to smaller scales. 

The background cosmological model is given by the energy density
$\rho_0(t)$, the pressure $p_0(t)$, and the recessional velocity
${\bf v}_0 = H(t) {\bf x}$, where ${\bf x}$ is the Euclidean spatial
coordinate vector (``physical coordinates''). The space- and time-dependent
fluctuating fields are defined as in the previous section:
\begin{eqnarray} \label{fluctansatz2}
\rho(t, {\bf x}) & = & \rho_0(t) \bigl(1 + \delta_{\epsilon}(t, {\bf x}) 
\bigr)\nonumber \\
{\bf v}(t, {\bf x}) & = & {\bf v}_0(t, {\bf x}) + \delta {\bf v}(t, {\bf x}) 
\\
p(t, {\bf x}) & = & p_0(t) + \delta p(t, {\bf x}) \, , \nonumber 
\end{eqnarray}
where $\delta_{\epsilon}$ is the fractional energy density perturbation
(we are interested in the fractional rather than in the absolute energy
density fluctuation!), and the pressure perturbation $\delta p$ is
defined as in (\ref{pressurepert}). In addition, there is the
possibility of a non-vanishing entropy perturbation defined as in
(\ref{rhbeq3}).

We now insert this ansatz into the basic hydrodynamical equations 
(\ref{rhbeq2}), linearize in the perturbation variables, and combine
the first order differential equations 
for $\delta_{\epsilon}$ and $\delta p$ into a single second order
differential equation for $\delta \rho_{\epsilon}$. The result simplifies
if we work in ``comoving coordinates'' ${\bf q}$ which are the coordinates
painted onto the expanding background, i.e. 
${\bf x}(t) \, = \, a(t) {\bf q}(t) \, .$
After some algebra, we obtain the following equation
which describes the time evolution of density fluctuations:
\begin{equation} \label{Newtoneq}
\ddot{\delta_{\epsilon}} + 2 H \dot{\delta_{\epsilon}} 
- {{c_s^2} \over {a^2}} \nabla_q^2 \delta_{\epsilon} 
- 4 \pi G \rho_0 \delta_{\epsilon} \, 
= \, {{\sigma} \over {\rho_0 a^2}} \delta S \, ,
\end{equation}
where the subscript $q$ on the $\nabla$ operator indicates that derivatives
with respect to comoving coordinates are used.
In addition, we have the equation of entropy conservation
$\dot{\delta S} \, = \, 0 \, .$

Comparing with the equations (\ref{rhbeq4}) obtained in the absence of
an expanding background, we see that the only difference is the presence
of a Hubble damping term in the equation for $\delta_{\epsilon}$. This
term will moderate the exponential instability of the background to
long wavelength density fluctuations. In addition, it will lead to a
damping of the oscillating solutions on short wavelengths. More specifically,
for physical wavenumbers $k_p \ll k_J$ (where $k_J$ is again given by
(\ref{Jeans})), and in a matter-dominated background cosmology, the
general solution of (\ref{Newtoneq}) in the absence of any entropy
fluctuations is given by
\begin{equation} \label{Newtonsol}
\delta_k(t) \, = \, c_1 t^{2/3} + c_2 t^{-1} \, ,
\end{equation}
where $c_1$ and $c_2$ are two constants determined by the initial
conditions, and we have dropped the subscript $\epsilon$ in expressions
involving $\delta_{\epsilon}$. 
There are two fundamental solutions, the first is a
growing mode with $\delta_k(t) \sim a(t)$, the second a decaying
mode with $\delta_k(t) \sim t^{-1}$.
On short wavelengths, one obtains damped oscillatory motion:
\begin{equation} \label{Newtonsolosc}
\delta_k(t) \, \sim \, a^{-1/2}(t) exp \bigl( \pm i c_s k \int dt' a^{-1}(t')
\bigr) \, .
\end{equation}

The above analysis applies for $t > t_{eq}$, when we are following the
fluctuations of the dominant component of matter. In the radiation era,
cold matter fluctuations only grow logarithmically in time since the
dominant component of matter (the relativistic radiation) does not
cluster and hence the term in the equation of motion for fluctuations
which represents the gravitational attraction force is suppressed.
In this sense, we can say that inhomogeneities begin to cluster
at the time $t_{eq}$.

As a simple application of the Newtonian equations for cosmological
perturbations derived above, let us compare the predicted cosmic
microwave background (CMB) anisotropies in a spatially
flat universe with only baryonic matter - Model A -
to the corresponding anisotropies
in a flat Universe with mostly cold dark matter (pressure-less non-baryonic
dark matter) - Model B. We start with the observationally known amplitude
of the relative density fluctuations today (time $t_0$), 
and we use the fact that
the amplitude of the CMB anisotropies on the angular scale $\theta(k)$
corresponding to the comoving wavenumber $k$ is set by the primordial
value of the gravitational potential $\phi$ - introduced in the
following section - which in turn is related to the primordial value of the
density fluctuations at Hubble radius crossing (and {\bf not}
to its value at the time $t_{rec}$ - see e.g. Chapter 17 of \cite{MFB}).

In Model A, the dominant component of the pressure-less matter is
coupled to radiation between $t_{eq}$ and $t_{rec}$, the time of
last scattering. Thus, the Jeans length is comparable to the Hubble
radius. Therefore, for comoving galactic scales, $k \gg k_J$ in this
time interval, and thus the fractional density contrast decreases
as $a(t)^{-1/2}$. In contrast, in Model B, the dominant component of
pressure-less matter couples only weakly to radiation, and hence
the Jeans length is negligibly small. Thus, in Model B, the
relative density contrast grows as $a(t)$ between $t_{eq}$ and $t_{rec}$.
In the time interval $t_{rec} < t < t_0$, the fluctuations scale
identically in Models A and B. Summarizing,
we conclude, working backwards in time from a fixed amplitude
of $\delta_k$ today, that the amplitudes of $\delta_k(t_{eq})$ in Models
A and B (and thus their primordial values) are related by
\begin{equation}
\delta_k(t_{eq})|_{A} \, \simeq \, 
\bigl({{a(t_{rec})} \over {a(t_{eq})}} \bigr)^{3/2} \delta_k(t_{eq})|_{B} \, .
\end{equation}
Hence, in Model A (without non-baryonic dark matter) the CMB anisotropies
are predicted to be a factor of about 30 larger 
\cite{SW} than in Model B, way
in excess of the recent observational results. This is one of the
strongest arguments for the existence of non-baryonic dark matter
\footnote{In my opinion no proposed alternative to particle
dark matter should be taken serious unless their proponents can
demonstrate that their model can reproduce the agreement between
the amplitudes of the CMB anisotropies on one hand and of the
large-scale matter power spectrum on the other hand.}. Note that
the precise value of the enhancement factor depends on the value of the
cosmological constant $\Lambda$ - 
the above result holds for $\Lambda = 0$.

\subsection{Observables}

Let us consider perturbations on a fixed
comoving length scale given by a comoving wavenumber $k$. 
The corresponding physical length increases
as $a(t)$. This is to be compared with the Hubble radius $H^{-1}(t)$
which scales as $t$ provided $a(t)$ grows as a power of $t$. In
the late time Universe, $a(t) \sim t^{1/2}$ in the radiation-dominated
phase (i.e. for $t < t_{eq}$), and $a(t) \sim t^{2/3}$ in the 
matter-dominated period ($t_{eq} < t < t_0$). 
Thus, we see that at sufficiently early times, all comoving scales
had a physical length larger than the Hubble radius. If we consider 
large cosmological scales (e.g. those corresponding to the observed
CMB anisotropies or to galaxy clusters), the time $t_H(k)$ of 
``Hubble radius crossing'' (when the physical length was equal to the
Hubble radius) was in fact later than $t_{eq}$.

Cosmological fluctuations can be described either in position space
or in momentum space. In position space, we compute the root mean
square mass fluctuation $\delta M / M(k, t)$ in a sphere of radius
$l = 2 \pi / k$ at time $t$. A scale-invariant spectrum of fluctuations is
defined by the relation
\begin{equation} \label{scaleinv}
{{\delta M} \over M}(k, t_H(k)) \, = \, {\rm const.} \, .
\end{equation}
Such a spectrum was first suggested by Harrison \cite{Harrison}
and Zeldovich \cite{Zeldovich} as a reasonable choice for the
spectrum of cosmological fluctuations. We can introduce the ``spectral
index'' $n$ of cosmological fluctuations by the relation
\begin{equation} \label{specindex}
\bigl({{\delta M} \over M}\bigr)^2(k, t_H(k)) \, \sim \, k^{n - 1} \, ,
\end{equation}
and thus a scale-invariant spectrum corresponds to $n = 1$.

To make the transition to the (more frequently used) momentum space
representation, we Fourier decompose the fractional spatial density
contrast
\begin{equation} \label{Fourier}
\delta_{\epsilon}({\bf x}, t) \, = \, V^{1/2}
\int d^3k {\tilde{\delta_{\epsilon}}}({\bf k}, t) e^{i {\bf k} \cdot {\bf x}}
\, ,
\end{equation}
where $V$ is a cutoff volume.
The {\bf power spectrum} $P_{\delta}$ of density fluctuations is defined by
\begin{equation} \label{densspec}
P_{\delta}(k) \, = \, k^3 |{\tilde{\delta_{\epsilon}}}(k)|^2 \, ,
\end{equation}
where $k$ is the magnitude of ${\bf k}$, and we have assumed for simplicity
a Gaussian distribution of fluctuations in which the amplitude of the
fluctuations only depends on $k$.

We can also define the power spectrum of the gravitational potential $\varphi$:
\begin{equation} \label{gravspec}
P_{\varphi}(k) \, = \, k^3 |{\tilde{\delta \varphi}}(k)|^2 \, .
\end{equation}
These two power spectra are related by the Poisson equation (\ref{rhbeq2})
\begin{equation} \label{relspec}
P_{\varphi}(k) \, \sim \, k^{-4} P_{\delta}(k) \, .
\end{equation}

In general, the condition of scale-invariance is expressed in momentum
space in terms of the power spectrum evaluated at a fixed time. To obtain
this condition, we first use the time dependence of the 
fractional density fluctuation from (\ref{Newtonsol}) to determine
the mass fluctuations at a fixed time $t > t_H(k) > t_{eq}$ (the last
inequality is a condition on the scales considered)
\begin{equation} \label{timerel}
\bigl({{\delta M} \over M}\bigr)^2(k, t) \, = \,
\bigl({t \over {t_H(k)}}\bigr)^{4/3} 
\bigl({{\delta M} \over M}\bigr)^2(k, t_H(k)) \, .
\end{equation}
The time of Hubble radius crossing is given by
\begin{equation} \label{Hubble}
a(t_H(k)) k^{-1} \, = \, 2 t_H(k) \, ,
\end{equation}
and thus
\begin{equation} \label{Hubble2}
t_H(k)^{1/3} \, \sim \, k^{-1} \, .
\end{equation}
Inserting this result into (\ref{timerel}) and making use of (\ref{specindex})
we find 
\begin{equation} \label{spec2}
\bigl({{\delta M} \over M}\bigr)^2(k, t) \, \sim \, k^{n + 3} \, .
\end{equation}
Since, for reasonable values of the index of the power spectrum, 
$\delta M / M (k, t)$ is dominated by the Fourier modes with
wavenumber $k$, we find that (\ref{spec2}) implies
\begin{equation} \label{spec3}
|{\tilde{\delta_{\epsilon}}}|^2 \, \sim \, k^{n} \, ,
\end{equation}
or, equivalently,
\begin{equation} \label{spec4}
P_\varphi(k) \, \sim \, k^{n - 1} \, .
\end{equation}

\subsection{Classical Relativistic Theory}

\subsubsection{Introduction}

The Newtonian theory of cosmological fluctuations discussed in the
previous section breaks down on scales larger than the Hubble radius
because it neglects perturbations of the metric, and because on large
scales the metric fluctuations dominate the dynamics.

Let us begin with a heuristic argument to show why metric fluctuations
are important on scales larger than the Hubble radius. For such
inhomogeneities, one should be able to approximately describe the
evolution of the space-time by applying the first 
FRW equation (\ref{FRW1}) of homogeneous
and isotropic cosmology to the local Universe (this approximation is
made more rigorous in \cite{Afshordi}).
Based on this equation, a large-scale fluctuation of the
energy density will lead to a fluctuation (``$\delta a$'') of
the scale factor $a$ which grows in time. This is due to the fact
that self gravity amplifies fluctuations even on length scales $\lambda$
greater than the Hubble radius.

This argument is made rigorous in the following analysis of cosmological
fluctuations in the context of general relativity, where both metric
and matter inhomogeneities are taken into account. We will consider
fluctuations about a homogeneous and isotropic background cosmology,
given by the metric (\ref{metric}), which can be written in
conformal time $\eta$ (defined by $dt = a(t) d\eta$) as
\begin{equation} \label{background2}
ds^2 \, = \, a(\eta)^2 \bigl( d\eta^2 - d{\bf x}^2 \bigr) \, .
\end{equation}

The theory of cosmological perturbations is based on expanding the Einstein
equations to linear order about the background metric. The theory was
initially developed in pioneering works by Lifshitz \cite{Lifshitz}. 
Significant progress in the understanding of the physics of cosmological
fluctuations was achieved by Bardeen \cite{Bardeen} who realized the
importance of subtracting gauge artifacts (see below) from the
analysis (see also \cite{PV,BKP}). The following discussion is based on
Part I of the comprehensive review article \cite{MFB}. Other reviews - in
some cases emphasizing different approaches - are 
\cite{Kodama,Ellis,Hwang,Durrer}.

\subsubsection{Classifying Fluctuations}

The first step in the analysis of metric fluctuations is to
classify them according to their transformation properties
under spatial rotations. There are scalar, vector and second rank
tensor fluctuations. In linear theory, there is no coupling
between the different fluctuation modes, and hence they evolve
independently (for some subtleties in this classification, see
\cite{Stewart}). 

We begin by expanding the metric about the FRW background metric
$g_{\mu \nu}^{(0)}$ given by (\ref{background2}):
\begin{equation} \label{pertansatz}
g_{\mu \nu} \, = \, g_{\mu \nu}^{(0)} + \delta g_{\mu \nu} \, .
\end{equation}
The background metric depends only on time, whereas the metric
fluctuations $\delta g_{\mu \nu}$ depend on both space and time.
Since the metric is a symmetric tensor, there are at first sight
10 fluctuating degrees of freedom in $\delta g_{\mu \nu}$.

There are four degrees of freedom which correspond to scalar metric
fluctuations (the only four ways of constructing a metric from
scalar functions):
\begin{equation} \label{scalarfl}
\delta g_{\mu \nu} \, = \, a^2 \left(
\begin{array} {cc}
2 \phi & -B_{,i} \\
-B_{,i} & 2\bigl(\psi \delta_{ij} - E_{,ij} \bigr) 
\end{array}
\right) \, ,
\end{equation}
where the four fluctuating degrees of freedom are denoted (following
the notation of \cite{MFB}) $\phi, B, E$, and $\psi$, a comma denotes
the ordinary partial derivative (if we had included spatial curvature
of the background metric, it would have been the covariant derivative
with respect to the spatial metric), and $\delta_{ij}$ is the Kronecker
symbol.

There are also four vector degrees of freedom of metric fluctuations,
consisting of the four ways of constructing metric fluctuations from
three vectors:
\begin{equation} \label{vectorfl}
\delta g_{\mu \nu} \, = \, a^2 \left(
\begin{array} {cc}
0 & -S_i \\
-S_i & F_{i,j} + F_{j,i} 
\end{array}
\right) \, ,
\end{equation}
where $S_i$ and $F_i$ are two divergence-less vectors (for a vector
with non-vanishing divergence, the divergence contributes to the
scalar gravitational fluctuation modes).

Finally, there are two tensor modes which correspond to the two
polarization states of gravitational waves:
\begin{equation} \label{tensorfl}
\delta g_{\mu \nu} \, = \, -a^2 \left(
\begin{array} {cc}
0 & 0 \\
0 & h_{ij} 
\end{array}
\right) \, ,
\end{equation}
where $h_{ij}$ is trace-free and divergence-less
\begin{equation}
h_i^i \, = \, h_{ij}^j \, = \, 0 \, .
\end{equation}

Gravitational waves do not couple at linear order to the matter
fluctuations. Vector fluctuations decay in an expanding background
cosmology and hence are not usually cosmologically important.
The most important fluctuations, at least in inflationary cosmology,
are the scalar metric fluctuations, the fluctuations which couple
to matter inhomogeneities and which are the relativistic generalization
of the Newtonian perturbations considered in the previous section.
Note that vector and tensor perturbations are important 
(see e.g. \cite{PST}) in topological defects models 
of structure formation and in bouncing cosmologies \cite{BB}.

\subsubsection{Gauge Transformation}

The theory of cosmological perturbations is at first sight complicated
by the issue of gauge invariance (at the final stage, however, we will
see that we can make use of the gauge freedom to substantially simplify
the theory). The coordinates $t, {\bf x}$ of space-time carry no
independent physical meaning. They are just labels to designate points
in the space-time manifold. By performing a small-amplitude 
transformation of the space-time coordinates
(called ``gauge transformation'' in the following), we can easily
introduce ``fictitious'' fluctuations in a homogeneous and isotropic
Universe. These modes are ``gauge artifacts''.

We will in the following take an ``active'' view of gauge transformation.
Let us consider two space-time manifolds, one of them a homogeneous
and isotropic Universe ${\cal M}_0$, the other a physical Universe 
${\cal M}$ with inhomogeneities. A choice of coordinates can be considered
to be a mapping ${\cal D}$ between the manifolds ${\cal M}_0$ and ${\cal M}$.
Let us consider a second mapping ${\tilde{\cal D}}$ which will take the
same point (e.g. the origin of a fixed coordinate system) in ${\cal M}_0$
to a different point in ${\cal M}$. Using the inverse of these maps
${\cal D}$ and ${\tilde{\cal D}}$, we can assign two different sets of
coordinates to points in ${\cal M}$. 

Consider now a physical quantity $Q$ (e.g. the Ricci scalar)
on ${\cal M}$, and the corresponding
physical quantity $Q^{(0)}$ on ${\cal M}_0$ Then, in the first coordinate
system given by the mapping ${\cal D}$, the perturbation $\delta Q$ of
$Q$ at the point $p \in {\cal M}$ is defined by
\begin{equation}
\delta Q(p) \, = \, Q(p) - Q^{(0)}\bigl({\cal D}^{-1}(p) \bigr) \, .
\end{equation}
Analogously, in the second coordinate system given by ${\tilde{\cal D}}$,
the perturbation is defined by
\begin{equation}
{\tilde{\delta Q}}(p) \, = \, Q(p) - 
Q^{(0)}\bigl({\tilde{{\cal D}}}^{-1}(p) \bigr) \, .
\end{equation}
The difference
\begin{equation}
\Delta Q(p) \, = \, {\tilde{\delta Q}}(p) - \delta Q(p)
\end{equation}
is obviously a gauge artifact and carries no physical significance.

Some of the degrees of freedom of metric perturbation introduced in
the first subsection are gauge artifacts. To isolate these, 
we must study how coordinate transformations act on the metric.
There are four independent gauge degrees of freedom corresponding
to the coordinate transformation
\begin{equation}
x^{\mu} \, \rightarrow \, {\tilde x}^{\mu} = x^{\mu} + \xi^{\mu} \, .
\end{equation}
The zero (time) component $\xi^0$ of $\xi^{\mu}$ leads to a scalar metric
fluctuation. The spatial three vector $\xi^i$ can be decomposed as
\begin{equation}
\xi^i \, = \, \xi^i_{tr} + \gamma^{ij} \xi_{,j}
\end{equation}
(where $\gamma^{ij}$ is the spatial background metric)
into a transverse piece $\xi^i_{tr}$ which has two degrees of freedom
which yield vector perturbations, and the second term (given by
the gradient of a scalar $\xi$) which gives
a scalar fluctuation. To summarize this paragraph, there are
two scalar gauge modes given by $\xi^0$ and $\xi$, and two vector
modes given by the transverse three vector $\xi^i_{tr}$. Thus,
there remain two physical scalar and two vector
fluctuation modes. The gravitational waves are gauge-invariant. 

Let us now focus on how the scalar gauge transformations (i.e. the
transformations given by $\xi^0$ and $\xi$) act on the scalar
metric fluctuation variables $\phi, B, E$, and $\psi$. An immediate
calculation yields:
\begin{eqnarray}
{\tilde \phi} \, &=& \, \phi - {{a'} \over a} \xi^0 - (\xi^0)^{'} \nonumber \\
{\tilde B} \, &=& \, B + \xi^0 - \xi^{'} \\
{\tilde E} \, &=& \, E - \xi \nonumber \\
{\tilde \psi} \, &=& \, \psi + {{a'} \over a} \xi^0 \, , \nonumber
\end{eqnarray}
where a prime indicates the derivative with respect to conformal time $\eta$.

There are two approaches to deal with the gauge ambiguities. The first is
to fix a gauge, i.e. to pick conditions on the coordinates which
completely eliminate the gauge freedom, the second is to work with a
basis of gauge-invariant variables.
If one wants to adopt the gauge-fixed approach, there are many
different gauge choices. Note that the often used synchronous gauge
determined by $\delta g^{0 \mu} = 0$ does not totally fix the
gauge. A convenient system which completely fixes the coordinates
is the so-called {\bf longitudinal} or {\bf conformal Newtonian gauge}
defined by $B = E = 0$. This is the gauge we will use in the
following.

Note that the above analysis was strictly at linear order in 
perturbation theory.  Beyond linear order, the structure of 
perturbation theory becomes much
more involved. In fact, one can show \cite{SteWa} that the only
fluctuation variables which are invariant under all coordinate
transformations are perturbations of variables which are constant
in the background space-time.

\subsubsection{Equations of Motion}
\label{flucteom}

We begin with the Einstein equations
\begin{equation} \label{Einstein}
G_{\mu\nu} \, = \, 8 \pi G T_{\mu\nu} \, , 
\end{equation}
where $G_{\mu\nu}$ is the Einstein tensor associated with the space-time
metric $g_{\mu\nu}$, and $T_{\mu\nu}$ is the energy-momentum tensor of matter,
insert the ansatz for metric and matter perturbed about a FRW 
background $\bigl(g^{(0)}_{\mu\nu}(\eta) ,\, \varphi^{(0)}(\eta)\bigr)$:
\begin{eqnarray} \label{pertansatz2}
g_{\mu\nu} ({\bf x}, \eta) & = & g^{(0)}_{\mu\nu} (\eta) + \delta g_{\mu\nu}
({\bf x}, \eta) \\
\varphi ({\bf x}, \eta) & = & \varphi_0 (\eta) + \delta \varphi
({\bf x}, \eta) \, , 
\end{eqnarray}
(where we have for simplicity replaced general matter by a scalar
matter field $\varphi$)
and expand to linear order in the fluctuating fields, obtaining the
following equations:
\begin{equation} \label{linein}
\delta G_{\mu\nu} \> = \> 8 \pi G \delta T_{\mu\nu} \, .
\end{equation}
In the above, $\delta g_{\mu\nu}$ is the perturbation in the metric and $\delta
\varphi$ is the fluctuation of the matter field $\varphi$.

Inserting the longitudinal gauge metric
\begin{equation} \label{longit}
ds^2 \, = \, a^2 \bigl[(1 + 2 \phi) d\eta^2
- (1 - 2 \psi)\gamma_{ij} dx^i dx^j \bigr] \,
\end{equation}
into the general perturbation equations
(\ref{linein}) yields the following set of equations of motion:
\begin{eqnarray} \label{perteom1}
- 3 {\cal H} \bigl( {\cal H} \phi + \psi^{'} \bigr) + \nabla^2 \psi \,
&=& \, 4 \pi G a^2 \delta T^{ 0}_0 \nonumber \\
\bigl( {\cal H} \phi + \psi^{'} \bigr)_{, i} \,
&=& 4 \pi G a^2 \delta T^{0}_i \nonumber \\
\bigl[ \bigl( 2 {\cal H}^{'} + {\cal H}^2 \bigr) \phi + {\cal H} \phi^{'}
+ \psi^{''} + 2 {\cal H} \psi^{'} \bigr] \delta^i_j &&  \\
+ {1 \over 2} \nabla^2 D \delta^i_j - {1 \over 2} \gamma^{ik} D_{, kj} \,
&=& - 4 \pi G a^2 \delta T^{i}_j \, , \nonumber
\end{eqnarray}
where $D \equiv \phi - \psi$ and ${\cal H} = a'/a$. 

The first conclusion we can draw is that if no anisotropic stress
is present in the matter at linear order in fluctuating fields, i.e.
$\delta T^i_j = 0$ for $i \neq j$, then the two metric fluctuation
variables coincide:
\begin{equation} \label{constr}
\phi \, = \, \psi \, .
\end{equation}
This will be the case in most simple cosmological models, e.g. in
theories with matter described by a set of scalar fields with
canonical form of the action, and in the case of a perfect fluid
with no anisotropic stress.

Let us now restrict our attention to the case of matter described
in terms of a single scalar field $\varphi$ (in which case the
perturbations on large scales are automatically adiabatic)
which can be  expanded as
\begin{equation} \label{mfexp}
\varphi ({\bf x}, \eta) \, = \, \varphi_0(\eta) + \delta \varphi({\bf x}, \eta)
\end{equation}
in terms of background matter $\varphi_0$ and matter fluctuation 
$\delta \varphi({\bf x}, \eta)$. Then, in longitudinal gauge,
(\ref{perteom1}) reduce to the following
set of equations of motion
\begin{eqnarray} \label{perteom2}
\nabla^2 \phi - 3 {\cal H} \phi^{'} - 
\bigl( {\cal H}^{'} + 2 {\cal H}^2 \bigr) \phi \, &=& \, 
4 \pi G \bigl( \varphi^{'}_0 \delta \varphi^{'} + 
V^{'} a^2 \delta \varphi \bigr) \nonumber \\
\phi^{'} + {\cal H} \phi \, &=& \, 4 \pi G \varphi^{'}_0 \delta \varphi \\
\phi^{''} + 3 {\cal H} \phi^{'} + 
\bigl( {\cal H}^{'} + 2 {\cal H}^2 \bigr) \phi \, &=& \, 
4 \pi G \bigl( \varphi^{'}_0 \delta \varphi^{'} - 
V^{'} a^2 \delta \varphi \bigr) \, , \nonumber
\end{eqnarray}
where $V^{'}$ denotes the derivative of $V$ with respect to $\varphi$.
These equations can be combined to give the following second order
differential equation for the relativistic potential $\phi$:
\begin{equation} \label{finaleom}
\phi^{''} + 2 \left( {\cal H} - 
{{\varphi^{''}_0} \over {\varphi^{'}_0}} \right) \phi^{'} - \nabla^2 \phi
+ 2 \left( {\cal H}^{'} - 
{\cal H}{{\varphi^{''}_0} \over {\varphi^{'}_0}} \right) \phi \, = \, 0 \, .
\end{equation}

This is the final result for the classical evolution of
cosmological fluctuations. First of all, we note the similarities with
the equation (\ref{Newtoneq}) obtained in the Newtonian theory.
The final term in (\ref{finaleom}) is the force due to gravity leading
to the instability, the second to last term is the pressure force
leading to oscillations (relativistic since we are considering matter
to be a relativistic field), and the second term represents the Hubble 
friction. For each wavenumber there are two fundamental solutions. On
small scales ($k > H$), the solutions correspond to damped oscillations,
on large scales ($k < H$) the oscillations freeze out and the dynamics
is governed by the gravitational force competing with the Hubble friction
term. Note, in particular, how the Hubble radius naturally emerges as
the scale where the nature of the fluctuating modes changes from oscillatory
to frozen.

Considering the equation in a bit more detail, we observe that if the
equation of state of the background is independent of time (which will be
the case if ${\cal H}^{'} = \varphi^{''}_0 = 0$), then in an
expanding background, the dominant mode of (\ref{finaleom}) is constant,
and the sub-dominant mode decays. If the equation of state is not constant,
then the dominant mode is not constant in time. Specifically, at the
end of inflation ${\cal H}^{'} < 0$, and this leads to a growth of 
$\phi$ (see the following subsection). In contrast, in a contracting
phase the dominant mode of $\phi$ is growing.

To study the quantitative implications of the equation of motion
(\ref{finaleom}), it is convenient to introduce \cite{BST,BK}
the variable $\zeta$ (which, up to a correction term of the order
$\nabla^2 \phi$ which is unimportant for large-scale fluctuations,
is equal to the curvature perturbation ${\cal R}$ in comoving gauge
\cite{Lyth}) by
\begin{equation} \label{zetaeq}
\zeta \, \equiv \, \phi + {2 \over 3} 
{{\bigl(H^{-1} {\dot \phi} + \phi \bigr)} \over { 1 + w}} \, ,
\end{equation}
where
\begin{equation} \label{wvar}
w \, = \, {p \over {\rho}}
\end{equation}
characterizes the equation of state of matter. In terms of $\zeta$,
the equation of motion (\ref{finaleom}) takes on the form
\begin{equation}
{3 \over 2} {\dot \zeta} H (1 + w) \, = \, {\cal O}(\nabla^2 \phi) \, .
\end{equation}
On large scales, the right hand side of the equation is negligible,
which leads to the conclusion that large-scale cosmological fluctuations
satisfy
\begin{equation} \label{zetacons}
{\dot \zeta} (1 + w) \, = \, 0 .
\end{equation}
This implies that $\zeta$ is constant
except possibly if $1 + w = 0$ at some point in time during the cosmological 
evolution (which occurs during reheating in inflationary
cosmology if the inflaton field undergoes oscillations 
\cite{Fabio1}).  In single matter field models it is indeed possible
to show that ${\dot \zeta} = 0$ on super-Hubble scales independent
of assumptions on the equation of state \cite{Weinberg2,Zhang}.
This ``conservation law'' makes it easy to relate
initial fluctuations to final fluctuations in inflationary cosmology,
as will be illustrated in the following subsection.

In the presence of non-adiabatic fluctuations (entropy fluctuations)
there is a source term on the righ-hand side of (\ref{zetacons}) which
is proportional to the amplitude of the entropy fluctuations.
Thus, as already seen in the case of the Newtonian theory of
cosmological fluctuations, non-adiabatic fluctuations
induce a growing adiabatic mode (see \cite{BaVi,Fabio2} for 
discussions of the consequences in double field inflationary models). 

\subsection{Quantum Theory}

\subsubsection{Overview}

As we have seen in the first section, in many models of the very early 
Universe, in particular in inflationary cosmology, in string gas
cosmology and in the matter bounce scenario, 
primordial inhomogeneities are generated in an initial phase
on sub-Hubble scales. The wavelength is then stretched
relative to the Hubble radius, becomes larger than the Hubble
radius at some time and then propagates on super-Hubble scales until
re-entering at late cosmological times. In a majority of the current
structure formation scenarios (string gas cosmology is an exception
in this respect), fluctuations are assumed to emerge as quantum
vacuum perturbations. Hence, to describe the generation and
evolution of the inhomogeneities a quantum treatment is
required.

In the context of a Universe
with a de Sitter phase, the quantum origin of cosmological fluctuations
was first discussed in \cite{Mukh}  and also \cite{Press,Sato} for
earlier ideas. In particular, Mukhanov \cite{Mukh} and Press 
\cite{Press} realized that
in an exponentially expanding background, the curvature fluctuations
would be scale-invariant, and Mukhanov provided a quantitative
calculation which also yielded the logarithmic deviation from
exact scale-invariance. 

The role of the Hubble radius has already been mentioned
repeatedly in these lectures. In particular, in the previous
subsection we saw that the Hubble radius separates scales
on which fluctuations oscillate (sub-Hubble scales) from those
where they are frozen in (super-Hubble scales). Another
way to see the role of the Hubble radius is to consider the 
equation of a free scalar matter field
$\varphi$ on an unperturbed expanding background:
\begin{equation}
\ddot{\varphi} + 3 H \dot{\varphi} - {{\nabla^2} \over {a^2}} \varphi
\, = \, 0 \, .
\end{equation}
The second term on the left hand side of this equation leads to damping
of $\varphi$ with a characteristic decay rate given by $H$. As a
consequence, in the absence of the spatial gradient term, $\dot{\varphi}$
would be of the order of magnitude $H \varphi$. Thus, comparing the
second and the third terms on the left hand side, we immediately
see that the microscopic (spatial gradient) term dominates on length
scales smaller than the Hubble radius, leading to oscillatory motion,
whereas this term is negligible on scales larger than the Hubble radius,
and the evolution of $\varphi$ is determined primarily by gravity. Note
that in general cosmological models the Hubble radius is much smaller than the
horizon (the forward light cone calculated from the initial time). In
an inflationary universe, the horizon is larger by a factor of at least 
${\rm exp}(N)$, where $N$ is the number of e-foldings of inflation, and the
lower bound is taken on if the Hubble radius and horizon coincide until
inflation begins. It is very important to realize this difference, a
difference which is obscured in most articles on cosmology in which the
term ``horizon'' is used when ``Hubble radius'' is meant. Note, in
particular, that the homogeneous inflaton field contains causal information
on super-Hubble but sub-horizon scales. Hence, it is completely consistent
with causality \cite{Fabio1}
to have a microphysical process related to the background
scalar matter field lead to exponential amplification of the amplitude
of fluctuations during reheating on such scales, as it does in models
in which entropy perturbations are present and not suppressed during
inflation \cite{BaVi,Fabio2}. Note that also in string gas cosmology
and in the matter bounce scenario the Hubble radius and horizon
are completely different. 

There are general relativistic
conservation laws \cite{Traschen} which imply that adiabatic fluctuations
produced locally must be Poisson-statistic suppressed on scales larger than
the Hubble radius. For example, fluctuations produced by the formation of
topological defects at a phase transition in the early universe are
initially isocurvature (entropy) in nature (see e.g. \cite{TTB} for
a discussion). Via the source term in the
equation of motion (\ref{rhbeq4}), a growing adiabatic mode is induced, but
at any fixed time the spectrum of the curvature fluctuation on scales larger
than the Hubble radius has index $n = 4$ (Poisson). A similar conclusion
applies to models \cite{Dvali,Kofman} of modulated
reheating (see \cite{Vernizzi} for
a nice discussion), and to models in which moduli fields obtain masses after
some symmetry breaking, their quantum fluctuations then inducing cosmological
perturbations. A prototypical example is given by axion fluctuations in an
inflationary universe (see e.g. \cite{ABT} and references therein).

To understand the generation and evolution of fluctuations in current
models of the very early Universe, we need both Quantum Mechanics
and General Relativity, i.e. quantum gravity. At first sight, we
are thus faced with an intractable problem, since the theory of quantum
gravity is not yet established. We are saved by the fact that today
on large cosmological scales the fractional amplitude of the fluctuations
is smaller than 1. Since gravity is a purely attractive force, the
fluctuations had to have been - at least in the context of an eternally
expanding background cosmology - very small in the early Universe. Thus,
a linearized analysis of the fluctuations (about a classical
cosmological background) is self-consistent.

From the classical theory of cosmological perturbations discussed in the
previous section, we know that the analysis of scalar metric inhomogeneities
can be reduced - after extracting gauge artifacts -
to the study of the evolution of a single fluctuating
variable. Thus, we conclude that the quantum theory of cosmological
perturbations must be reducible to the quantum theory of a single
free scalar field which we will denote by $v$. 
Since the background in which this scalar field
evolves is time-dependent, the mass of $v$ will be time-dependent. The
time-dependence of the mass will lead to quantum particle production
over time if we start the evolution in the vacuum state for $v$. As
we will see, this quantum particle production corresponds to the
development and growth of the cosmological fluctuations. Thus,
the quantum theory of cosmological fluctuations provides a consistent
framework to study both the generation and the evolution of metric
perturbations. The following analysis is based on Part II of \cite{MFB}.
 
\subsubsection{Outline of the Analysis}

In order to obtain the action for linearized cosmological
perturbations, we expand the action to quadratic order in
the fluctuating degrees of freedom. The linear terms cancel
because the background is taken to satisfy the background
equations of motion.

We begin with the Einstein-Hilbert action for gravity and the
action of a scalar matter field (for the more complicated
case of general hydrodynamical fluctuations the reader is
referred to \cite{MFB})
\begin{equation} \label{action}
S \, = \,  \int d^4x \sqrt{-g} \bigl[ - {1 \over {16 \pi G}} R
+ {1 \over 2} \partial_{\mu} \varphi \partial^{\mu} \varphi - V(\varphi)
\bigr] \, ,
\end{equation}
where $R$ is the Ricci curvature scalar.

The simplest way to proceed is to work in 
longitudinal gauge, in which the metric and matter take the form
(assuming no anisotropic stress)
\begin{eqnarray} \label{long}
ds^2 \, &=& \, a^2(\eta)\bigl[(1 + 2 \phi(\eta, {\bf x}))d\eta^2
- (1 - 2 \phi(t, {\bf x})) d{\bf x}^2 \bigr] \nonumber \\
\varphi(\eta, {\bf x}) \, 
&=& \, \varphi_0(\eta) + \delta \varphi(\eta, {\bf x}) \, .
\end{eqnarray}
The two fluctuation variables
$\phi$ and $\varphi$ must be linked by the Einstein constraint
equations since there cannot be matter fluctuations without induced
metric fluctuations. 

The two nontrivial tasks of the lengthy \cite{MFB} computation 
of the quadratic piece of the action is to find
out what combination of $\varphi$ and $\phi$ gives the variable $v$
in terms of which the action has canonical kinetic term, and what the form
of the time-dependent mass is. This calculation involves inserting
the ansatz (\ref{long}) into the action (\ref{action}),
expanding the result to second order in the fluctuating fields, making
use of the background and of the constraint equations, and dropping
total derivative terms from the action. In the context of
scalar field matter, the quantum theory of cosmological
fluctuations was developed by Mukhanov \cite{Mukh2,Mukh3} and
Sasaki \cite{Sasaki}. The result is the following
contribution $S^{(2)}$ to the action quadratic in the
perturbations:
\begin{equation} \label{pertact}
S^{(2)} \, = \, {1 \over 2} \int d^4x \bigl[v'^2 - v_{,i} v_{,i} + 
{{z''} \over z} v^2 \bigr] \, ,
\end{equation}
where the canonical variable $v$ (the ``Sasaki-Mukhanov variable'' introduced
in \cite{Mukh3} - see also \cite{Lukash}) is given by
\begin{equation} \label{Mukhvar}
v \, = \, a \bigl[ \delta \varphi + {{\varphi_0^{'}} \over {\cal H}} \phi
\bigr] \, ,
\end{equation}
with ${\cal H} = a' / a$, and where
\begin{equation} \label{zvar}
z \, = \, {{a \varphi_0^{'}} \over {\cal H}} \, .
\end{equation}

As long as
the equation of state does not change over time)
\begin{equation} \label{zaprop}
z(\eta) \, \sim \, a(\eta) \, .
\end{equation}
Note that the variable $v$ is related to the curvature
perturbation ${\cal R}$ in comoving coordinates introduced
in \cite{Lyth} and closely related to the variable $\zeta$ used
in \cite{BST,BK}:
\begin{equation} \label{Rvar}
v \, = \, z {\cal R} \, .
\end{equation}

The equation of motion which follows from the action (\ref{pertact}) is
(in momentum space)
\begin{equation} \label{pertEOM2}
v_k^{''} + k^2 v_k - {{z^{''}} \over z} v_k \, = \, 0 \, ,
\end{equation}
where $v_k$ is the k'th Fourier mode of $v$. 
The mass term in the above equation is in general
given by the Hubble scale. Thus, it immediately follows that on small
length scales, i.e. for
$k > k_H$, the solutions for $v_k$ are constant amplitude oscillations . 
These oscillations freeze out at Hubble radius crossing,
i.e. when $k = k_H$. On longer scales ($k \ll k_H$), there is
a mode of  $v_k$ which scales as $z$. This mode is the dominant
one in an expanding universe, but not in a contracting one.

Given the action (\ref{pertact}), the quantization of the cosmological
perturbations can be performed by canonical quantization (in the same
way that a scalar matter field on a fixed cosmological background
is quantized \cite{BD}). 

The final step in the quantum theory of cosmological perturbations
is to specify an initial state. Since in inflationary cosmology
all pre-existing classical fluctuations are red-shifted by the
accelerated expansion of space, one usually assumes (we will
return to a criticism of this point when discussing the
trans-Planckian problem of inflationary cosmology) that the field
$v$ starts out at the initial time $t_i$ mode by mode in its vacuum
state. Two questions immediately emerge: what is the initial time $t_i$,
and which of the many possible vacuum states should be chosen. It is
usually assumed that since the fluctuations only oscillate on sub-Hubble
scales, the choice of the initial time is not important, as long
as it is earlier than the time when scales of cosmological interest
today cross the Hubble radius during the inflationary phase. The
state is usually taken to be the Bunch-Davies vacuum (see e.g. \cite{BD}),
since this state is empty of particles at $t_i$ in the coordinate frame
determined by the FRW coordinates  Thus, we choose the initial
conditions
\begin{eqnarray} \label{incond}
v_k(\eta_i) \, = \, {1 \over {\sqrt{2 k}}} \\
v_k^{'}(\eta_i) \, = \, {{\sqrt{k}} \over {\sqrt{2}}} \, \, \nonumber
\end{eqnarray} 
where $\eta_i$ is the conformal time corresponding
to the physical time $t_i$.
 
Returning to the case of an expanding universe, the scaling
\begin{equation} \label{squeezing}
v_k \, \sim \, z \, \sim \, a \,
\end{equation}
implies that the wave function of the quantum variable $v_k$ which
performs quantum vacuum fluctuations on sub-Hubble scales,
stops oscillating on super-Hubble scales and instead is
squeezed (the amplitude increases in configuration space
but decreases in momentum space). This squeezing corresponds
to quantum particle production. It is also one of the two
conditions which are required for the classicalization of
the fluctuations. The second condition is decoherence which
is induced by the non-linearities in the dynamical system
which are inevitable since the Einstein action leads to
highly nonlinear equatiions (see \cite{Starob3} for a recent
discussion of this point, and \cite{Martineau} for related
work).

Note that the squeezing of cosmological fluctuations on
super-Hubble scales occurs in all models, in particular
in string gas cosmology and in the bouncing universe
scenario since also in these scenarios perturbations
propagate on super-Hubble scales for a long period of
time. In a contracting phase, the dominant
mode of $v_k$ on super-Hubble scales is not the one given
in (\ref{squeezing}) (which in this case is a decaying
mode), but the second mode which scales as $z^{-p}$
with an exponent $p$ which is positive and whose
exact value depends on the background equation of state. 

Applications of this theory in inflationary cosmology, in
the matter bounce scenario and in string gas cosmology
will be considered in the respective sections of these
lecture notes.

\subsubsection{Quantum Theory of Gravitational Waves}

The quantization of gravitational waves parallels the
quantization of scalar metric fluctuations, but is
more simple because there are no gauge ambiguities. Note
that at the level of linear fluctuations, scalar metric
fluctuations and gravitational waves are independent. Both
can be quantized on the same cosmological background determined
by the background scale factor and the background matter. However, in
contrast to the case of scalar metric fluctuations, the tensor
modes are also present in pure gravity (i.e. in the absence of
matter).

Starting point is the action (\ref{action}). Into this
action we insert the metric which corresponds to a 
classical cosmological background plus tensor metric
fluctuations:
\begin{equation}
ds^2 \, = \, a^2(\eta) \bigl[ d\eta^2 - (\delta_{ij} + h_{ij}) dx^i dx^j 
\bigr]\, ,
\end{equation}
where the second rank tensor $h_{ij}(\eta, {\bf x})$ represents the
gravitational waves, and in turn can be decomposed as
\begin{equation}
h_{ij}(\eta, {\bf x}) \, = \, h_{+}(\eta, {\bf x}) e^+_{ij}
+ h_{x}(\eta, {\bf x}) e^x_{ij}
\end{equation}
into the two polarization states. Here, $e^{+}_{ij}$ and $e^{x}_{ij}$ are
two fixed polarization tensors, and $h_{+}$ and $h_{x}$ are the two 
coefficient functions.

To quadratic order in the fluctuating fields, the action consists of
separate terms involving $h_{+}$ and $h_{x}$. Each term is of the form
\begin{equation} \label{actgrav}
S^{(2)} \, = \, \int d^4x {{a^2} \over 2} \bigl[ h'^2 - (\nabla h)^2 \bigr]  \, ,
\end{equation}
leading to the equation of motion
\begin{equation}
h_k^{''} + 2 {{a'} \over a} h_k^{'} + k^2 h_k \, = \, 0 \, .
\end{equation}
The variable in terms of which the action (\ref{actgrav}) has canonical
kinetic term is
\begin{equation} \label{murel}
\mu_k \, \equiv \, a h_k \, ,
\end{equation}
and its equation of motion is
\begin{equation}
\mu_k^{''} + \bigl( k^2 - {{a''} \over a} \bigr) \mu_k \, = \, 0 \, .
\end{equation}
This equation is very similar to the corresponding equation (\ref{pertEOM2}) 
for scalar gravitational inhomogeneities, except that in the mass term
the scale factor $a(\eta)$ replaces $z(\eta)$, which leads to a
very different evolution of scalar and tensor modes during the reheating
phase in inflationary cosmology during which the equation of state of the
background matter changes dramatically.
 
Based on the above discussion we have the following theory for the
generation and evolution of gravitational waves in an accelerating
Universe (first developed by Grishchuk \cite{Grishchuk}): 
waves exist as quantum vacuum fluctuations at the initial time
on all scales. They oscillate until the length scale crosses the Hubble
radius. At that point, the oscillations freeze out and the quantum state
of gravitational waves begins to be squeezed in the sense that
\begin{equation}
\mu_k(\eta) \, \sim \, a(\eta) \, ,
\end{equation}
which, from (\ref{murel}) corresponds to constant amplitude of $h_k$.
The squeezing of the vacuum state leads to the emergence of classical
properties of this state, as in the case of scalar metric fluctuations.

\section{Inflationary Cosmology}

\subsection{Mechanism of Inflation}

The idea of inflationary cosmology is to assume that there was
a period in the very early Universe during which the scale factor was
accelerating, i.e. ${\ddot a} > 0$. This implies that the Hubble
radius was shrinking in comoving coordinates, or, equivalently, that fixed
comoving scales were ``exiting'' the Hubble radius. In the simplest models of
inflation, the scale factor increases nearly exponentially.
\begin{figure}
\centering
\includegraphics[height=8cm]{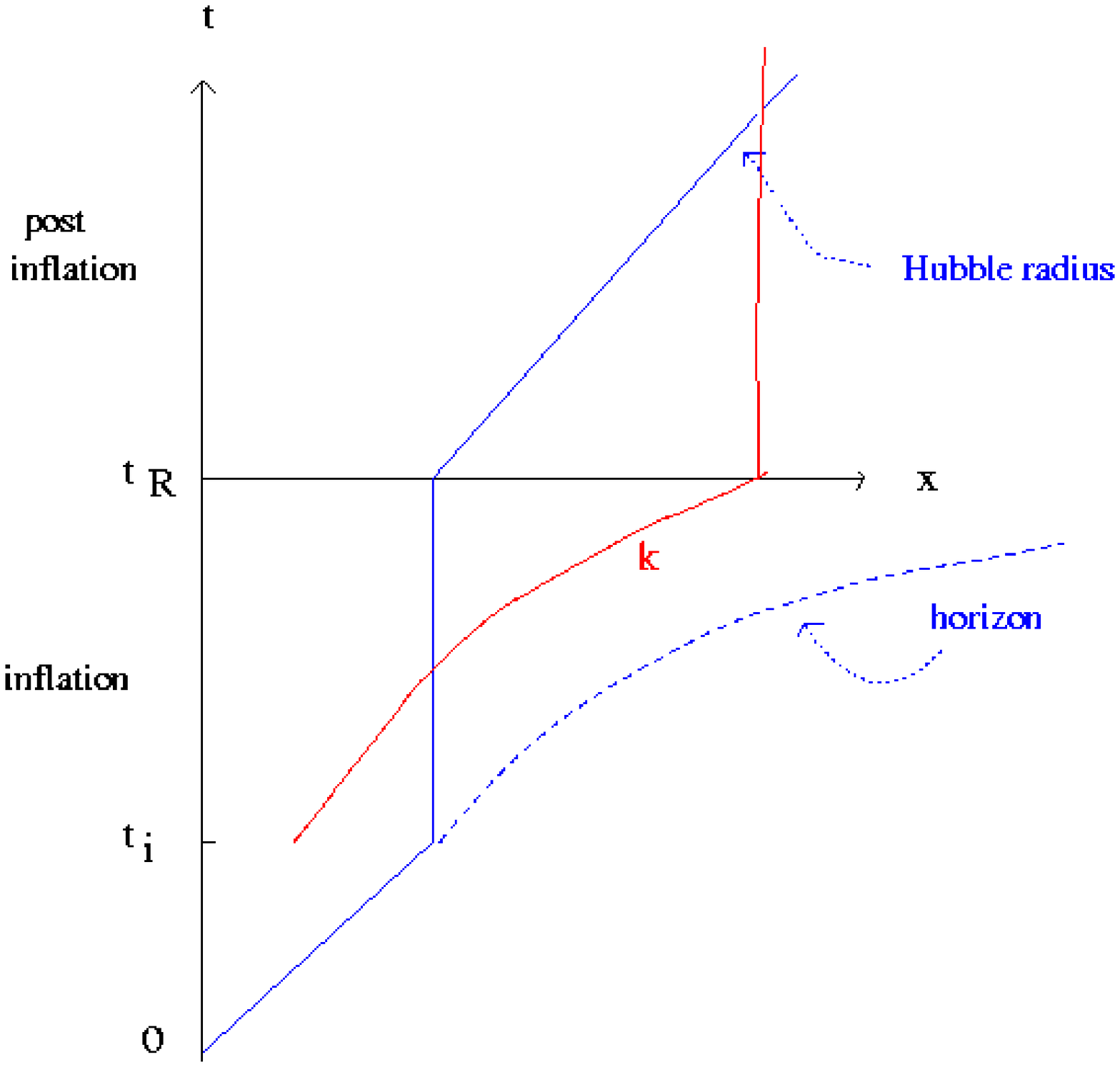}
\caption{Space-time diagram (sketch) showing the evolution
of scales in inflationary cosmology. The vertical axis is
time, and the period of inflation lasts between $t_i$ and
$t_R$, and is followed by the radiation-dominated phase
of standard big bang cosmology. During exponential inflation,
the Hubble radius $H^{-1}$ is constant in physical spatial coordinates
(the horizontal axis), whereas it increases linearly in time
after $t_R$. The physical length corresponding to a fixed
comoving length scale labelled by its wavenumber $k$ increases
exponentially during inflation but increases less fast than
the Hubble radius (namely as $t^{1/2}$), after inflation.}
\label{infl11}       
\end{figure}
As illustrated in Figure \ref{infl11}, 
the basic geometry of inflationary cosmology
provides a solution of the fluctuation problem. As long as the phase
of inflation is sufficiently long, all length scales within our present
Hubble radius today originate at the beginning of inflation with a 
wavelength smaller than the Hubble radius at that time. Thus, it
is possible to create perturbations locally using physics obeying
the laws of special relativity (in particular causality). As will be
discussed later, it is quantum vacuum fluctuations of matter fields
and their associated curvature perturbations which are responsible for
the structure we observe today.

Postulating a phase of inflation in the very early universe solves
the {\it horizon problem} of the SBB, namely it explains why the causal horizon
at the time $t_{rec}$ when photons last scatter is larger than the radius of the
past light cone at $t_{rec}$, the part of the last scattering surface which is 
visible today in CMB experiments. Inflation also explains the near flatness of
the universe: in a decelerating universe spatial flatness is an unstable fixed
point of the dynamics, whereas in an accelerating universe it becomes an
attractor. Another important aspect of the inflationary solution of the 
{\it flatness problem} is that inflation exponentially increases the volume 
of space. Without inflation, it is not possible that a Planck scale universe 
at the Planck time evolves into a sufficiently large universe today.

\subsection{Fluctuations in Inflationary Cosmology}

We will now use the quantum theory of cosmological
perturbations developed in the previous section 
to calculate the spectrum
of curvature fluctuations in inflationary cosmology. 
The starting point are quantum vacuum initial conditions
for the canonical fluctuation variable $v_k$:
\be \label{IC}
v_k(\eta_i) \, = \, \frac{1}{\sqrt{2 k}} \, 
\ee
for all $k$ for which the wavelength is smaller than
the Hubble radius at the initial time $t_i$. 

The amplitude remains unchanged until the modes
exit the Hubble radius at the respective times
$t_H(k)$ given by
\begin{equation} \label{Hubble3}
a^{-1}(t_H(k)) k \, = \, H \, .
\end{equation}

We need to compute the power spectrum ${\cal P}_{\cal R}(k)$ 
of the curvature fluctuation ${\cal R}$ defined in (\ref{Rvar}) at
some late time $t$ when the modes are super-Hubble.
We first relate the power spectrum via the growth rate (\ref{squeezing})
of $v$ on super-Hubble scales to the power spectrum at the time 
$t_H(k)$ and then use the constancy of the amplitude
of $v$ on sub-Hubble scales to relate it to the initial conditions
(\ref{IC}). Thus
\begin{eqnarray} \label{finalspec1}
{\cal P}_{\cal R}(k, t) \, \equiv  \, k^3 {\cal R}_k^2(t) \, 
&=& \, k^3 z^{-2}(t) |v_k(t)|^2 \\
&=& \, k^3 z^{-2}(t) \bigl( {{a(t)} \over {a(t_H(k))}} \bigr)^2
|v_k(t_H(k))|^2 \nonumber \\
&=& \, k^3 z^{-2}(t_H(k)) |v_k(t_H(k))|^2 \nonumber \\
&\sim& \, k^3 \bigl( \frac{a(t)}{z(t)} \bigr)^2 a^{-2}(t_H(k)) |v_k(t_i)|^2 \, , \nonumber
\end{eqnarray}
where in the final step we have used (\ref{zaprop}) and the
constancy of the amplitude of $v$ on sub-Hubble scales. 
Making use of the condition (\ref{Hubble3}) 
for Hubble radius crossing, and of the
initial conditions (\ref{IC}), we immediately see that
\begin{equation} \label{finalspec2}
{\cal P}_{\cal R}(k, t) \, \sim \, \bigl( \frac{a(t)}{z(t)} \bigr)^2 
k^3 k^{-2} k^{-1} H^2 \, ,
\end{equation}
and that thus a scale invariant power spectrum with amplitude
proportional to $H^2$ results, in agreement with what was
argued on heuristic grounds in the overview of inflation in the
the first section. To obtain
the precise amplitude, we need to make use of the relation
between $z$ and $a$. We obtain
\be \label{amplitude}
{\cal P}_{\cal R}(k, t) \, \sim \, \frac{H^4}{\dot{\varphi}_0^2} \,
\ee
which for any given value of $k$ is to be evaluated 
at the time $t_H(k)$ (before the end of inflation). For
a scalar field potential (see following subsection)
\be
V(\varphi) \, = \, \lambda \varphi^4
\ee
the resulting amplitude in (\ref{amplitude}) is $\lambda$.
Thus, in order to obtain the observed value of the
power spectrum of the order of $10^{-10}$, 
the coupling constant $\lambda$ must
be tuned to a very small value.

\subsection{Models of Inflation}

Let us now consider how it is possible to obtain a phase of cosmological
inflation. We will assume that space-time is described using the
equations of General Relativity \footnote{Note, however, that the first model
of exponential expansion of space \cite{Starob1} made use of a higher derivative
gravitational action.}. In this case, the dynamics of the scale factor
$a(t)$ is determined by the Friedmann-Robertson-Walker (FRW) equations
\begin{equation} \label{FRW1}
({{\dot a} \over a})^2 \, = \, 8 \pi G \rho 
\end{equation}
and
\begin{equation} \label{FRW2}
{{\ddot a} \over a} \, = \, - \frac{4 \pi G}{3} (\rho + 3p)
\end{equation}
where for simplicity we have omitted the contributions of spatial curvature
(since spatial curvature is diluted during inflation) and of the cosmological
constant (since any small cosmological constant which might be present today
has no effect in the early Universe since the associated energy density does
not increase when going into the past). From (\ref{FRW2}) it is clear
that in order to obtain an accelerating universe, matter with sufficiently
negative pressure
\begin{equation}
p \, < \, - {1 \over 3} \rho
\end{equation}
is required. Exponential inflation is obtained for $p = - \rho$.

Conventional perfect fluids have positive semi-definite pressure and thus
cannot yield inflation. Quantum field theory can come to the rescue.
We know that a description of matter in
terms of classical perfect fluids must break down at early times. An improved
description of matter will be given in terms of quantum fields. Matter
which we observe today consists of spin 1/2 and spin 1 fields. Such
fields cannot yield inflation in the context of the renormalizable quantum field
theory framework. The existence of scalar matter fields is postulated to
explain the generation of mass via spontaneous symmetry
breaking. Scalar matter fields (denoted by $\varphi$) 
are special in that they allow - at the level of a renormalizable
action - the presence of a potential energy term $V(\varphi)$. The energy density and
pressure of a scalar field $\varphi$ with canonically normalized action
\footnote{See \cite{Eva,kinflation} for  discussions of fields with non-canonical
kinetic terms.}
\begin{equation} \label{scalarlag}
{\cal L} \, = \, \sqrt{-g}
\bigl[{1 \over 2} \partial_{\mu} \varphi \partial^{\mu} \varphi
- V(\varphi)\bigr] \,  
\end{equation}
(where Greek indices are space-time indices and $g$ is the determinant of the metric) 
are given by
\begin{eqnarray}
\rho \, &=& \, {1 \over 2} ({\dot \varphi})^2 + 
{1 \over 2} a^{-2} (\nabla \varphi)^2 + V(\varphi) \nonumber \\
p \, &=& \, {1 \over 2} ({\dot \varphi})^2 - 
{1 \over 6} a^{-2} (\nabla \varphi)^2
- V(\varphi) \, .
\end{eqnarray}
Thus, it is possible to obtain an almost exponentially expanding universe 
provided the scalar field configuration \footnote{The scalar field
yielding inflation is called the {\it inflaton}.} satisfies
\begin{eqnarray}
{1 \over 2} (\nabla_p \varphi)^2 \, &\ll& \, V(\varphi) \, , \label{gradcond} \\
{1 \over 2} ({\dot \varphi})^2 \, &\ll& \, V(\varphi) \, . \label{tempcond}
\end{eqnarray}
In the above, $\nabla_p \equiv a^{-1} \nabla$ is the gradient with respect
to physical as opposed to comoving coordinates.
Since spatial gradients redshift as the universe expands, the first condition
will (for single scalar field models) always be satisfied if it is satisfied at
the initial time \footnote{In fact, careful studies \cite{Kung}
show that since the gradients decrease even in a non-inflationary 
backgrounds, they can become subdominant even if they are not initially 
subdominant.}. It is the second condition which is harder to satisfy. In
particular, this condition is in general not preserved in time even it is
initially satisfied. 

In his original model \cite{Guth}, Guth assumed that inflation was generated
by a scalar field $\varphi$ sitting in a false vacuum with positive
vacuum energy. Hence, the conditions
(\ref{gradcond}) and (\ref{tempcond}) are automatically satisfied. Inflation
ends when $\varphi$ tunnels to its true vacuum state with (by assumption)
vanishing potential energy. However, as Guth realized himself, this model
is plagued by a ``graceful exit" problem: the nucleation produces a
bubble of true vacuum whose initial size is microscopic \cite{Coleman} (see
also \cite{RHBRMP} for a review of field theory methods which are
useful in inflationary cosmology). After inflation, this bubble will expand
at the speed of light, but it is too small to explain the observed size of
the universe. Since space between the bubbles expands exponentially, the
probability of bubble percolation is negligible, at least in Einstein gravity.
For this reason, the focus in inflationary model building shifted to
the ``slow-roll" paradigm \cite{new}.

It is sufficient to obtain a period of cosmological inflation that
the {\it slow-roll conditions} for $\varphi$ are satisfied. Recall that
the equation of motion for a homogeneous scalar field in a cosmological
space-time is (as follows from (\ref{scalarlag})) is
\begin{equation} \label{scalareom}
{\ddot \varphi} + 3 H {\dot \varphi} \, = \, - V^{\prime}(\varphi) \, ,
\end{equation}
where a prime indicates the derivative with respect to $\varphi$. In order
that the scalar field roll slowly, it is necessary that
\begin{equation} \label{roll}
{\ddot \varphi} \, \ll \, 3 H {\dot \varphi}
\end{equation}
such that the first term in the scalar field equation of motion 
(\ref{scalareom}) is negligible. In this case, the condition (\ref{tempcond})
becomes
\begin{equation} \label{SRcond1}
({{V^{\prime}} \over V})^2 \, \ll \, 48 \pi G \,
\end{equation}
and (\ref{roll}) becomes
\begin{equation} \label{SRcond2}
{{V^{\prime \prime}} \over V} \, \ll \, 24 \pi G \, .
\end{equation}

There are many models of scalar field-driven slow-roll inflation. 
Many of them can be divided into three groups: 
small-field inflation, large-field 
inflation and hybrid inflation. {\it Small-field inflationary models} are based
on ideas from spontaneous symmetry breaking in particle physics. We take the
scalar field to have a potential of the form
\begin{equation} \label{Mexican}
V(\varphi) \, = \, {1 \over 4} \lambda (\varphi^2 - \sigma^2)^2 \, ,
\end{equation}
where $\sigma$ can be interpreted as a symmetry breaking scale, and
$\lambda$ is a dimensionless coupling constant. The hope of initial
small-field models (``new inflation'' \cite{new}) was that the scalar
field would begin rolling close to its symmetric point $\varphi = 0$,
where thermal equilibrium initial conditions would localize it in the
early universe. At sufficiently high temperatures, $\varphi = 0$ 
is a stable ground state of the one-loop finite temperature effective 
potential $V_T(\varphi)$ (see e.g. \cite{RHBRMP} for a review). 
Once the temperature drops to a value 
smaller than the critical temperature $T_c$, $\varphi = 0$ turns into 
an unstable local maximum of $V_T(\varphi)$, and $\varphi$ is free to 
roll towards a ground state of the zero temperature potential (\ref{Mexican}). 
The direction of the initial rolling is triggered by quantum fluctuations. 
The reader can
easily check that for the potential (\ref{Mexican}) the slow-roll conditions
cannot be satisfied if $\sigma \ll m_{pl}$, where $m_{pl}$ is the
Planck mass which is related to $G$. 
If the potential is modified to a Coleman-Weinberg \cite{CW} form 
\begin{equation} \label{CWpot}
V(\varphi) \, = \, {{\lambda} \over 4} \varphi^4
\bigl[ {\rm ln}  {{|\varphi|} \over v} - {1 \over 4} \bigr]
+ {1 \over {16}} \lambda v^4
\end{equation}
(where $v$ denotes the value of the minimum of the potential) 
then the slow-roll conditions can be satisfied. However, this
corresponds to a severe fine-tuning of the shape of the potential.
A further problem for most small-field models of inflation (see 
e.g. \cite{Goldwirth} for a review) is that in order to end up close to the
slow-roll trajectory, the initial field velocity must be constrained
to be very small. This {\it initial condition problem} of small-field
models of inflation effects a number of recently proposed brane inflation
scenarios, see e.g. \cite{GhazalScott} for a discussion. 

There is another reason for abandoning small-field inflation models: in
order to obtain a sufficiently small amplitude of density fluctuations,
the interaction coefficients of $\varphi$ must be very small. 
This makes it inconsistent to assume that $\varphi$ started
out in thermal equilibrium \cite{MUW}. In the absence of thermal 
equilibrium, the phase space of initial conditions is much larger for 
large values of $\varphi$. 

This brings us to the discussion of large-field inflation
models, initially proposed in \cite{chaotic} under the name
``chaotic inflation''. The simplest example is
provided by a massive scalar field with potential
\begin{equation} \label{mpot}
V(\varphi) \, = \, {1 \over 2} m^2 \varphi^2 \, ,
\end{equation}
where $m$ is the mass. It is assumed that the scalar field rolls towards
the origin from large values of $|\varphi|$. It is a simple exercise 
for the reader to verify that the slow-roll conditions (\ref{SRcond1}) and
(\ref{SRcond2}) are satisfied provided
\begin{equation}
|\varphi| \, > \, {1 \over {\sqrt{12 \pi}}} m_{pl} \, .
\end{equation}
Values of $|\varphi|$ comparable or greater than $m_{pl}$ are also
required in other realizations of large-field inflation. Hence, one
may worry whether such a toy model can consistently be embedded in
a realistic particle physics model, e.g. supergravity. In many such
models $V(\varphi)$ receives supergravity-induced correction terms
which destroy the flatness of the potential for $|\varphi| > m_{pl}$.
As can be seen by applying the formulas of the previous subsection
to the potential (\ref{mpot}), a value
of $m \sim 10^{13}$GeV is required in order to obtain the observed
amplitude of density fluctuations. Hence, the configuration space range
with $|\varphi| > m_{pl}$ but $V(\varphi) < m_{pl}^4$ 
dominates the measure of field values. 
It can also be verified that the slow-roll trajectory is a local
attractor in field initial condition space \cite{Kung}, even including
metric fluctuations at the perturbative level \cite{Feldman}.

With two scalar fields it is possible to construct a class of models
which combine some of the nice features of large-field inflation
(large phase space of initial conditions yielding inflation) and of
small-field inflation (better contact with conventional particle physics).
These are models of hybrid inflation \cite{hybrid}. To give a prototypical
example, consider two scalar fields $\varphi$ and $\chi$ with a potential
\begin{equation} \label{hybridpot}
V(\varphi, \chi) \, = \, {1 \over 4} \lambda_{\chi} (\chi^2 - \sigma^2)^2
+ {1 \over 2} m^2 \varphi^2 - {1 \over 2} g^2 \varphi^2 \chi^2 \, .
\end{equation}
In the absence of thermal equilibrium, it is natural to assume that 
$|\varphi|$ begins at large values, values for which the effective mass
of $\chi$ is positive and hence $\chi$ begins at $\chi = 0$. The parameters
in the potential (\ref{hybridpot}) are now chosen such that $\varphi$
is slowly rolling for values of $|\varphi|$ somewhat smaller than $m_{pl}$,
but that the potential energy for these field values is dominated by
the first term on the right-hand side of (\ref{hybridpot}). The reader
can easily verify that for this model it is no longer required to have
values of $|\varphi|$ greater than $m_{pl}$ in order to obtain slow-rolling
\footnote{Note that the slow-roll conditions (\ref{SRcond1}) and (\ref{SRcond2})
were derived assuming that $H$ is given by the contribution of
$\varphi$ to $V$ which is not the case here.}
The field $\varphi$ is slowly rolling whereas the potential energy is
determined by the contribution from $\chi$. Once $|\varphi|$ drops to the
value
\begin{equation}
|\varphi_c| \, = \, {{{\sqrt{\lambda_{\chi}}}} \over g} \sigma 
\end{equation}
the configuration $\chi = 0$ becomes unstable and decays to its ground
state $|\chi| = \sigma$, yielding a graceful exit from inflation.
Since in this example the ground state of $\chi$ is not unique, there
is the possibility of the formation of topological defects at the end
of inflation (see \cite{ShellVil,HK,RHBtoprev} for reviews of topological
defects in cosmology).

\subsection{Reheating in Inflationary Cosmology}

During the inflationary period the energy density of matter
becomes completely dominated by the energy of the
inflaton field. The density of regular matter (the kind of
matter which is observed today) is exponentially diluted.
An essential part of any inflationary model is the mechanism
by which the energy density is transferred to regular matter
at the end of the period of inflation.  This is the so-called
``reheating" process. It typically begins with a non-perturbative
decay process called ``preheating". We will give a brief overview
of reheating in this subsection (for an in-depth recent review
the reader is referred to \cite{Allahverdi}.

\subsubsection{Preheating}

After the slow-roll conditions break down, the period of inflation ends,
and the inflaton $\varphi$ begins to oscillate around its ground state.
Due to couplings of $\varphi$ to other matter fields, the energy of the
universe, which at the end of the period of inflation is stored completely
in $\varphi$, gets transferred to the matter fields of the particle
physics Standard Model. Initially, the energy transfer was described
perturbatively \cite{initial}. Later, it was realized \cite{TB,KLS,STB,KLS2}
that through a parametric resonance instability, particles are very
rapidly produced, leading to a fast energy transfer (``preheating'').
The quanta later thermalize, and thereafter the universe evolves as described
by SBB cosmology.

Let us give a brief overview of the theory of reheating in inflationary
cosmology. We assume that the inflaton $\varphi$
is coupled to another scalar field $\chi$ and take
the interaction Lagrangian to be
\be \label{toy}
{\cal L}_{\rm int} \, = \, - \frac{1}{2} g^2 \varphi^2 \chi^2 \, ,
\ee
where $g$ is a dimensionless coupling constant. 

In the initial perturbative analysis of reheating \cite{initial}
the decay of the coherently oscillating inflaton field
$\varphi$ is treated perturbatively. The
interaction rate $\Gamma$ of processes in which
two $\varphi$ quanta interact to produce a pair of
$\chi$ particles is taken as the decay rate of the
inflaton. The inflaton dynamics is modeled via an
effective equation
\be \label{effeom}
{\ddot \varphi} + 3 H {\dot \varphi} + \Gamma {\dot \varphi} \, = \, - V^{'}(\varphi) \, .
\ee
For small coupling constant, the interaction rate $\Gamma$ is
typically much smaller than the Hubble parameter at the end
of inflation. Thus, at the beginning of the phase of inflaton
oscillations, the energy loss into particles is initially negligible
compared to the energy loss due to the expansion of space.
It is only once the Hubble expansion rate decreases to a
value comparable to $\Gamma$ that $\chi$ particle production
becomes effective. It is the energy density at the time when
$H = \Gamma$ which determines how much energy ends up
in $\chi$ particles and thus determines the ``reheating temperature",
the temperature of the SM fields after energy transfer.
\be
T_R \, \sim \, \left( \Gamma m_{pl} \right)^{1/2} \, .
\ee
Since $\Gamma$ is proportional to the square of the coupling
constant $g$ which is generally very small, perturbative
reheating is slow and produces a reheating temperature
which can be very low compared to the energy scale
at which inflation takes place.

There are two main problems with the perturbative decay
analysis described above. First of all, even if the inflaton
decay were perturbative, it is not justified to use the heuristic
equation (\ref{effeom}) since it violates the fluctuation-dissipation
theorem: in systems with dissipation, there are always fluctuations,
and these are missing in (\ref{effeom}). For an improved
effective equation of motion see e.g. \cite{Ramos}.

The main problem with the perturbative analysis is that it does
not take into account the coherent nature of the inflaton field $\varphi$.
At the beginning of the period of oscillations $\varphi$
is not a superposition of free asymptotic single inflaton states,
but rather a coherently oscillating homogeneous field. The large
amplitude of oscillation implies that it is well justified to treat
the inflaton classically. However, the 
matter fields can be assumed to start off in their vacuum state
(the red-shifting during the period of inflation will remove any
matter particles present at the beginning of inflation). Thus,
matter fields $\chi$ must be treated quantum mechanically.
The improved approach to reheating initiated in \cite{TB}
(see also \cite{DK}) is to consider reheating as a quantum
production of $\chi$ particles in a classical $\varphi$ background.

We expand the field $\chi$ is terms of the usual creation and
annihilation operators. The mode functions satisfy the
equation
\be\label{math3}
\ddot{\chi}_k + 3 H \dot{\chi} + 
\left(\frac{k^2}{a^2} + m_{\chi}^2 + g^2\Phi(t)^2\sin^2{(mt)}\right)\chi_k \, = \, 0 \, ,
\ee
where $\Phi(t)$ is the amplitude of oscillation of $\varphi$.

The first level of approximation \cite{TB} is to neglect the
expansion of space, to estimate the efficiency of reheating
in this approximation, and to check self-consistency of the
approximation. The equation (\ref{math3}) then reduces to
\be\label{math4}
\ddot{\chi}_k + \left(k^2 + m_{\chi}^2 + g^2\Phi^2\sin^2{(mt)}\right)\chi_k \, = \, 0 \, ,
\ee
and $\Phi(t)$ is constant. The equation is the Mathieu equation
\cite{Mathieu} which is conventionally written in the form
\be\label{mat}
\chi''_k+(A_k-2q\cos{2z})\chi_k \, = \, 0 \, ,
\ee
where we have introduced the dimensionless time variable $z = m t$
and a prime now denotes the derivative with respect to $z$.
Comparing the coefficients, we see that
\be\label{param1}
A_k=\frac{k^2+m_{\chi}^2}{m^2}+2q\qquad q=\frac{g^2\Phi^2}{4m^2} \, .
\ee

The equation (\ref{mat}) describes parametric resonance in classical mechanics.
From the examples in classical mechanics in which this equation
arises we know that there are bands of $k$ values (depending on the
frequency $m$ of the external source) which exhibit exponential
instability. For the general theory of the Mathieu equation the reader is
referred to \cite{Mathieu}. For values of $k$ in a resonance band,
the growth of $\chi_k$ can be written as
\be
\chi_k \, \sim \, e^{\mu_k z} \, ,
\ee
where $\mu_k$ is called the Floquet exponent. In the model we
are considering in this subsection, resonance occurs for all
long wavelength modes - roughly speaking all modes with
\be
k^2 \, \ll \, g m \Phi \, .
\ee
This is called {\it broad-band resonance} \cite{KLS,KLS2}. 
The Floquet exponent is of order $1$, and this implies very
efficient energy transfer from the coherently oscillating
inflaton field to a gas of $\chi$ particles. This initial
stage of energy transfer is called {\it preheating}.
The time scale of the energy transfer is short compared
to the cosmological time scale $H^{-1}$ at the end
of inflation \cite{TB}. Hence, the approximation of neglecting
the expansion of space is self-consistent.

It is possible to perform an improved analysis which keeps
track of the expansion of space. The first step \cite{KLS2} is to
rescale the field variable to extract the cosmological
red-shifting
\be
X_k(t) \, \equiv \, a(t) \chi_k(t) \, ,
\ee
and to work in terms of conformal time.
The field $X_k$ obeys an equation without cosmological
damping term which is very similar to (\ref{mat}), except
that the oscillatory correction term to the mass is replaced
by a more general periodic function of the conformal time
$\eta$. The Floquet theory of \cite{Mathieu} also applies
to the resulting equation, and we reach the conclusion
that the exponential instability of $\chi$ (modulated by
$a(t)$) persists when including the effects of the expansion
of space. For details the reader is referred to \cite{KLS2}
or to the recent review \cite{Allahverdi}.

\subsubsection{Preheating of Metric Perturbations}

A parametric resonance instability should be expected
for all fields which couple to the oscillating inflaton
condensate. In particular, metric fluctuations should
also be effected. As was shown in \cite{Fabio1}, there
is no parametric resonance instability for long
wavelength modes in the case of purely adiabatic
perturbations (see also \cite{Afshordi}). However, in
the presence of entropy fluctuations, parametric
amplification of cosmological perturbations driven
by the oscillating inflaton condensate is possible
\cite{BaVi,Fabio2} (see also \cite{BKM}).

To study the preheating of metric fluctuations, we refer
back to the formalism of cosmological perturbations
developed in Subsection (\ref{flucteom}). In the
presence of an entropy perturbation $\delta S$,
the ``conservation" equation (\ref{zetacons})
for the variable $\zeta$ (which describes the
curvature perturbation in comoving gauge) gets replaced by
\be \label{zetaent}
{\dot{\zeta}} \, = \, \frac{\dot{p}}{p + \rho} \delta S \, .
\ee
If long wavelength fluctuations of a matter field
$\chi$ which coupled to the inflaton $\varphi$
($\chi$ corresponds to an entropy mode) 
undergo exponential
growth during reheating as they do in the case of
a model with broad parametric resonance, then
this will induce an exponential growth of $\delta S$
which will via (\ref{zetaent}) induce
an exponentially growing curvature perturbation.

For an elegant formalism to compute the source term in
(\ref{zetaent}) in terms of the scalar fields $\varphi$ and
$\chi$ the reader is referred to \cite{Gordon}.
This resonant growth of entropy fluctuations is only important
in models in which the entropy fluctuations are not suppressed
during inflation. Some recent examples were studied in
\cite{BFL,ABD,Laurence}. The source for this instability need not be
the oscillations of the inflaton field. In SUSY
models, the decay of flat directions can also induce this
instability \cite{Francis}.

\subsection{Problems of Inflation}

In spite of the phenomenological success of the
inflationary paradigm, conventional scalar field-driven
inflation suffers from several important conceptual problems.

The first problem concern the nature of the inflaton, the scalar
field which generates the inflationary expansion. No particle
corresponding to the excitation of a scalar field has yet been
observed in nature, and the Higgs field which is introduced
to give elementary particles masses in the Standard Model of
particle physics does not have the required flatness of the
potential to yield inflation, unless it is non-minimally coupled
to gravity \cite{Shaposh}. In particle physics theories beyond
the Standard Model there are often many scalar fields,
but it is in general very hard to obtain the required flatness
properties on the potential 

The second problem (the {\bf amplitude problem})
relates to the amplitude of the spectrum of
cosmological perturbations. In a wide class of inflationary
models, obtaining the correct amplitude requires the introduction
of a hierarchy in scales, namely \cite{Adams}
\be
{{V(\varphi)} \over {\Delta \varphi^4}}
\, \leq \, 10^{-12} \, ,
\ee
where $\Delta \varphi$ is the change in the inflaton field during
the minimal length of the inflationary period, and $V(\varphi)$ is
the potential energy during inflation.

\begin{figure}
\includegraphics[height=9cm]{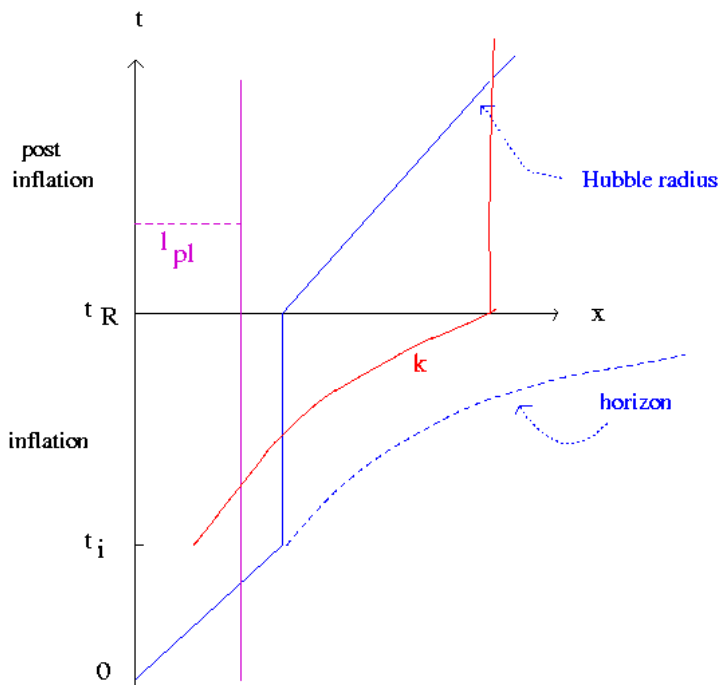}
\caption{Space-time diagram (sketch) 
of inflationary cosmology where we have added an extra
length scale, namely the Planck length $l_{pl}$ (majenta vertical line).
The symbols have the same meaning as in Figure 5.
Note, specifically, that - as long as the period of inflation
lasts a couple of e-foldings longer than the minimal value
required for inflation to address the problems of standard
big bang cosmology - all wavelengths of cosmological interest
to us today start out at the beginning of the period of inflation
with a wavelength which is smaller than the Planck length.}
\label{infl2}       
\end{figure}

A more serious problem is the {\bf trans-Planckian problem} \cite{RHBrev3}.
Returning to the space-time diagram of inflation (see Figure \ref{infl2}), 
we can immediately
deduce that, provided that the period of inflation lasted sufficiently
long (for GUT scale inflation the number is about 70 e-foldings),
then all scales inside the Hubble radius today started out with a
physical wavelength smaller than the Planck scale at the beginning of
inflation. Now, the theory of cosmological perturbations is based
on Einstein's theory of General Relativity coupled to a simple
semi-classical description of matter. It is clear that these
building blocks of the theory are inapplicable on scales comparable
and smaller than the Planck scale. Thus, the key
successful prediction of inflation (the theory of the origin of
fluctuations) is based on suspect calculations since 
new physics {\it must} enter
into a correct computation of the spectrum of cosmological perturbations.
The key question is as to whether the predictions obtained using
the current theory are sensitive to the specifics of the unknown
theory which takes over on small scales.
 
One approach to study the sensitivity of the usual predictions of
inflationary cosmology to the unknown physics on trans-Planckian scales
is to study toy models of ultraviolet physics which allow explicit
calculations. The first approach which was used \cite{Jerome1,Niemeyer}
is to replace the usual linear dispersion relation for the Fourier
modes of the fluctuations by a modified dispersion relation, a
dispersion relation which is linear for physical wavenumbers smaller
than the scale of new physics, but deviates on larger scales. Such
dispersion relations were used previously to test the sensitivity
of black hole radiation on the unknown physics of the UV 
\cite{Unruh,CJ}. It was found \cite{Jerome1} that if the evolution of modes on
the trans-Planckian scales is non-adiabatic, then substantial
deviations of the spectrum of fluctuations from the usual results
are possible. Non-adiabatic evolution turns an initial state
minimizing the energy density into a state which is excited once
the wavelength becomes larger than the cutoff scale. Back-reaction
effects of these excitations may limit the magnitude of the
trans-Planckian effects if we assume that inflation took place
\cite{Jerome2,Tanaka,Starob4}. On the other hand, large
trans-Planckian effects may also prevent the onset of inflation.

A fourth problem is the {\bf singularity problem}. It was known for a long
time that standard Big Bang cosmology cannot be the complete story of
the early universe because of
the initial singularity, a singularity which is unavoidable when basing
cosmology on Einstein's field equations in the presence of a matter
source obeying the weak energy conditions (see e.g. \cite{HE} for
a textbook discussion). Recently, the singularity theorems have been
generalized to apply to Einstein gravity coupled to scalar field
matter, i.e. to scalar field-driven inflationary cosmology \cite{Borde}.
It was shown that, in this context, a past singularity at some point
in space is unavoidable. Thus we know, from the outset, that scalar
field-driven inflation cannot be the ultimate theory of the very
early universe.

The Achilles heel of scalar field-driven inflationary cosmology may
be the {\bf cosmological constant problem}. We know from
observations that the large quantum vacuum energy of field theories
does not gravitate today. However, to obtain a period of inflation
one is using the part of the energy-momentum tensor of the scalar field
which looks like the vacuum energy. In the absence of a 
solution of the cosmological constant problem it is unclear whether
scalar field-driven inflation is robust, i.e. whether the
mechanism which renders the quantum vacuum energy gravitationally
inert today will not also prevent the vacuum energy from
gravitating during the period of slow-rolling of the inflaton 
field. Note that the
approach to addressing the cosmological constant problem making use
of the gravitational back-reaction of long range fluctuations
(see \cite{RHBrev4} for a summary of this approach) does not prevent
a long period of inflation in the early universe.

A final problem which we will mention here is the concern that the
energy scale at which inflation takes place is too high to justify
an effective field theory analysis based on Einstein gravity. In
simple toy models of inflation, the energy scale during the period
of inflation is about $10^{16} \rm{GeV}$, very close to the string scale
in many string models, and not too far from the Planck scale. Thus,
correction terms in the effective action for matter and gravity may 
already be important at the energy scale of inflation, and the 
cosmological dynamics may be rather different from what is
obtained when neglecting the correction terms.

\begin{figure}
\includegraphics[height=10cm]{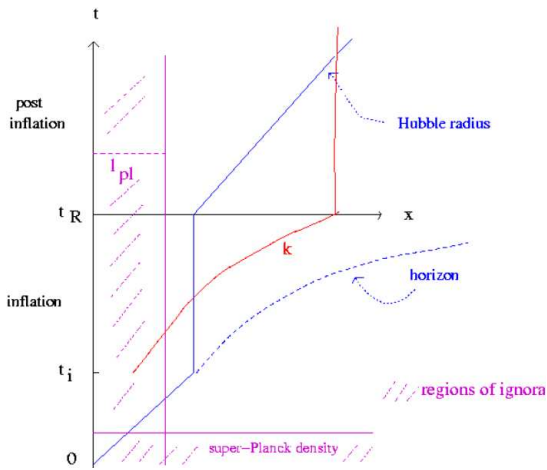}
\caption{Space-time diagram (sketch) 
of inflationary cosmology including the two zones of
ignorance - sub-Planckian wavelengths and trans-Planckian 
densities. The symbols have the same meaning as in Figure 5.
Note, specifically, that - as long as the period of inflation
lasts a couple of e-foldings longer than the minimal value
required for inflation to address the problems of standard
big bang cosmology - all wavelengths of cosmological interest
to us today start out at the beginning of the period of inflation
with a wavelength which is in the zone of ignorance.}
\label{infl3}       
\end{figure}

In Figure \ref{infl3} we show once again the space-time sketch
of inflationary cosmology. In addition to the length scales
which appear in the previous versions of this figure,
we have now shaded the ``zones of ignorance", zones
where the Einstein gravity effective action is sure to
break down. As described above, fluctuations emerge
from the short distance zone of ignorance (except if
the period of inflation is very short), and the energy
scale of inflation might put the period of inflation too
close to the high energy density zone of ignorance to
trust the predictions based on using the Einstein
action. 

The arguments in this subsection provide a motivation
for considering alternative scenarios of early universe
cosmology. Below we will focus on two scenarios, the
{\it matter bounce} and {\it string gas cosmology}. It is
important to emphasize that these are not the only
alternative scenarios. Others include the 
{\it Pre-Big-Bang scenario} \cite{PBB}
and the {\it Ekpyrotic paradigm} \cite{Ekp}. 

\section{Matter Bounce}

The {\it matter bounce} scenario is a non-singular cosmology
in which time runs from $- \infty$ to $+ \infty$. Negative times
correspond to a contracting phase, positive times to expansion.
It is assumed that the bounce phase is short \footnote{Short means short
compared to the time it takes light to travel over distances of
about $1 {\rm mm}$, the physical length which corresponds to
to the largest comoving scales we observe today (assuming that
the energy scale of the bounce is about $10^{16} {\rm GeV}$).}.

As indicated in Figure \ref{bounce}, the fluctuations which we observe
today have exited the Hubble radius early in the contracting
phase. The word {\it matter} in {\it matter bounce} should
emphasize that we are assuming that the phase during which
the fluctuations which we observe today are exiting the
Hubble radius is dominated by cold pressure-less matter.
Viewing the contracting phase as the time reverse of the
expanding phase we are living in, this last assumption we
have made does not seem very restrictive. In fact, one could
expect that some entropy is generated during the bounce,
in which case the matter phase during the contracting period
would last up to higher energy densities than it does in the
expanding phase.

\subsection{Models for a Matter Bounce and Background Cosmology}

It follows from the singularity theorems of General Relativity 
(see e.g. \cite{HE}) that in order to obtain a non-singular bounce
we must either invoke matter which violates the weak energy
condition, or go beyond General Relativity. A review of ways
to obtain a non-singular cosmology is given in \cite{Novello}.
The major challenges in constructing a non-singular bouncing
cosmology include the following: first, the model should be free
of ghost-like excitations, since a model with ghosts will be
unstable \cite{ghost}. Next, the background cosmology should be
stable towards the addition of radiation to the matter sector.
Finally, the homogeneous and isotropic background trajectory
should also be stable against anisotropic stress. 

In the following we will only mention a few recent attempts which
the author of these lecture notes has been involved in. We will
first discuss models obtained by modifying the matter sector
of the theory, and then models where the gravitational sector
of the theory is changed.

Introducing quintom matter \cite{quintom} yields a way of obtaining
a non-singular bouncing cosmology, as discussed in
\cite{Cai1}. Quintom matter is a set of two matter
fields, one of them regular matter (obeying the weak energy
condition), the second a field with opposite sign kinetic
term, a field which violates the energy conditions. We can 
\cite{Cai1} model both matter components with
scalar fields. Let us denote the mass of the regular one 
($\varphi$) by $m$, 
and by $M$ that of the field ${\tilde{\varphi}}$ with 
wrong sign kinetic term. We assume that early in the contracting
phase both fields are oscillating, but that the amplitude $\cal{A}$
of $\varphi$ greatly exceeds the corresponding amplitude
${\tilde{\cal A}}$ of ${\tilde{\varphi}}$ such that the energy density
is dominated by $\varphi$. Both fields will initially be oscillating 
during the contracting phase, and both amplitudes grow at the
same rate. At some point, $\cal{A}$ will become so large 
that the oscillations of $\varphi$ freeze out \footnote{This
corresponds to the time reverse of
entering a region of large-field inflation.}.
Then, $\cal{A}$ will grow only slowly, whereas ${\tilde{\cal A}}$
will continue to increase. Thus, the (negative) energy density
in ${\tilde{\varphi}}$ will grow in absolute values relative to that
of $\varphi$. The total energy density will decrease towards $0$.
At that point, $H = 0$ by the Friedmann equations. It can in
fact easily be seen that ${\dot{H}} > 0$ when $H = 0$. Hence,
a non-singular bounce occurs. The Higgs sector of the
Lee-Wick model \cite{LW} provides a concrete realization of 
the quintom bounce model, as studied in \cite{LWbounce}.
Quintom models like all other models with negative sign kinetic
terms suffer from an instability problem \cite{ghost} in the
matter sector and are hence problematic. In addition, they are
unstable against the addition of radiation (see e.g. \cite{Karouby})
and anisotropic stress.

An improved way of obtaining a non-singular bouncing cosmology
using modified matter is by using a ghost condensate field \cite{Chunshan}.
The ghost condensation mechanism is the analog of the Higgs
mechanism in the kinetic sector of the theory. In the Higgs
mechanism we take a field $\phi$ whose mass when evaluated
at $\phi = 0$ is tachyonic, add higher powers of $\phi^2$ to the
potential term in the Lagrangian such that there is a stable
fixed point $\phi = v \neq 0$, and thus when expanded about
$\phi = v$ the mass term has the ``safe" non-tachyonic sign.
In the ghost condensate construction we take a field $\phi$
whose kinetic term
\be
X \, \equiv \,  - g^{\mu \nu} \partial_{\mu} \phi \partial_{\nu} \phi 
\ee
appears with the wrong sign in the Lagrangian. Then, we add
higher powers of $X$ to the kinetic Lagrangian such that there
is a stable fixed point $X = c^2$ and such that when expanded
about $X = c^2$ the fluctuations have the regular sign of the kinetic
term:
\be
{\cal L} \, = \, \frac{1}{8} M^4 \bigl( X - c^2 \bigr)^2 - V(\phi) \, ,
\ee
where $V(\phi)$ is a usual potential function, $M$ is a 
characteristic mass scale and the dimensions of $\phi$
are chosen such that $X$ is dimensionless.

In the context of cosmology, the ghost condensate is
\be
\phi \, = \, c t 
\ee
and breaks local Lorentz invariance. Now let us expand the
homogeneous component of $\phi$ about the ghost condensate:
\be
\phi(t) \, = \, c t + \pi(t) \, .
\ee
If ${\dot \pi} < 0$ then the gravitational energy density is negative,
and a non-singular bounce is possible. Thus, in \cite{Chunshan}
we constructed a model in which the ghost condensate field
starts at negative values and the potential $V(\phi)$ is
negligible. As $\phi$ approaches $\phi = 0$ it encounters a
positive potential which slows it down, leading to ${\dot \pi} < 0$
and hence to negative gravitational energy density. Thus,
a non-singular bounce can occur. We take the potential
to be of the form
\be
V(\phi) \, \sim \, \phi^{- \alpha}
\ee
 for $|\phi| \gg M$, where $M$ is the mass scale above which
 the higher derivative kinetic terms are important. For
 sufficiently large values of $\alpha$, namely
 \be
 \alpha \, \geq \, 6 \, ,
 \ee
 the energy density in the ghost condensate increases faster
 than that of radiation and anisotropic stress at the universe
 contracts . Hence, this bouncing cosmology is stable
 against the addition of radiation and anisotropic stress.

Turning to the second way of obtaining a non-singular bounce,
namely by modifying the gravitational action, we should emphasize
that modifications of the gravitational action are expected at
the high densities at which the bounce will occur. Concrete
examples were already mentioned in the Introduction: the
higher derivative Lagrangian resulting from the nonsingular 
universe construction of \cite{MBS}, the model of \cite{Biswas1}
which is based on a non-local higher derivative action which
is ghost-free about Minkowski space-time, mirage cosmologies
\cite{mirage} resulting from the effective action of gravity on
a brane which is moving into and out of a high-curvature
throat in a higher-dimensional space-time. 

More recently,
interest has focused on a non-singular bouncing cosmology
which emerges from Horava-Lifshitz gravity \cite{Horava}.
Horava-Lifshitz gravity is a power-counting renormalizable
quantum theory of gravity which is based on anisotropic
scaling between space and time. In one sense it is a
very conservative approach to quantum gravity in the 
sense that the dynamical degrees of freedom are
the usual ones: the spatial metric $g_{i j}$, the lapse
function $N$ and the shift vector $N_i$. There are no
extra dimensions or new degrees of freedom such as
strings. On the other hand, it is a very radical approach
in the sense that it gives up local Lorentz invariance
and space-time diffeomorphism invariance. The residual
symmetries are local rotational invariance and spatial
diffeomorphism. On the other hand, there is a scaling
symmetry under
\be
x \, \rightarrow \, l x  \,\,\,\, , \,\,\, t \, \rightarrow \, l^3 t \, ,
\ee
where $l$ is the scaling parameter. The Lagrangian
is constructed to contain all terms which are consistent
with the residual symmetries and with power-counting
renormalizability with respect to the above scaling
symmetry. The action does not contain any higher time 
derivative terms, but it does contain higher space derivative 
terms. These terms dominate in the ultraviolet. In the infrared, the
extra terms are sub-dominant. Thus, in the infrared
the theory is supposed to flow towards General Relativity.
 
The extra spatial derivative terms in the gravitational action
have an important effect in the very early universe at high
energy densities. As shown in \cite{HLbounce}, if the spatial sections have 
a non-vanishing spatial curvature constant $k$, then
the higher-derivative terms in the action will lead to
terms in the Friedmann equations which act as ghost
radiation and ghost anisotropic stress, i.e. terms of
gravitational origin and negative effective energy
density which scale as $a^{-4}$ and $a^{-6}$, respectively.
Starting with a contracting universe dominated by
regular matter, eventually the
ghost terms will catch up and yield a non-singular
bounce in analogy to how the ghost matter in the quintom
model does.

The analysis of the spectrum of cosmological perturbations
on scales of current observational interest is, however,
independent of the details of the bouncing phase, as long
as that phase is short compared to the time it takes light
to travel over the length scales of current interest, and
as long as no extra degrees of freedom (extra when compared
to those arising in Einstein gravity) become important
\footnote{This last condition is very relevant in the case
of the Horava-Lifshitz bounce where we do have to worry
about extra gravitational degrees of freedom. We will
come back to this issue at the end of the following
subsection.}. We
now turn to a discussion of the evolution of fluctuations
in a generic matter bounce model.

\subsection{Fluctuations in a Matter Bounce}

First we will consider fluctuations in a matter bounce without
extra degrees of freedom. In this case, we need only focus
on the usual fluctuation variable $v$.
The equation of motion its Fourier mode $v_k$ is
\be \label{EOM}
v_k^{''} + \bigl( k^2 - \frac{z^{''}}{z} \bigr) v_k \, = \, 0 \, .
\ee
If the equation of state of the background is time-independent, then
$z \sim a$ and hence the negative square mass term in (\ref{EOM})
is $H^2$. Thus, on length scales smaller than the Hubble radius,
the solutions of (\ref{EOM}) are oscillating, whereas on larger
scales they are frozen in, and their amplitude
depends on the time evolution of $z$. 

In the case of an expanding universe the dominant mode of $v$ scales as
$z$. However, in a contracting universe it is the second of the
two modes which dominates. 
If the contracting phase is matter-dominated, i.e. $a(t) \sim t^{2/3}$
and $\eta(t) \sim t^{1/3}$ the dominant mode of $v$ scales as $\eta^{-1}$
and hence
\be
v_k(\eta) \, = \, c_1 \eta^2 + c_2 \eta^{-1} \, ,
\ee
where $c_1$ and $c_2$ are constants. The $c_1$ mode is the 
mode for which $\zeta$ is constant on super-Hubble scales. However,
in a contracting universe it is the $c_2$ mode which dominates and
leads to a scale-invariant spectrum \cite{Wands,FB2,Wands2}:
\bea
P_{\zeta}(k, \eta) \, &\sim& k^3 |v_k(\eta)|^2 a^{-2}(\eta) \\
&\sim& \, k^3 |v_k(\eta_H(k))|^2 \bigl( \frac{\eta_H(k)}{\eta} \bigr)^2 \, 
\sim \, k^{3 - 1 - 2} \nonumber \\
&\sim& \, {\rm const}  \, , \nonumber 
\eea
where we have made use of the scaling of the dominant mode of $v_k$, the 
Hubble radius crossing condition $\eta_H(k) \sim k^{-1}$, and
the assumption that we have a vacuum spectrum at Hubble radius crossing.

Up to this point, the analysis shows that in the contracting phase the
curvature fluctuations are scale-invariant. In non-singular bouncing
cosmologies, the fluctuations can be followed in an unambiguous
way through the bounce. This was done in the case of the higher
derivative bounce model of \cite{Biswas1} in \cite{ABB}, and in a 
non-singular bouncing mirage cosmology model the analysis was
performed in \cite{BFS}. In the case of the Quintom and Lee-Wick bounces,
respectively,  the fluctuations were followed through the bounce
in \cite{Cai2} and \cite{LWbounce}, respectively. Finally, in the case of
the non-projectable version \footnote{See below for the specification
of this case.} of the HL bounce, the evolution of the fluctuations through the bounce
was recently studied in \cite{HLbounce2}.

The equations of motion can be solved numerically without approximation.
Alternatively, we can solve them approximately using analytical techniques.
Key to the analytical analysis are the General Relativistic matching
conditions for fluctuations across a phase transition in the background
\cite{HV,DM}. These conditions imply that both $\Phi$ and $\tilde{\zeta}$
are conserved at the bounce, where
\be
\tilde{\zeta} \, = \, \zeta + \frac{1}{3} \frac{k^2 \Phi}{{\cal{H}}^2 - {\cal{H}}^{'}} \, .
\ee
However, as stressed in \cite{Durrer2}, these matching conditions
can only be used at a transition when the background metric
obeys the matching conditions. This is not the case if we
were to match directly between the contracting matter phase
and the expanding matter phase. 

In the case of a non-singular
bounce we have three phases: the initial contracting phase with
a fixed equation of state (e.g. $w = 0$), a bounce phase during
which the universe smoothly transits between contraction
and expansion, and finally the expanding phase with constant $w$. 
We need to apply  the matching conditions twice: 
first at the transition between the
contracting phase and the bounce phase (on both sides of
the matching surface the universe is contracting), and then between
the bouncing phase and the expanding phase. The bottom line
of the studies of  \cite{ABB,BFS,Cai2,LWbounce,HLbounce2} is that
on length scales large compared to the time of the bounce, the
spectrum of curvature fluctuations is not changed during the
bounce phase. Since typically the bounce time is set by
a microphysical scale whereas the wavelength of fluctuations
which we observe today is macroscopic (about $1 {\rm mm}$
if the bounce scale is set by the particle physics GUT scale),
we conclude that for scales relevant to current observations
the spectrum is unchanged during the bounce. This completes
the demonstration that a non-singular matter bounce leads
to a scale-invariant spectrum of cosmological perturbations
after the bounce provided that the initial spectrum on
sub-Hubble scales is vacuum. Initial thermal fluctuations
were followed through the bounce in \cite{Thermalflucts}.
Note that perturbations are processed during a bounce
and this implies that even if the background cosmology
were cyclic, the full evolution including linear fluctuations
would be non-cyclic \cite{processing}.

The above analysis is applicable only as long as no new
degrees of freedom become relevant at high energy
densities, in particular during the bounce phase. In
non-singular bounce models obtained by modifying
the matter sector, new degrees of freedom
arise from the extra matter fields. They can thus
give entropy fluctuations which may compete with
the adiabatic mode studied above. In the quintom
bounce model this issue has recently been studied
in \cite{Yifu}. It was found that fluctuations in the ghost
field which yields the bounce are unimportant on large
scales since they have a blue spectrum. However,
entropy fluctuations due to extra low-mass fields
can be important. They yield the ``matter bounce
curvaton" mechanism. Their spectrum is also
scale-invariant.

In non-singular bouncing models obtained by modifying
the gravitational sector of the theory the identification
of potential extra degrees of freedom is more difficult.
As an example, let us mention the situation in the
case of the Horava-Lifshitz bounce. The theory has
the same number of geometric degrees of freedom
as General Relativity, but less symmetries. Thus, more
of the degrees of freedom are physical. Recall from
the discussion of the theory of cosmological perturbations
in Section 2 that there are ten total geometrical degrees
of freedom for linear cosmological perturbations, four
of them being scalar, four vector and two tensor. In Einstein
gravity the symmetry group of space-time diffeomorphisms is
generated at the level of linear fluctuations by four functions,
leaving six of the ten geometrical variables as physical -
two scalar, two vector and two tensor modes.
In the absence of anisotropic stress the number of scalar variables
is reduced by one, and the Hamiltonian constraint relates
the remaining scalar metric fluctuation to matter.

In Horava-Lifshitz gravity one loses one scalar gauge
degree of freedom, namely that of space-dependent
time reparametrizations. Thus, one expects an extra
physical degree of freedom. It has been recently
been shown \cite{Cerioni1} that in the projectable
version of the theory (in which the lapse function
$N(t)$ is constrainted to be a function of time only) the
extra degree of scalar cosmological perturbations is
either ghost-like or tachyonic, depending on parameters
in the Lagrangian. Thus, the theory appears to be
ill-behaved in the context of cosmology. However, in
the full non-projectable version (in which the lapse
$N(t, {\bf x})$ is a function of both space and time,
the extra degree of freedom is well behaved. It is
important on ultraviolet scales but decouples in the
infrared \cite{Cerioni2}.


\subsection{Key Predictions of the Matter Bounce}

Canonical single field inflation models predict very small
non-Gaussianities in the spectrum of fluctuations. One
way to characterize the non-Gaussianities is via the
three point function of the curvature fluctuation, also
called the ``bispectrum". As
realized in \cite{Xue}, the bispectrum induced
in the minimal matter bounce scenario 
(no entropy modes considered) has an amplitude which
is at the borderline of what the Planck satellite experiment
will be able to detect, and it has a special form.
These are specific predictions of the matter bounce
scenario with which the matter bounce scenario can
be distinguished from those of standard inflationary models
(see \cite{Xingang} for a recent detailed review of
non-Gaussianities in inflationary cosmology and a list
of references). In the following we give a very brief
summary of the analysis of non-Gaussianities
in the matter bounce scenario.

Non-Gaussianities are induced in any cosmological model
simply because the Einstein equations are non-linear.
In momentum space, the bispectrum contains amplitude
and shape information. The bispectrum is a function of
the three momenta. Momentum conservation implies
that the three momenta have to add up to zero. However,
this still leaves a rich shape information in the bispectrum
in addition to the information about the overall amplitude.

A formalism to compute the non-Gaussianities for the
curvature fluctuation variable $\zeta$ was developed in 
\cite{Malda}. Working in the interaction representation,
the three-point function of $\zeta$ is given to leading
order by
\begin{eqnarray}\label{tpf}
&& <\zeta(t,\vec{k}_1)\zeta(t,\vec{k}_2)\zeta(t,\vec{k}_3)>  \\
&& \,\,\,\, = \, i \int_{t_i}^{t} dt'
<[\zeta(t,\vec{k}_1)\zeta(t,\vec{k}_2)\zeta(t,\vec{k}_3),{L}_{int}(t')]>~,
\nonumber
\end{eqnarray}
where $t_i$ corresponds to the initial time before which
there are any non-Gaussianities. The square parentheses indicate
the commutator, and $L_{int}$ is the interaction Lagrangian

The interaction Lagrangian contains many terms. In particular,
there are terms containing the time derivative of $\zeta$. Each
term leads to a particular shape of the bispectrum. In an
expanding universe such as in inflationary cosmology
$\dot{\zeta} = 0$. However, in a contracting phase the time
derivative of $\zeta$ does not vanish since the dominant
mode is growing in time. Hence, there are new contributions
to the shape which have a very different form from the shape
of the terms which appear in inflationary cosmology.
The larger value of the amplitude of the bispectrum follows
again from the fact that there is a mode function which grows
in time in the contracting phase.

The three-point function can be expressed in the following
general form:
\bea
<\zeta(\vec{k}_1)\zeta(\vec{k}_2)\zeta(\vec{k}_3)> \, &=& \, 
(2\pi)^7 \delta(\sum\vec{k}_i) \frac{P_{\zeta}^2}{\prod k_i^3} \nonumber \\
&& \times {\cal A}(\vec{k}_1,\vec{k}_2,\vec{k}_3)~,
\eea
where $k_i=|\vec{k}_i|$ and ${\cal A}$ is the shape function. In
this expression we have factored out the dependence on
the power spectrum $P_{\zeta}$. In inflationary cosmology
it has become usual to express the bispectrum in terms
of a non-Gaussianity parameter $f_{NL}$. However, 
this is only useful if the shape of the three point function is
known. As a generalization, we here use \cite{Xue}
\be
{\cal |B|}_{NL}(\vec{k}_1,\vec{k}_2,\vec{k}_3) \, = \, \frac{10}{3}\frac{{\cal
A}(\vec{k}_1,\vec{k}_2,\vec{k}_3)}{\sum_ik_i^3}~.
\ee

The computation of the bispectrum is tedious. In the
case of the matter bounce (no entropy fluctuations)
the result is
\begin{eqnarray} \label{result}
{\cal A} \, &=& \, \frac{3}{256\prod k_i^2} \bigg\{ 3\sum k_i^9 +
\sum_{i\neq j}k_i^7k_j^2 \nonumber \\
&& - 9 \sum_{i\neq j}k_i^6k_j^3 +5\sum_{i \neq j}k_i^5k_j^4 \\
&& -66 \sum_{i\neq j\neq k}k_i^5k_j^2k_k^2
+9\sum_{i\neq j\neq k}k_i^4k_j^3k_k^2 \bigg\}~. \nonumber
\end{eqnarray}
This equation describes the shape which is predicted.
Some of the terms (such as the last two) are the same
as those which occur in single field slow-roll inflation,
but the others are new. Note, in particular, that
the new terms are not negligible. 

If we project the
resulting shape function ${\cal A}$ onto some
popular shape masks we 
\begin{eqnarray}
{\cal |B|}_{NL}^{\rm local} \, = \, -\frac{35}{8}~,
\end{eqnarray}
for the local shape ($k_1 \ll k_2 = k_3$). This
is negative and of order $O(1)$.
For the equilateral form
($k_1= k_2= k_3$) the result is
\begin{eqnarray}
{\cal |B|}_{NL}^{\rm equil} \, = \, -\frac{255}{64}~\, ,
\end{eqnarray}
and for the folded form ($k_1=2k_2=2k_3$) one
obtains the value
\begin{eqnarray}
{\cal |B|}_{NL}^{\rm folded} \, = \, -\frac{9}{4}~.
\end{eqnarray}
These amplitudes are close to what the
Planck CMB satellite experiment will be
able to detect.

The matter bounce scenario also predicts a change in the
slope of the primordial power spectrum on small
scales \cite{LiHong}: scales which exit the Hubble
radius in the radiation phase retain a blue
spectrum since the squeezing rate on scales larger
than the Hubble radius is insufficient to give longer
wavelength modes a sufficient boost relative to
the shorter wavelength ones.

\subsection{Problems of the Matter Bounce}

The main problem of the matter bounce scenario
is that the physics which yields the non-singular
bounce is not (yet) well established. It is clearly
required to go beyond conventional quantum
field theory coupled to General Relativity to
obtain such a bounce. New Planck-scale physics
needs to be invoked to obtain the required
background cosmology. We remind the
reader that it is precisely the absence of such
new physics which needs to be invoked to
argue for an inflationary background evolution.

In terms of the trans-Planckian problem for
cosmological fluctuations, the bouncing
cosmology has a clear advantage: the wavelength
of the fluctuations which we are interested in
always remains in the far infrared compared to
the ultraviolet scales of the new physics. It
can be shown that the correction terms due
to the new ultraviolet physics on the evolution
of fluctuations of interest to observers are
negligible (see e.g. \cite{HLbounce2}).

The second main problem of the matter
bounce scenario may be the sensitivity
of the bounce to assumptions on the initial
conditions in the far past. The amplitude
of the classical initial fluctuations must be
small in order for the homogeneous
background dynamics not to be perturbed,
and for the vacuum fluctuations to dominate
on the scales relevant in the infrared
(see e.g. \cite{Lindecrit} for a criticism
of initial conditions in a related scenario,
the Pre-Big-Bang scenario). In addition,
the initial shear must be very small. In some
bouncing cosmology backgrounds the
bouncing solution is unstable to the smallest
addition of shear. This, however, is not the
case in other models such as the Horava-Lifshitz
bounce \cite{HLbounce} or the 
ghost condensate bounce \cite{Chunshan}.

The matter bounce scenario does not
address the cosmological constant problem,
but on the other hand the scenario does
not depend on how the cosmological
constant problem is solved. Note that
inflationary cosmology is NOT robust
in this respect.

\section{String Gas Cosmology}

Another scenario of early universe cosmology which can
provide an explanation for the current data is string
gas cosmology. It is an ``emergent universe" \cite{Ellis2}
scenario in which the universe begins in a long hot
and almost static phase. The key input from string
theory is that matter is a gas of closed fundamental strings
compared to a gas of point particles as is assumed in
Standard Big Bang cosmology.

\subsection{Principles and Background}

\subsubsection{Principles}

String theory may be the best candidate we have at the
present time for a quantum theory of gravity which unifies
all four forces of nature. This theory is, however, currently
not yet well understood beyond perturbation theory
\footnote{Note that there are concrete proposals for a
non-perturbative formulation such as the AdS/CFT
correspondence \cite{Malda2}.}. For
applications to early universe cosmology, however, a
non-perturbative understanding will be essential.

In the absence of a non-perturbative formulation of string theory,
the approach to string cosmology which we have suggested,
{\it string gas cosmology} \cite{BV} (see also \cite{Perlt},
and \cite{RHBSGrev,RHBrev6,BattWatrev} for reviews and 
more complete list of references),
is to focus on symmetries and degrees of freedom which are new to
string theory (compared to point particle theories) and which will
be part of any non-perturbative string theory, and to use
them to develop a new cosmology. The symmetry we make use of is
{\bf T-duality}, and the new degrees of freedom are the {\bf string
oscillatory modes} and the {\bf string winding modes}.

String gas cosmology is based on coupling a classical background
which includes the graviton and the dilaton fields to a gas of
closed strings (and possibly other basic degrees of freedom of
string theory such as ``branes" \cite{ABE}). All dimensions of space are taken
to be compact, for reasons which will become clear later.
For simplicity, we take all spatial directions to be toroidal and
 denote the radius of the torus by $R$. Strings have three types
of states: {\it momentum modes} which represent the center
of mass motion of the string, {\it oscillatory modes} which
represent the fluctuations of the strings, and {\it winding
modes} counting the number of times a string wraps the torus.

Since the number of string oscillatory states increases exponentially
with energy, there is a limiting  temperature for a gas of strings in
thermal equilibrium, the {\it Hagedorn temperature} \cite{Hagedorn}
$T_H$. Thus, if we take a box of strings and adiabatically decrease the box
size, the temperature will never diverge. This is the first indication that
string theory has the potential to resolve the cosmological singularity
problem.

The second key feature of string theory upon which string gas cosmology
is based is {\it T-duality}. To introduce this symmetry, let us discuss the
radius dependence of the energy of the basic string states:
The energy of an oscillatory mode is independent of $R$, momentum
mode energies are quantized in units of $1/R$, i.e.
\be
E_n \, = \, n \mu \frac{{l_s}^2}{R} \, ,
\ee
where $l_s$ is the string length and $\mu$ is the mass per unit length of
a string. The winding mode energies are 
quantized in units of $R$, i.e.
\be
E_m \, = \, m \mu R \, ,
\ee
where both $n$ and $m$ are integers. Thus, a new symmetry of
the spectrum of string states emerges: Under the change
\be
R \, \rightarrow \, 1/R
\ee
in the radius of the torus (in units of  $l_s$)
the energy spectrum of string states is
invariant if winding
and momentum quantum numbers are interchanged
\be
(n, m) \, \rightarrow \, (m, n) \, .
\ee
The above symmetry is the simplest element of a larger
symmetry group, the T-duality symmetry group which in
general also mixes fluxes and geometry.
The string vertex operators are consistent with this symmetry, and
thus T-duality is a symmetry of perturbative string theory. Postulating
that T-duality extends to non-perturbative string theory leads
\cite{Pol} to the need of adding D-branes to the list of fundamental
objects in string theory. With this addition, T-duality is expected
to be a symmetry of non-perturbative string theory.
Specifically, T-duality will take a spectrum of stable Type IIA branes
and map it into a corresponding spectrum of stable Type IIB branes
with identical masses \cite{Boehm}. 

As discussed in \cite{BV}, the above T-duality symmetry leads to
an equivalence between small and large spaces, an equivalence
elaborated on further in \cite{Hotta,Osorio}.

\subsubsection{Background Cosmology}

\begin{figure} 
\includegraphics[height=6cm]{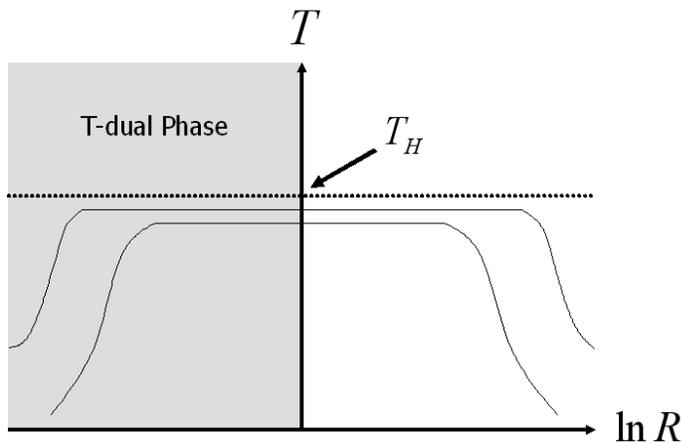}
\caption{The temperature (vertical axis) as a function of
radius (horizontal axis) of a gas of closed strings in thermal
equilibrium. Note the absence of a temperature singularity. The
range of values of $R$ for which the temperature is close to
the Hagedorn temperature $T_H$ depends on the total entropy
of the universe. The upper of the two curves corresponds to
a universe with larger entropy.}
\label{jirofig1}
\end{figure}

That string gas cosmology will lead to a dynamical evolution of the
early universe very different from what is obtained in standard and
inflationary cosmology can already be seen by combining the
basic ingredients from string theory discussed in the previous
subsection. As the radius of a box of strings decreases from an
initially very large value - maintaining thermal
equilibrium - , the temperature first rises as in
standard cosmology since the string states which are occupied
(the momentum modes) get heavier. However, as the temperature
approaches the Hagedorn temperature, the energy begins to
flow into the oscillatory modes and the increase in temperature
levels off. As the radius $R$ decreases below the string scale,
the temperature begins to decrease as the energy begins to
flow into the winding modes whose energy decreases as $R$
decreases (see Figure \ref{jirofig1}). Thus, as argued in \cite{BV},  
the temperature singularity of early universe cosmology
should be resolved  in string gas cosmology.

The equations that govern the background of string gas cosmology
are not known. The Einstein equations are not the correct
equations since they do not obey the T-duality symmetry of
string theory. Many early studies of string gas cosmology were
based on using the dilaton gravity equations \cite{TV,Ven,Tseytlin}. However,
these equations are not satisfactory, either. Firstly,
we expect that string theory correction terms to the
low energy effective action of string theory become dominant
in the Hagedorn phase. Secondly, the dilaton gravity
equations yield a
rapidly changing dilaton during the Hagedorn phase (in the
string frame). Once the dilaton becomes large, it becomes
inconsistent to focus on fundamental string states rather
than brane states. In other words, using dilaton gravity as a
background for string gas cosmology does not correctly
reflect the S-duality symmetry of string theory. Recently, a
background for string gas cosmology including a rolling
tachyon was proposed \cite{Kanno1} which allows a background
in the Hagedorn phase with constant scale factor and constant
dilaton. Another study of this problem was given in \cite{Sduality}.

\begin{figure} 
  \includegraphics[height=6cm]{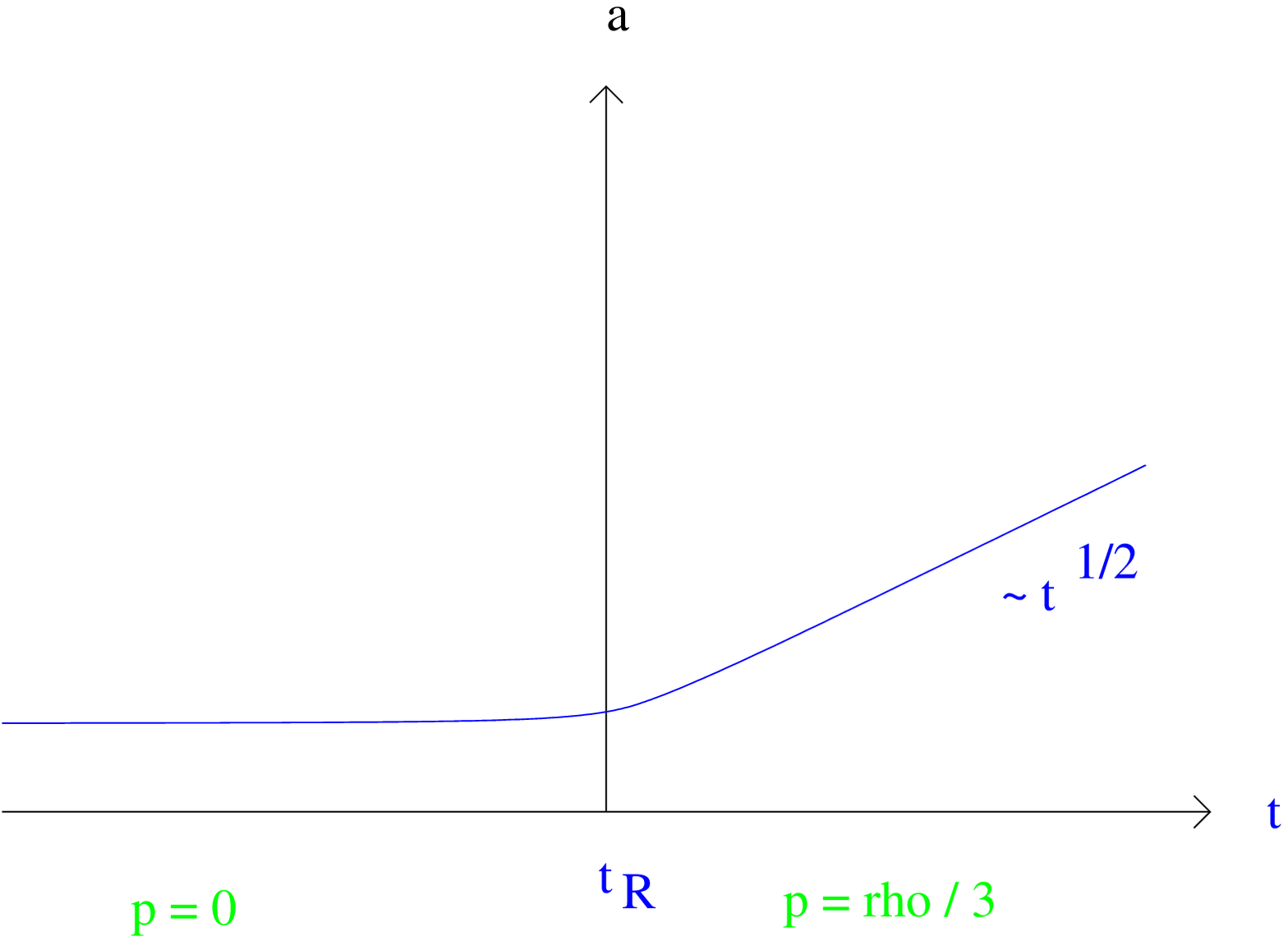}
\caption{The dynamics of string gas cosmology. The vertical axis
represents the scale factor of the universe, the horizontal axis is
time. Along the horizontal axis, the approximate equation of state
is also indicated. During the Hagedorn phase the pressure is negligible
due to the cancellation between the positive pressure of the momentum
modes and the negative pressure of the winding modes, after time $t_R$
the equation of state is that of a radiation-dominated universe.}
\label{timeevol2}
\end{figure}

Some conclusions about the time-temperature relation in string
gas cosmology can be derived based on thermodynamical
considerations alone. One possibility is that  $R$ starts out
much smaller than the self-dual value and increases monotonically.
From Figure \ref{jirofig1} it then follows that the time-temperature curve
will correspond to that of a bouncing cosmology. Alternatively,
it is possible that the universe starts out in a meta-stable state
near the Hagedorn temperature, the {\it Hagedorn phase}, and
then smoothly evolves into an expanding phase dominated by
radiation like in standard cosmology (Figure \ref{timeevol2}). 
Note that we are assuming that not only the scale factor but
also the dilaton is constant in time.  

\begin{figure} 
  \includegraphics[height=2.8cm]{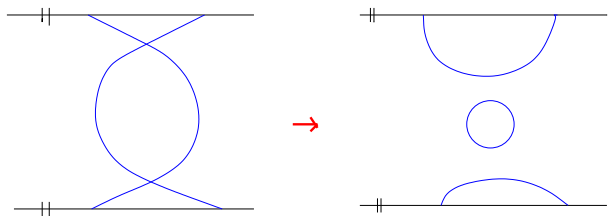}
\caption{The process by which string loops are produced via the
intersection of winding strings. The top and bottom lines
are identified and the space between these lines represents
space with one toroidal dimension un-wrapped.}
\label{decay}
\end{figure}

The transition between the quasi-static Hagedorn phase and the
radiation phase at the time $t_R$ is a consequence of 
the annihilation of string winding modes into string loops (see Figure \ref{decay}). 
Since this process corresponds to the production of radiation, we denote
this time by the same symbol $t_R$ as the time of reheating in inflationary
cosmology. As pointed out in \cite{BV}, this annihilation process
only is possible in at most three large spatial dimensions. This is
a simple dimension counting argument: string world sheets have
measure zero intersection probability in more than four large 
space-time dimensions. Hence, string gas cosmology
may provide a natural mechanism for explaining why there are
exactly three large spatial dimensions. This argument was
supported by numerical studies of string evolution in three and
four spatial dimensions \cite{Mairi} (see also \cite{Cleaver}). 
The flow of energy from
winding modes to string loops can be modelled by effective
Boltzmann equations \cite{BEK} analogous to those used to
describe the flow of energy between infinite cosmic strings and
cosmic string loops (see e.g. \cite{ShellVil,HK,RHBtoprev} for
reviews). 

Several comments are in place concerning the above mechanism.
First, in the analysis of \cite{BEK} it was assumed that the
string interaction rates were time-independent. If the dynamics of
the Hagedorn phase is modelled by dilaton gravity, the dilaton is
rapidly changing during the phase in which the string frame scale
factor is static. As discussed in \cite{Col2,Danos} (see also \cite{Kabat}), 
in this case the
mechanism which tells us that exactly three spatial dimensions 
become macroscopic does not work. Another comment concerns
the isotropy of the three large dimensions. In contrast to the
situation in Standard cosmology, in string gas cosmology the
anisotropy decreases in the expanding phase \cite{Watson1}.
Thus, there is a natural isotropization mechanism for the three
large spatial dimensions.

\subsubsection{Moduli Stabilization}

In the following, we shall assume that either the mechanism of \cite{BV}
for setting in motion the preferential expansion of exactly three
spatial dimensions works, or, alternatively, that three dimensions
are distinguished from the beginning as being large. In either case,
we must address the moduli stabilization problem, i.e. we must
show that the radii (radions) and shapes of the extra dimensions are stabilized.
This is a major challenge in string-motivated field theory models
of higher dimensions. The situation in string gas cosmology is
much better in comparison: all moduli except for the dilaton
are stabilized without the need of introducing extra ingredients
such as fluxes or special non-perturbative effects.

Radion stabilization in the string frame was initially studied in
\cite{Watson2}. The basic idea is that the winding modes about
the extra spatial dimensions provide a confining force which prevents
the radii from increasing whereas the momentum modes provide
a force which resists the complete contraction. Thus, there will
be a stable minimum of the effective potential for the
radion. 

In order to make contact with late time cosmology, it is important to
consider the issue of radion stabilization when the dilaton is frozen, 
or, more generally, in the Einstein frame. As was discussed in 
\cite{Subodh1,Subodh2} (see also earlier comments in \cite{Watson2}),
the existence of string modes which are massless at the self-dual radius
is crucial in obtaining radion stabilization in the Einstein frame
(for more general studies of the importance of
massless modes in string cosmology see \cite{Watson3,Eva2}). Such
massless modes do not exist in all known string theories. They
exist in the Heterotic theory, but not in Type II theories \cite{Pol}.
The following discussion is taken from \cite{RHBrev5}.

Let us consider the equations of motion which arise from coupling
the Einstein action to a string gas. 
In the anisotropic setting when the metric is taken to be
\be
ds^2 \, = \, dt^2 - a(t)^2 d{\bf x}^2 - 
\sum_{\alpha = 1}^6 b_{\alpha}(t)^2 dy_{\alpha}^2 \, ,
\ee
where the $y_{\alpha}$ are the internal coordinates, the
equation of motion for $b_{\alpha}$ becomes
\be \label{extra}
{\ddot b_{\alpha}} + 
\bigl( 3 H + \sum_{\beta = 1, \beta \neq \alpha}^6 {{{\dot b_{\beta}}} \over {b_{\beta}}} \bigr) {\dot b_{\alpha}} \, = \, 
\sum_{n, m} 8 \pi G {{\mu_{m,n}} \over {\sqrt{g} \epsilon_{m,n}}}{\cal S} \,
\ee
where $\mu_{m,n}$ is the number density of $(m,n)$ strings, $\epsilon_{m,n}$
is the energy of an individual $(m,n)$ string, and $g$ is the determinant of
the metric. The source term ${\cal S}$ depends on the quantum numbers of the
string gas, and the sum runs over all momentum and winding
number vectors $m$ and $n$, respectively (note that $n$ and $m$ are
six-vectors, one component for each internal dimension). If the number
of right-moving oscillator modes is given by $N$, then the source term
for fixed $m$ and $n$ vectors is
\be \label{source}
{\cal S} \, = \, \sum_{\alpha} \bigl( {{m_{\alpha}} \over {b_{\alpha}}} \bigr)^2
- \sum_{\alpha} n_{\alpha}^2 b_{\alpha}^2 
+ {2 \over {D - 1}} \bigl[ (n,n) + (n, m) + 2(N - 1) \bigr] \, .
\ee
To obtain this equation, we have made use of the mass spectrum of
string states and of the level matching conditions. In the case of
the bosonic superstring, the mass $M$ of a string state with 
fixed $m, n, N$ and  ${\tilde N}$,
where $N$ and ${\tilde N}$ are the number of right- and left-moving oscillator states,
on a six-dimensional torus whose radii are given by $b_{\alpha}$ is
\be
M^2 \, = \,  \bigl( {{m_{\alpha}} \over {b_{\alpha}}} \bigr)^2
- \sum_{\alpha} n_{\alpha}^2 b_{\alpha}^2 + 2 (N + {\tilde N} - 2) \, ,
\ee
and the level matching condition reads
\be
{\tilde N} \, = \, (n,m) + N \, ,
\ee
where $(n,m)$ indicates the scalar product of $n$ and $m$ in the 
trivial metric.

There are modes which are massless at the self-dual radius $b_{\alpha} = 1$.
One such mode is the graviton with $n = m = 0$ and $N = 1$. The modes of
interest to us are modes which contain winding and momentum, namely 
\begin{itemize}
\item{} $N = 1$, $(m,m) = 1$, $(m, n) = -1$ and $(n,n) = 1$;
\item{} $N = 0$, $(m,m) = 1$, $(m, n) = 1$ and $(n,n) =  1$;
\item{} $N = 0$  $(m,m) = 2$, $(m, n) = 0$ and $(n,n) =  2$.
\end{itemize}
Note that these modes survive in the Heterotic string theory, but
do not survive the GSO \cite{Pol} truncation in Type II string
theories.

In string theories which admit massless states (i.e. states
which are massless at the self-dual radius), these states
will dominate the initial partition function. The background
dynamics will then also be dominated by these states. To understand
the effect of these strings, consider the equation of motion (\ref{extra})
with the source term (\ref{source}). The first two terms in the
source term correspond to an effective potential with a stable
minimum at the self-dual radius. However, if the third term in the
source does not vanish at the self-dual radius, it will lead to
a positive potential which causes the radion to increase. Thus,
a condition for the stabilization of $b_{\alpha}$ at the self-dual
radius is that the third term in (\ref{source}) vanishes at the
self-dual radius. This is the case if and only if the string state
is a massless mode.

The massless modes have other nice features which are explored in
detail in \cite{Subodh2}. They act as radiation from the
point of view of our three large dimensions and hence do not
lead to a over-abundance problem. As our three spatial dimensions
grow, the potential which confines the radion becomes shallower.
However, rather surprisingly, it turns out the the potential
remains steep enough to avoid fifth force constraints. 

In the presence of massless string states, the shape moduli also
can be stabilized, at least in the simple toroidal backgrounds
considered so far \cite{Edna}. To study this issue, we consider
a metric of the form
\be
ds^2 \, = \, dt^2 - d{\bf x}^2 - G_{mn}dy^mdy^n \, ,
\ee
where the metric of the internal space (here for simplicity
considered to be a two-dimensional torus) contains a shape
modulus, the angle between the two cycles of the torus:
\be
G_{11} \, = \, G_{22} \, = \, 1
\ee
and
\be
G_{12} \, = \, G_{21} \, = \, {\rm sin}\theta \, ,
\ee
where $\theta = 0$ corresponds to a rectangular torus. The ratio
between the two toroidal radii is a second shape modulus. However,
from the discussion of the previous subsection we already know
that each radion individually is stabilized at the self-dual
radius. Thus, the shape modulus corresponding to the ratio of
the toroidal radii is fixed, and the angle is the only shape modulus which is
yet to be considered.

Combining the $00$ and the $12$ Einstein equations, we obtain
a harmonic oscillator equation for $\theta$ with $\theta = 0$
as the stable fixed point.
\be
{\ddot \theta} + 8K^{-1/2} e^{-2 \phi} \theta \, = \, 0 \, ,
\ee
where $K$ is a constant whose value depends on the quantum numbers
of the string gas. In the case of an expanding three-dimensional
space we would have obtained an additional damping term in the
above equation of motion.
We thus conclude that the shape modulus is dynamically stabilized at a value
which maximizes the area to circumference ratio. 

The only modulus which is not stabilized by string winding and momentum modes
is the dilaton. Recently, it has been show \cite{Danos2} that a gaugino
condensation mechanism (similar to those used in string inflation model
building) can be introduced which generates a stabilizing potential for
the dilaton without interfering with the radion stabilization force provided
by the string winding and momentum modes.

In the next subsection we turn to the predictions of string gas cosmology
for observations. These predictions do not depend on
the details of the theory, but only on three inputs. The first
is the existence of a quasi-static initial phase in thermal
equilibrium. The second condition is the applicability of
the Einstein field equations for fluctuations on infrared
scales (scales of the order of $1 {\rm mm}$), many orders
of magnitude larger than the string scale, and the third is a
holographic scaling of the specific heat capacity. The role
of the last condition will become manifest below.

\subsection{Fluctuations in String Gas Cosmology}

\subsubsection{Overview}

At the outset of this section, let us recall the mechanism
by which inflationary cosmology leads to the possibility of
a causal generation mechanism for cosmological fluctuations
which yields an almost scale-invariant spectrum of perturbations.
The space-time diagram of inflationary cosmology is sketched
in Figure \ref{infl1}. 

During the period of inflation, the Hubble radius
\be
l_H(t) \, = \, \frac{a}{{\dot a}}
\ee
is approximately constant. In contrast, the physical length
of a fixed co-moving scale (labelled by $k$ in the figure)
is expanding exponentially.
In this way, in inflationary cosmology scales which have
microscopic sub-Hubble wavelengths at the beginning of
inflation are red-shifted to become super-Hubble-scale
fluctuations at the end of the period of inflation.
After inflation, the Hubble radius increases linearly in
time, faster than the physical wavelength corresponding
to a fixed co-moving scale. Thus, scales re-enter the
Hubble radius at late times.

Since inflation red-shifts any classical fluctuations which might
have been present at the beginning of the inflationary phase,
fluctuations in inflationary cosmology are generated by
quantum vacuum perturbations. The fluctuations begin
in their quantum vacuum state at the onset of inflation. Once the
wavelength exceeds the Hubble radius, squeezing of the
wave-function of the fluctuations sets in (see \cite{MFB,RHBrev1}).
This squeezing plus the de-coherence of the fluctuations due
to the interaction between short and long wavelength modes
generated by the intrinsic non-linearities in both the gravitational and
matter sectors of the theory (see \cite{Martineau,Starob3} for
recent discussions of this aspect and references to previous work)
lead to the classicalization of the fluctuations on super-Hubble
scales.

\begin{figure}  
 \includegraphics[height=.5\textheight]{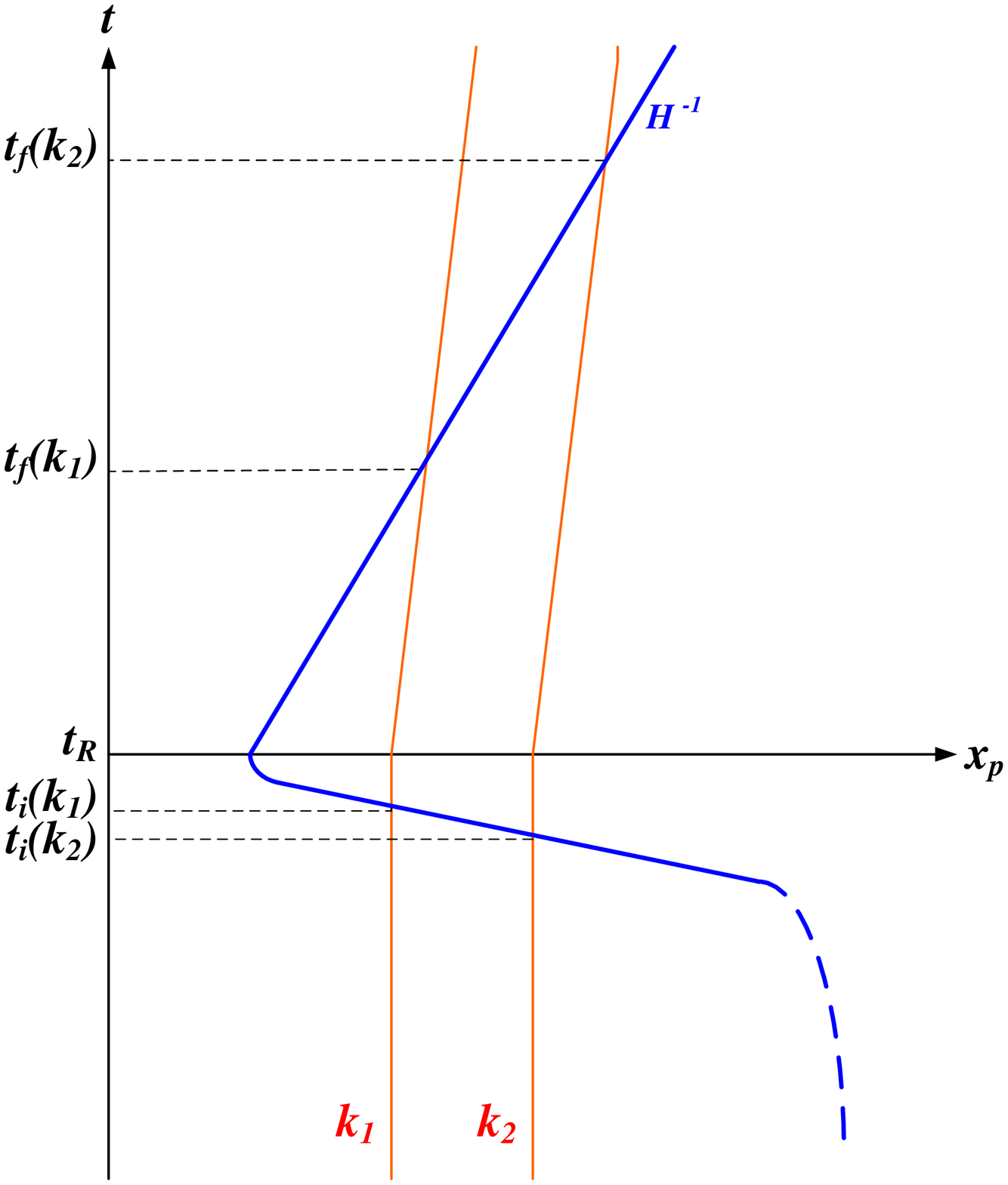}
\caption{Space-time diagram (sketch) showing the evolution of fixed 
co-moving scales in string gas cosmology. The vertical axis is time, 
the horizontal axis is physical distance.  
The solid curve represents the Einstein frame Hubble radius 
$H^{-1}$ which shrinks abruptly to a micro-physical scale at $t_R$ and then 
increases linearly in time for $t > t_R$. Fixed co-moving scales (the 
dotted lines labeled by $k_1$ and $k_2$) which are currently probed 
in cosmological observations have wavelengths which are smaller than 
the Hubble radius before $t_R$. They exit the Hubble 
radius at times $t_i(k)$ just prior to $t_R$, and propagate with a 
wavelength larger than the Hubble radius until they reenter the 
Hubble radius at times $t_f(k)$.}
\label{spacetimenew2}
\end{figure}

Let us now turn to the cosmological background of string gas
cosmology represented in Figure \ref{timeevol2}.  This string gas cosmology
background yields the space-time diagram sketched in Figure \ref{spacetimenew2}.
As in Figure \ref{infl1}, the vertical axis is time and 
the horizontal axis denotes the
physical distance. For times $t < t_R$, 
we are in the static Hagedorn phase and the Hubble radius is
infinite. For $t > t_R$, the Einstein frame 
Hubble radius is expanding as in standard cosmology. The time
$t_R$ is when the string winding modes begin to decay into
string loops, and the scale factor starts to increase, leading to the
transition to the radiation phase of standard cosmology.

Let us now compare the evolution of the physical 
wavelength corresponding to a fixed co-moving scale  
with that of the Einstein frame Hubble radius $H^{-1}(t)$.
The evolution of scales in string gas cosmology is identical
to the evolution in standard and in inflationary cosmology
for $t > t_R$. If we follow the physical wavelength of the
co-moving scale which corresponds to the current Hubble
radius back to the time $t_R$, then - taking the Hagedorn
temperature to be of the order $10^{16}$ GeV - we obtain
a length of about 1 mm. Compared to the string scale and the
Planck scale, this is in the far infrared.

The physical wavelength is constant in the Hagedorn phase
since space is static. But, as we enter the Hagedorn
phase going back in time, the Hubble radius diverges to
infinity. Hence, as in the case of inflationary cosmology,
fluctuation modes begin sub-Hubble during the Hagedorn
phase, and thus a causal generation mechanism
for fluctuations is possible.

However, the physics of the generation mechanism is
very different. In the case of inflationary cosmology,
fluctuations are assumed to start as quantum vacuum
perturbations because classical inhomogeneities are
red-shifting. In contrast, in the Hagedorn phase of string gas
cosmology there is no red-shifting of classical matter.
Hence, it is the fluctuations in the classical matter which
dominate. Since classical matter is a string gas, the
dominant fluctuations are string thermodynamic fluctuations. 

Our proposal for string gas structure formation is the following 
\cite{NBV} (see \cite{BNPV2} for a more detailed description).
For a fixed co-moving scale with wavenumber $k$ we compute the matter
fluctuations while the scale in sub-Hubble (and therefore gravitational
effects are sub-dominant). When the scale exits the Hubble radius
at time $t_i(k)$ we use the gravitational constraint equations to
determine the induced metric fluctuations, which are then propagated
to late times using the usual equations of gravitational perturbation
theory. Since the scales we are interested
in are in the far infrared, we use the Einstein constraint equations for
fluctuations.

\subsubsection{Spectrum of Cosmological Fluctuations}

We write the metric including cosmological perturbations
(scalar metric fluctuations) and gravitational waves in the
following form (using conformal time $\eta$) 
\be \label{pertmetric}
d s^2 \, = \, a^2(\eta) \left\{(1 + 2 \Phi)d\eta^2 - [(1 - 
2 \Phi)\delta_{ij} + h_{ij}]d x^i d x^j\right\} \, . 
\ee 
We have fixed the gauge (i.e. coordinate) freedom for the scalar metric
perturbations by adopting the longitudinal gauge and we have taken 
matter to be free of anisotropic stress. The spatial
tensor $h_{ij}({\bf x}, t)$ is transverse and traceless and represents the
gravitational waves. 

Note that in contrast to the case of slow-roll inflation, scalar metric
fluctuations and gravitational waves are generated by matter
at the same order in cosmological perturbation theory. This could
lead to the expectation that the amplitude of gravitational waves
in string gas cosmology should be generically larger than in inflationary
cosmology. This expectation, however, is not realized \cite{BNPV1}
since there is a different mechanism which suppresses the gravitational
waves relative to the density perturbations (namely the fact
that the gravitational wave amplitude is set by the amplitude of
the pressure, and the pressure is suppressed relative to the
energy density in the Hagedorn phase).

Assuming that the fluctuations are described by the perturbed Einstein
equations (they are {\it not} if the dilaton is not fixed 
\cite{Betal,KKLM}), then the spectra of cosmological perturbations
$\Phi$ and gravitational waves $h$ are given by the energy-momentum 
fluctuations in the following way \cite{BNPV2}
\be \label{scalarexp} 
\langle|\Phi(k)|^2\rangle \, = \, 16 \pi^2 G^2 
k^{-4} \langle\delta T^0{}_0(k) \delta T^0{}_0(k)\rangle \, , 
\ee 
where the pointed brackets indicate expectation values, and 
\be 
\label{tensorexp} \langle|h(k)|^2\rangle \, = \, 16 \pi^2 G^2 
k^{-4} \langle\delta T^i{}_j(k) \delta T^i{}_j(k)\rangle \,, 
\ee 
where on the right hand side of (\ref{tensorexp}) we mean the 
average over the correlation functions with $i \neq j$, and
$h$ is the amplitude of the gravitational waves \footnote{The
gravitational wave tensor $h_{i j}$ can be written as the
amplitude $h$ multiplied by a constant polarization tensor.}.
 
Let us now use (\ref{scalarexp}) to determine the spectrum of
scalar metric fluctuations. We first calculate the 
root mean square energy density fluctuations in a sphere of
radius $R = k^{-1}$. For a system in thermal equilibrium they 
are given by the specific heat capacity $C_V$ via 
\be \label{cor1b}
\langle \delta\rho^2 \rangle \,  = \,  \frac{T^2}{R^6} C_V \, . 
\ee 
The specific  heat of a gas of closed strings
on a torus of radius $R$ can be derived from the partition
function of a gas of closed strings. This computation was
carried out in \cite{Deo} with the result
\be \label{specheat2b} 
C_V  \, \approx \, 2 \frac{R^2/\ell^3}{T \left(1 - T/T_H\right)}\, . 
\ee 
The specific heat capacity scales holographically with the size
of the box. This result follows rigorously from evaluating the
string partition function in the Hagedorn phase. The result, however,
can also be understood heuristically: in the Hagedorn phase the
string winding modes are crucial. These modes look like point
particles in one less spatial dimension. Hence, we expect the
specific heat capacity to scale like in the case of point particles
in one less dimension of space \footnote{We emphasize that it was
important for us to have compact spatial dimensions in order to
obtain the winding modes which are crucial to get the holographic
scaling of the thermodynamic quantities.}.

With these results, the power spectrum $P(k)$ of scalar metric fluctuations can
be evaluated as follows
\bea \label{power2} 
P_{\Phi}(k) \, & \equiv & \, {1 \over {2 \pi^2}} k^3 |\Phi(k)|^2 \\
&=& \, 8 G^2 k^{-1} <|\delta \rho(k)|^2> \, . \nonumber \\
&=& \, 8 G^2 k^2 <(\delta M)^2>_R \nonumber \\ 
               &=& \, 8 G^2 k^{-4} <(\delta \rho)^2>_R \nonumber \\
&=& \, 8 G^2 {T \over {\ell_s^3}} {1 \over {1 - T/T_H}} 
\, , \nonumber 
\eea 
where in the first step we have used (\ref{scalarexp}) to replace the 
expectation value of $|\Phi(k)|^2$ in terms of the correlation function 
of the energy density, and in the second step we have made the 
transition to position space 

The first conclusion from the result (\ref{power2}) is that the spectrum
is approximately scale-invariant ($P(k)$ is independent of $k$). It is
the `holographic' scaling $C_V(R) \sim R^2$ which is responsible for the
overall scale-invariance of the spectrum of cosmological perturbations.
However, there is a small $k$ dependence which comes from the fact
that in the above equation for a scale $k$ 
the temperature $T$ is to be evaluated at the
time $t_i(k)$. Thus, the factor $(1 - T/T_H)$ in the 
denominator is responsible 
for giving the spectrum a slight dependence on $k$. Since
the temperature slightly decreases as time increases at the
end of the Hagedorn phase, shorter wavelengths for which
$t_i(k)$ occurs later obtain a smaller amplitude. Thus, the
spectrum has a slight red tilt.

\subsubsection{Key Prediction of String Gas Cosmology}

As discovered in \cite{BNPV1}, the spectrum of gravitational
waves is also nearly scale invariant. However, in the expression
for the spectrum of gravitational waves the factor $(1 - T/T_H)$
appears in the numerator, thus leading to a slight blue tilt in
the spectrum. This is a prediction with which the cosmological
effects of string gas cosmology can be distinguished from those
of inflationary cosmology, where quite generically a slight red
tilt for gravitational waves results. The physical reason is that
large scales exit the Hubble radius earlier when the pressure
and hence also the off-diagonal spatial components of $T_{\mu \nu}$
are closer to zero.

Let us analyze this issue in a bit more detail and
estimate the dimensionless power spectrum of gravitational waves.
First, we make some general comments. In slow-roll inflation, to
leading order in perturbation theory matter fluctuations do not
couple to tensor modes. This is due to the fact that the matter
background field is slowly evolving in time and the leading order
gravitational fluctuations are linear in the matter fluctuations. In
our case, the background is not evolving (at least at the level of
our computations), and hence the dominant metric fluctuations are
quadratic in the matter field fluctuations. At this level, matter
fluctuations induce both scalar and tensor metric fluctuations.
Based on this consideration we might expect that in our string gas
cosmology scenario, the ratio of tensor to scalar metric
fluctuations will be larger than in simple slow-roll inflationary
models. However, since the amplitude $h$ of the gravitational
waves is proportional to the pressure, and the pressure is suppressed
in the Hagedorn phase, the amplitude of the gravitational waves
will also be small in string gas cosmology.

The method for calculating the spectrum of gravitational waves
is similar to the procedure outlined in the last section
for scalar metric fluctuations. For a mode with fixed co-moving
wavenumber $k$, we compute the correlation function of the
off-diagonal spatial elements of the string gas energy-momentum
tensor at the time $t_i(k)$ when the mode exits the Hubble radius
and use (\ref{tensorexp}) to infer the amplitude of the power
spectrum of gravitational waves at that time. We then
follow the evolution of the gravitational wave power spectrum
on super-Hubble scales for $t > t_i(k)$ using the equations
of general relativistic perturbation theory.

The power spectrum of the tensor modes is given by (\ref{tensorexp}):
\be \label{tpower1}
P_h(k) \, = \, 16 \pi^2 G^2 k^{-4} k^3
\langle\delta T^i{}_j(k) \delta T^i{}_j(k)\rangle
\ee
for $i \neq j$. Note that the $k^3$ factor multiplying the momentum
space correlation function of $T^i{}_j$ gives the position space
correlation function $C^i{}_j{}^i{}_j(R)$ , namely the root mean 
square fluctuation of $T^i{}_j$ in a region of radius $R = k^{-1}$ 
(the reader who is skeptical about this point is invited to check 
that the dimensions work out properly). Thus,
\be \label{tpower2}
P_h(k) \, = \, 16 \pi^2 G^2 k^{-4} C^i{}_j{}^i{}_j(R) \, .
\ee
The correlation function $C^i{}_j{}^i{}_j$ on the right hand side
of the above equation follows from the thermal closed string
partition function and was computed explicitly in
\cite{Ali} (see also \cite{RHBSGrev}). We obtain
\be \label{tpower3}
P_h(k) \, \sim \, 16 \pi^2 G^2 {T \over {l_s^3}}
(1 - T/T_H) \ln^2{\left[\frac{1}{l_s^2 k^2}(1 -
T/T_H)\right]}\, ,
\ee
which, for temperatures close to the Hagedorn value reduces to
\be \label{tresult}
P_h(k) \, \sim \,
\left(\frac{l_{Pl}}{l_s}\right)^4 (1 -
T/T_H)\ln^2{\left[\frac{1}{l_s^2 k^2}(1 - T/T_H)\right]} \, .
\ee
This shows that the spectrum of tensor modes is - to a first
approximation, namely neglecting the logarithmic factor and
neglecting the $k$-dependence of $T(t_i(k))$ - scale-invariant. 

On super-Hubble scales, the amplitude $h$ of the gravitational waves
is constant. The wave oscillations freeze out when the wavelength
of the mode crosses the Hubble radius. As in the case of scalar metric
fluctuations, the waves are squeezed. Whereas the wave amplitude remains
constant, its time derivative decreases. Another way to see this
squeezing is to change variables to 
\be
\psi(\eta, {\bf x}) \, = \, a(\eta) h(\eta, {\bf x})
\ee
in terms of which the action has canonical kinetic term. The action
in terms of $\psi$ becomes
\be
S \, = \, {1 \over 2} \int d^4x \left( {\psi^{\prime}}^2 -
\psi_{,i}\psi_{,i} + {{a^{\prime \prime}} \over a} \psi^2 \right)
\ee
from which it immediately follows that on super-Hubble scales
$\psi \sim a$. This is the squeezing of gravitational 
waves \cite{Grishchuk}.

Since there is no $k$-dependence in the squeezing factor, the
scale-invariance of the spectrum at the end of the Hagedorn phase
will lead to a scale-invariance of the spectrum at late times.

Note that in the case of string gas cosmology, the squeezing
factor $z(\eta)$ does not differ substantially from the
squeezing factor $a(\eta)$ for gravitational waves. In the
case of inflationary cosmology, $z(\eta)$ and $a(\eta)$
differ greatly during reheating, leading to a much larger
squeezing for scalar metric fluctuations, and hence to a
suppressed tensor to scalar ratio of fluctuations. In the
case of string gas cosmology, there is no difference in
squeezing between the scalar and the tensor modes. 

Let us return to the discussion of the spectrum of gravitational
waves. The result for the power spectrum is given in
(\ref{tresult}), and we mentioned that to a first approximation this
corresponds to a scale-invariant spectrum. As realized in
\cite{BNPV1}, the logarithmic term and the $k$-dependence of
$T(t_i(k))$ both lead to a small blue-tilt of the spectrum. This
feature is characteristic of our scenario and cannot be reproduced
in inflationary models. In inflationary models, the amplitude of
the gravitational waves is set by the Hubble constant $H$. The
Hubble constant cannot increase during inflation, and hence no
blue tilt of the gravitational wave spectrum is possible.

A heuristic way of understanding the origin of the slight blue tilt
in the spectrum of tensor modes
is as follows. The closer we get to the Hagedorn temperature, the
more the thermal bath is dominated by long string states, and thus
the smaller the pressure will be compared to the pressure of a pure
radiation bath. Since the pressure terms (strictly speaking the
anisotropic pressure terms) in the energy-momentum tensor are
responsible for the tensor modes, we conclude that the smaller the
value of the wavenumber $k$ (and thus the higher the temperature
$T(t_i(k))$ when the mode exits the Hubble radius, the lower the
amplitude of the tensor modes. In contrast, the scalar modes are
determined by the energy density, which increases at $T(t_i(k))$ as
$k$ decreases, leading to a slight red tilt.

The reader may ask about the predictions of string gas cosmology
for non-Gaussianities. The answer is \cite{SGNG} that the
non-Gaussianities from the thermal string gas perturbations
are Poisson-suppressed on scales larger than the thermal
wavelength in the Hagedorn phase. However, if the spatial
sections are initially large, then it is possible that a network
of cosmic superstrings \cite{Witten} will be left behind. These
strings - if stable - would achieve a scaling solution (constant
number of strings crossing each Hubble volume at each
time \cite{ShellVil,HK,RHBtoprev}). Such strings give rise
to linear discontinuities in the CMB temperature maps \cite{KS},
lines which can be searched for using edge detection
algorithms such as the Canny algorithm (see \cite{Amsel}
for recent feasibility studies).

\subsection{Problems of String Gas Cosmology}

The main problem of the current implementation of string gas cosmology
is that it does not provide a quantitative model for the Hagedorn phase. The
criticisms of string gas cosmology raised in \cite{KKLM,Kaloper} are
based on assuming that dilaton gravity should be the background.
However, as mentioned earlier, this is not a reasonable assumption
since the coupling of string gas matter to dilaton gravity is not S-duality
invariant. The quasi-static early Hagedorn phase should also
have constant dilaton. In addition, we do not expect that at high
densities such as the Hagedorn density an simple effective action such
as that of dilaton gravity will apply. Nevertheless, to make the
string gas cosmology scenario into a real theory, it is crucial to
obtain a good understanding of the background dynamics. For
some initial steps in this direction see \cite{Kanno1}.

A second problem of string gas cosmology is the size problem (and
the related entropy problem). If the string scale is about $10^{17} {\rm GeV}$
as is preferred in early heterotic superstring models, then the radius
of the universe during the Hagedorn phase must be many orders
of magnitude larger than the string scale. Without embedding string
gas cosmology into a bouncing cosmology, it seems unnatural to
demand such a large initial size. This problem disappears if the
Hagedorn phase is preceded by a phase of contraction, as in the
model of \cite{Biswas2}. In this case, however, it is non-trivial to arrange
that the Hagedorn phase lasts sufficiently long to maintain thermal
equilibrium over the required range of scales.

It should be noted, however, that some of the conceptual problems
of inflationary cosmology such as the trans-Planckian problem for
fluctuations, do not arise in string gas cosmology. As in the case
of the matter bounce scenario, the basic mechanism of the scenario
is insensitive to what sets the cosmological constant to its
observed very small value.

\section{Conclusions}

In these lectures I have argued that
it is possible to explore physics of the very early universe
using current cosmological observations. Given the large amount
of new data which is expected over the next decade, early universe
cosmology will be a vibrant field of research. The information about
the very early universe is transferred to the current time mainly via
imprints on the spectrum of cosmological fluctuations.
Thus, the theory of cosmological fluctuations is the key tool of modern
cosmology. It links current observations with early
universe cosmology. The theory can be applied to any background
cosmology, not just inflationary cosmology. I have illustrated
the application of the theory of cosmological perturbations
to three scenarios of the very early universe, the inflationary
scenario, the matter bounce paradigm, and string gas
cosmology.

The Hubble radius plays a key role in the evolution of fluctuations.
On sub-Hubble scales the microphysical forces dominate, whereas on
super-Hubble scales matter forces freeze out and gravity dominates.
In order to have a causal mechanism of structure formation, it is therefore
crucial that scales of current interest in cosmology originate at very early
times inside the Hubble radius, and that they then propagate over an
extended period of time on super-Hubble scales. This propagation on
super-Hubble scales is required in order to obtain the squeezing of
the perturbations which is required to explain the acoustic oscillations
in the angular power spectrum of CMB anisotropies.

The inflationary scenario is the current paradigm of early universe
cosmology. It explains important conceptual problems of Standard
Big Bang Cosmology, and most importantly it provided a mechanism 
to explain the origin of structure in the universe via causal physics.
The perturbations begin as quantum vacuum fluctuations on
sub-Hubble scales. They exit the Hubble radius during the phase
of inflationary expansion of space and are then squeezed.  
Inflationary cosmology has been predictive: it predicted the detailed
shape of the angular power spectrum of CMB anisotropies more
than 15 years before the observations. However, it is important
to realize that any theory which provides a scale-invariant spectrum
of primordial curvature fluctuations and which admits a period during
which the scales which are observed today evolve outside of the 
Hubble radius will have the same predictions, as already realized
more than a decade before the development of inflationary
cosmology by Sunyaev and Zel'dovich \cite{SZ} and by
Peebles and Yu \cite{Peebles} (see also \cite{Harrison,Zeldovich}). 
Thus, it is incorrect to claim
that the current observations confirm the inflationary scenario.
In fact, current realizations of inflationary cosmology suffer from
some basic conceptual problems. This motivates the search
for an improved understanding of the very early universe.

I have described two alternative early universe scenarios, the
``matter bounce", and ``string gas cosmology". Both lead
to a scale-invariant spectrum of curvature fluctuations and
involve squeezing of the perturbations on super-Hubble
scales and are thus in agreement with current observations.
The challenge for alternative scenarios is to identify clean 
predictions with which the new scenarios can be distinguished 
from those of inflation. In the case of
the matter bounce the prediction we have identified is a particular
shape of the bispectrum, in the case of string gas cosmology the
key prediction is a slight blue tilt in the spectrum of gravitational
waves.

It is important to realize that physics of the very early universe
is determined by physics of the highest energy scales,
scales many orders of magnitude higher than those which
are probed in accelerator experiments. Thus, a better
understanding of quantum gravity (e.g. string theory) might
lead to completely new possibilities for cosmology. In
turn, it is only through cosmology that physics of the highest
energies can be probed.


\section*{Acknowledgments}

I wish to thank the organizers of this conference, in particular
Maria Emilia Guimaraes, for the invitation  to give these lectures 
and for their wonderful hospitality in Brazil. I wish to thank the
students for their many good questions.
This research is supported in part by a NSERC Discovery Grant,
by the Canada Research Chairs program and by a Killam Foundation
Research Fellowship.



\begin{thebibliography}{99}

\bibitem{ON}
R.~H.~Brandenberger,
  ``Cosmology of the Very Early Universe,''
  AIP Conf.\ Proc.\  {\bf 1268}, 3-70 (2010).
  [arXiv:1003.1745 [hep-th]].
  
\bibitem{Guth}
Guth AH, 
 ``The Inflationary Universe: A Possible Solution To The Horizon And Flatness
 Problems,''
  Phys.\ Rev.\  D {\bf 23}, 347 (1981).
  
\bibitem{Brout}
R.~Brout, F.~Englert and E.~Gunzig,
  ``The Creation Of The Universe As A Quantum Phenomenon,''
  Annals Phys.\  {\bf 115}, 78 (1978).
  
\bibitem{Starob1}
A.~A.~Starobinsky,
  ``A New Type Of Isotropic Cosmological Models Without Singularity,''
  Phys.\ Lett.\ B {\bf 91}, 99 (1980).
  
\bibitem{Sato}
K.~Sato,
  ``First Order Phase Transition Of A Vacuum And Expansion Of The Universe,''
  Mon.\ Not.\ Roy.\ Astron.\ Soc.\  {\bf 195}, 467 (1981).

\bibitem{Mukh}
V. Mukhanov and G. Chibisov,
  ``Quantum Fluctuation And Nonsingular Universe. (In Russian),''
  JETP Lett.\  {\bf 33}, 532 (1981)
  [Pisma Zh.\ Eksp.\ Teor.\ Fiz.\  {\bf 33}, 549 (1981)].

\bibitem{Press}
W. Press,
``Spontaneous production of the Zel'dovich spectrum of cosmological 
fluctuations'',
 Phys. Scr. {\bf 21}, 702 (1980).

\bibitem{Starob2}
A.~A.~Starobinsky,
  ``Spectrum of relict gravitational radiation and the early state of the
  universe,''
  JETP Lett.\  {\bf 30}, 682 (1979)
  [Pisma Zh.\ Eksp.\ Teor.\ Fiz.\  {\bf 30}, 719 (1979)].

\bibitem{WMAP}
C.~L.~Bennett {\it et al.},
   ``First Year Wilkinson Microwave Anisotropy Probe (WMAP) Observations:
  Preliminary Maps and Basic Results,''
  Astrophys.\ J.\ Suppl.\  {\bf 148}, 1 (2003)
  [arXiv:astro-ph/0302207].

\bibitem{CosPA08}
R.~H.~Brandenberger,
  ``Alternatives to Cosmological Inflation,''
  arXiv:0902.4731 [hep-th].
  
\bibitem{RHBSGrev}
R.~H.~Brandenberger,
  ``String Gas Cosmology,''
  arXiv:0808.0746 [hep-th].

\bibitem{CMBblack}
H.~P.~Gush, M.~Halpern and E.~H.~Wishnow,
  ``Rocket Measurement Of The Cosmic-Background-Radiation Mm-Wave Spectrum,''
  Phys.\ Rev.\ Lett.\  {\bf 65}, 537 (1990);\\
J.~C.~Mather {\it et al.},
  ``Measurement of the Cosmic Microwave Background spectrum by the COBE FIRAS
  instrument,''
  Astrophys.\ J.\  {\bf 420}, 439 (1994).
  
\bibitem{Wands}
D.~Wands,
  ``Duality invariance of cosmological perturbation spectra,''
  Phys.\ Rev.\  D {\bf 60}, 023507 (1999)
  [arXiv:gr-qc/9809062].
  
 \bibitem{FB2}
F.~Finelli and R.~Brandenberger,
  ``On the generation of a scale-invariant spectrum of adiabatic  fluctuations
  in cosmological models with a contracting phase,''
  Phys.\ Rev.\  D {\bf 65}, 103522 (2002)
  [arXiv:hep-th/0112249].

\bibitem{Wands2}
L.~E.~Allen and D.~Wands,
  ``Cosmological perturbations through a simple bounce,''
  Phys.\ Rev.\  D {\bf 70}, 063515 (2004)
  [arXiv:astro-ph/0404441].

\bibitem{Biswas1}
T.~Biswas, A.~Mazumdar and W.~Siegel,
  ``Bouncing universes in string-inspired gravity,''
  JCAP {\bf 0603}, 009 (2006)
  [arXiv:hep-th/0508194].
  
\bibitem{MBS}
R.~H.~Brandenberger, V.~F.~Mukhanov and A.~Sornborger,
  ``A Cosmological theory without singularities,''
  Phys.\ Rev.\  D {\bf 48}, 1629 (1993)
  [arXiv:gr-qc/9303001].
  
\bibitem{BFS}
R.~Brandenberger, H.~Firouzjahi and O.~Saremi,
  ``Cosmological Perturbations on a Bouncing Brane,''
  JCAP {\bf 0711}, 028 (2007)
  [arXiv:0707.4181 [hep-th]].
  
\bibitem{Cai1}
Y.~F.~Cai, T.~Qiu, Y.~S.~Piao, M.~Li and X.~Zhang,
  ``Bouncing Universe with Quintom Matter,''
  JHEP {\bf 0710}, 071 (2007)
  [arXiv:0704.1090 [gr-qc]];\\
 Y.~F.~Cai, T.~T.~Qiu, J.~Q.~Xia and X.~Zhang,
  ``A Model Of Inflationary Cosmology Without Singularity,''
  arXiv:0808.0819 [astro-ph].

\bibitem{Novello}
M.~Novello and S.~E.~P.~Bergliaffa,
  ``Bouncing Cosmologies,''
  Phys.\ Rept.\  {\bf 463}, 127 (2008)
  [arXiv:0802.1634 [astro-ph]].
  
\bibitem{HLbounce}
R.~Brandenberger,
  ``Matter Bounce in Horava-Lifshitz Cosmology,''
  Phys.\ Rev.\  D {\bf 80}, 043516 (2009)
  [arXiv:0904.2835 [hep-th]].
  
\bibitem{Horava}
 P.~Horava,
  ``Quantum Gravity at a Lifshitz Point,''
  Phys.\ Rev.\  D {\bf 79}, 084008 (2009)
  [arXiv:0901.3775 [hep-th]].
   
\bibitem{ABB}
S.~Alexander, T.~Biswas and R.~H.~Brandenberger,
  ``On the Transfer of Adiabatic Fluctuations through a Nonsingular
  Cosmological Bounce,''
  arXiv:0707.4679 [hep-th].
  
\bibitem{Cai2}
Y.~F.~Cai, T.~Qiu, R.~Brandenberger, Y.~S.~Piao and X.~Zhang,
  ``On Perturbations of Quintom Bounce,''
  JCAP {\bf 0803}, 013 (2008)
  [arXiv:0711.2187 [hep-th]];\\
    Y.~F.~Cai and X.~Zhang,
  ``Evolution of Metric Perturbations in Quintom Bounce model,''
  arXiv:0808.2551 [astro-ph].

\bibitem{LWbounce} 
Y.~F.~Cai, T.~Qiu, R.~Brandenberger and X.~Zhang,
  ``A Nonsingular Cosmology with a Scale-Invariant Spectrum of Cosmological
  Perturbations from Lee-Wick Theory,''
  arXiv:0810.4677 [hep-th].

\bibitem{HLbounce2}
X.~Gao, Y.~Wang, W.~Xue and R.~Brandenberger,
  ``Fluctuations in a Ho\v{r}ava-Lifshitz Bouncing Cosmology,''
  arXiv:0911.3196 [hep-th];\\
  X.~Gao, Y.~Wang, R.~Brandenberger and A.~Riotto,
  ``Cosmological Perturbations in Ho\v{r}ava-Lifshitz Gravity,''
  arXiv:0905.3821 [hep-th].
  
\bibitem{Thermalflucts}
Y.~F.~Cai, W.~Xue, R.~Brandenberger and X.~m.~Zhang,
  ``Thermal Fluctuations and Bouncing Cosmologies,''
  JCAP {\bf 0906}, 037 (2009)
  [arXiv:0903.4938 [hep-th]].
  
\bibitem{BV}
R.~H.~Brandenberger and C.~Vafa,
  ``Superstrings in the Early Universe,''
  Nucl.\ Phys.\  B {\bf 316}, 391 (1989).
  
\bibitem{Perlt}
 J.~Kripfganz and H.~Perlt,
  ``Cosmological Impact Of Winding Strings,''
  Class.\ Quant.\ Grav.\  {\bf 5}, 453 (1988).

\bibitem{Hagedorn}
R.~Hagedorn,
  ``Statistical Thermodynamics Of Strong Interactions At High-Energies,''
  Nuovo Cim.\ Suppl.\  {\bf 3}, 147 (1965).
 
\bibitem{NBV}
A.~Nayeri, R.~H.~Brandenberger and C.~Vafa,
  ``Producing a scale-invariant spectrum of perturbations in a Hagedorn  phase
  of string cosmology,''
  Phys.\ Rev.\ Lett.\  {\bf 97}, 021302 (2006)
  [arXiv:hep-th/0511140].
     
\bibitem{processing}
R.~H.~Brandenberger,
  ``Processing of Cosmological Perturbations in a Cyclic Cosmology,''
  Phys.\ Rev.\  D {\bf 80}, 023535 (2009)
  [arXiv:0905.1514 [hep-th]].
  
\bibitem{RHBrev1}
R.~H.~Brandenberger,
  ``Lectures on the theory of cosmological perturbations,''
  Lect.\ Notes Phys.\  {\bf 646}, 127 (2004)
  [arXiv:hep-th/0306071].
  
\bibitem{MFB}
V.~F.~Mukhanov, H.~A.~Feldman and R.~H.~Brandenberger,
  ``Theory of cosmological perturbations. Part 1. Classical perturbations. Part
  2. Quantum theory of perturbations. Part 3. Extensions,''
  Phys.\ Rept.\  {\bf 215}, 203 (1992).
  
\bibitem{SW}
R.~K.~Sachs and A.~M.~Wolfe,
  ``Perturbations of a cosmological model and angular variations of the
  microwave background,''
  Astrophys.\ J.\  {\bf 147}, 73 (1967)
  [Gen.\ Rel.\ Grav.\  {\bf 39}, 1929 (2007)].
  
\bibitem{Harrison}
E.~R.~Harrison,
  ``Fluctuations at the threshold of classical cosmology,''
  Phys.\ Rev.\  D {\bf 1}, 2726 (1970).
  
\bibitem{Zeldovich}
Y.~B.~Zeldovich,
  ``A Hypothesis, unifying the structure and the entropy of the
  universe,''
  Mon.\ Not.\ Roy.\ Astron.\ Soc.\  {\bf 160}, 1P (1972).
  
\bibitem{Afshordi}
N.~Afshordi and R.~H.~Brandenberger,
  ``Super-Hubble nonlinear perturbations during inflation,''
  Phys.\ Rev.\  D {\bf 63}, 123505 (2001)
  [arXiv:gr-qc/0011075].
  
\bibitem{Lifshitz}
E.~Lifshitz,
  ``On the Gravitational stability of the expanding universe,''
  J.\ Phys.\ (USSR) {\bf 10}, 116 (1946);\\
  E.~M.~Lifshitz and I.~M.~Khalatnikov,
  ``Investigations in relativistic cosmology,''
  Adv.\ Phys.\  {\bf 12}, 185 (1963).
  
\bibitem{Bardeen}
J.~M.~Bardeen,
  ``Gauge Invariant Cosmological Perturbations,''
  Phys.\ Rev.\  D {\bf 22}, 1882 (1980).
  
\bibitem{PV}
W.~H.~Press and E.~T.~Vishniac,
  ``Tenacious myths about cosmological perturbations larger than the horizon
  size,''
  Astrophys.\ J.\  {\bf 239}, 1 (1980).
  
\bibitem{BKP}
R.~H.~Brandenberger, R.~Kahn and W.~H.~Press,
  ``Cosmological Perturbations In The Early Universe,''
  Phys.\ Rev.\  D {\bf 28}, 1809 (1983).
  
\bibitem{Kodama}
H.~Kodama and M.~Sasaki,
  ``Cosmological Perturbation Theory,''
  Prog.\ Theor.\ Phys.\ Suppl.\  {\bf 78}, 1 (1984).
  
\bibitem{Ellis}
M.~Bruni, G.~F.~R.~Ellis and P.~K.~S.~Dunsby,
  ``Gauge invariant perturbations in a scalar field dominated universe,''
  Class.\ Quant.\ Grav.\  {\bf 9}, 921 (1992).
  
\bibitem{Hwang}
J.~c.~Hwang,
  ``Evolution of ideal fluid cosmological perturbations,''
  Astrophys.\ J.\  {\bf 415}, 486 (1993).
  
\bibitem{Durrer}
R.~Durrer,
  ``Anisotropies in the cosmic microwave background: Theoretical
  foundations,''
  Helv.\ Phys.\ Acta {\bf 69}, 417 (1996)
  [arXiv:astro-ph/9610234].
  
\bibitem{Stewart}
J.~M.~Stewart,
  ``Perturbations of Friedmann-Robertson-Walker cosmological models,''
  Class.\ Quant.\ Grav.\  {\bf 7}, 1169 (1990).
  
\bibitem{PST}    
N.~Turok, U.~L.~Pen and U.~Seljak,
  ``The scalar, vector and tensor contributions to CMB anisotropies from
  cosmic defects,''
  Phys.\ Rev.\  D {\bf 58}, 023506 (1998)
  [arXiv:astro-ph/9706250].
  
\bibitem{BB} 
T.~J.~Battefeld and R.~Brandenberger,
  ``Vector perturbations in a contracting universe,''
  Phys.\ Rev.\  D {\bf 70}, 121302 (2004)
  [arXiv:hep-th/0406180].
  
\bibitem{SteWa}
J.~M.~Stewart and M.~Walker,
  ``Perturbations Of Spacetimes In General Relativity,''
  Proc.\ Roy.\ Soc.\ Lond.\  A {\bf 341}, 49 (1974).
  
\bibitem{BST}
J.~M.~Bardeen, P.~J.~Steinhardt and M.~S.~Turner,
  ``Spontaneous Creation Of Almost Scale - Free Density Perturbations In An
  Inflationary Universe,''
  Phys.\ Rev.\  D {\bf 28}, 679 (1983).
  
\bibitem{BK}
R.~H.~Brandenberger and R.~Kahn,
  ``Cosmological Perturbations In Inflationary Universe Models,''
  Phys.\ Rev.\  D {\bf 29}, 2172 (1984).
  
\bibitem{Lyth}
D.~H.~Lyth,
  ``Large Scale Energy Density Perturbations And Inflation,''
  Phys.\ Rev.\  D {\bf 31}, 1792 (1985).
  
\bibitem{Fabio1}
F.~Finelli and R.~H.~Brandenberger,
  ``Parametric amplification of gravitational fluctuations during  reheating,''
  Phys.\ Rev.\ Lett.\  {\bf 82}, 1362 (1999)
  [arXiv:hep-ph/9809490].
  
\bibitem{Weinberg2}
S.~Weinberg,
  ``Adiabatic modes in cosmology,''
  Phys.\ Rev.\  D {\bf 67}, 123504 (2003)
  [arXiv:astro-ph/0302326].
 
 \bibitem{Zhang}
W.~B.~Lin, X.~H.~Meng and X.~M.~Zhang,
  ``Adiabatic gravitational perturbation during reheating,''
  Phys.\ Rev.\  D {\bf 61}, 121301 (2000)
  [arXiv:hep-ph/9912510].
  
\bibitem{BaVi}
B.~A.~Bassett and F.~Viniegra,
  ``Massless metric preheating,''
  Phys.\ Rev.\  D {\bf 62}, 043507 (2000)
  [arXiv:hep-ph/9909353].
  
\bibitem{Fabio2}
F.~Finelli and R.~H.~Brandenberger,
  ``Parametric amplification of metric fluctuations during reheating in two
  field models,''
  Phys.\ Rev.\  D {\bf 62}, 083502 (2000)
  [arXiv:hep-ph/0003172].
  
\bibitem{Traschen}
J.~H.~Traschen,
  ``Constraints On Stress Energy Perturbations In General Relativity,''
  Phys.\ Rev.\  D {\bf 31}, 283 (1985).
  
\bibitem{TTB}
J.~H.~Traschen, N.~Turok and R.~H.~Brandenberger,
  ``Microwave Anisotropies from Cosmic Strings,''
  Phys.\ Rev.\  D {\bf 34}, 919 (1986).
  
\bibitem{Dvali}
G.~Dvali, A.~Gruzinov and M.~Zaldarriaga,
  ``Cosmological perturbations from inhomogeneous reheating, freezeout, and
  mass domination,''
  Phys.\ Rev.\  D {\bf 69}, 083505 (2004)
  [arXiv:astro-ph/0305548].
  
\bibitem{Kofman}
L.~Kofman,
  ``Probing string theory with modulated cosmological fluctuations,''
  arXiv:astro-ph/0303614.
  
\bibitem{Vernizzi}
F.~Vernizzi,
  ``Cosmological perturbations from varying masses and couplings,''
  Phys.\ Rev.\  D {\bf 69}, 083526 (2004)
  [arXiv:astro-ph/0311167].
  
\bibitem{ABT}
M.~Axenides, R.~H.~Brandenberger and M.~S.~Turner,
  ``Development Of Axion Perturbations In An Axion Dominated Universe,''
  Phys.\ Lett.\  B {\bf 126}, 178 (1983).
  
\bibitem{Mukh2}
V.~F.~Mukhanov,
  ``Quantum Theory of Gauge Invariant Cosmological Perturbations,''
  Sov.\ Phys.\ JETP {\bf 67}, 1297 (1988)
  [Zh.\ Eksp.\ Teor.\ Fiz.\  {\bf 94N7}, 1 (1988)].
  
\bibitem{Mukh3}
V.~F.~Mukhanov,
  ``Gravitational Instability Of The Universe Filled With A Scalar Field,''
  JETP Lett.\  {\bf 41}, 493 (1985)
  [Pisma Zh.\ Eksp.\ Teor.\ Fiz.\  {\bf 41}, 402 (1985)].
  
\bibitem{Sasaki}
M.~Sasaki,
  ``Large Scale Quantum Fluctuations in the Inflationary Universe,''
  Prog.\ Theor.\ Phys.\  {\bf 76}, 1036 (1986).
  
\bibitem{Lukash}
V.~N.~Lukash,
  ``Production of phonons in an isotropic universe,''
  Sov.\ Phys.\ JETP {\bf 52}, 807 (1980)
  [Zh.\ Eksp.\ Teor.\ Fiz.\  {\bf 79},  (19??)].
  
\bibitem{BD}   
N.~D.~Birrell and P.~C.~W.~Davies,
  ``Quantum Fields In Curved Space,''
{\it  Cambridge, Uk: Univ. Pr. ( 1982) 340p}

\bibitem{Starob3}    
C.~Kiefer, I.~Lohmar, D.~Polarski and A.~A.~Starobinsky,
  ``Pointer states for primordial fluctuations in inflationary cosmology,''
  Class.\ Quant.\ Grav.\  {\bf 24}, 1699 (2007)
  [arXiv:astro-ph/0610700].
  
\bibitem{Martineau}
 P.~Martineau,
  ``On the decoherence of primordial fluctuations during inflation,''
  Class.\ Quant.\ Grav.\  {\bf 24}, 5817 (2007)
  [arXiv:astro-ph/0601134].

\bibitem{Grishchuk}
L.~P.~Grishchuk,
  ``Amplification Of Gravitational Waves In An Istropic Universe,''
  Sov.\ Phys.\ JETP {\bf 40}, 409 (1975)
  [Zh.\ Eksp.\ Teor.\ Fiz.\  {\bf 67}, 825 (1974)].
  
\bibitem{Eva} 
E.~Silverstein and D.~Tong,
  ``Scalar Speed Limits and Cosmology: Acceleration from D-cceleration,''
  Phys.\ Rev.\  D {\bf 70}, 103505 (2004)
  [arXiv:hep-th/0310221].

\bibitem{kinflation}
C.~Armendariz-Picon, T.~Damour and V.~F.~Mukhanov,
  ``k-Inflation,''
  Phys.\ Lett.\  B {\bf 458}, 209 (1999)
  [arXiv:hep-th/9904075].
  
\bibitem{Kung}
R.~H.~Brandenberger and J.~H.~Kung,
  ``Chaotic Inflation As An Attractor In Initial Condition Space,''
  Phys.\ Rev.\  D {\bf 42}, 1008 (1990).
  
\bibitem{Coleman}
S.~R.~Coleman,
  ``The Fate Of The False Vacuum. 1. Semiclassical Theory,''
  Phys.\ Rev.\  D {\bf 15}, 2929 (1977)
  [Erratum-ibid.\  D {\bf 16}, 1248 (1977)].
  
\bibitem{RHBRMP}
R.~H.~Brandenberger,
  ``Quantum Field Theory Methods And Inflationary Universe Models,''
  Rev.\ Mod.\ Phys.\  {\bf 57}, 1 (1985).
  
\bibitem{new}
A. Linde,
  ``A New Inflationary Universe Scenario: A Possible Solution Of The Horizon,
  Flatness, Homogeneity, Isotropy And Primordial Monopole Problems,''
  Phys.\ Lett.\  B {\bf 108}, 389 (1982);\\
A. Albrecht and P. Steinhardt,
  ``Cosmology For Grand Unified Theories With Radiatively Induced Symmetry
  Breaking,''
  Phys.\ Rev.\ Lett.\  {\bf 48}, 1220 (1982).

\bibitem{CW}
S. Coleman and E. Weinberg,
  ``Radiative Corrections As The Origin Of Spontaneous Symmetry Breaking,''
  Phys.\ Rev.\  D {\bf 7}, 1888 (1973).

\bibitem{Goldwirth}
D. Goldwirth and T. Piran,
  ``Initial conditions for inflation,''
  Phys.\ Rept.\  {\bf 214}, 223 (1992);\\
A. Albrecht and R. Brandenberger,
  ``On The Realization Of New Inflation,''
  Phys.\ Rev.\  D {\bf 31}, 1225 (1985).

\bibitem{GhazalScott}
R.~Brandenberger, G.~Geshnizjani and S.~Watson,
  ``On the initial conditions for brane inflation,''
  Phys.\ Rev.\  D {\bf 67}, 123510 (2003)
  [arXiv:hep-th/0302222].
  
\bibitem{MUW}
G.~F.~Mazenko, R.~M.~Wald and W.~G.~Unruh,
  ``Does A Phase Transition In The Early Universe Produce The Conditions Needed
  For Inflation?,''
  Phys.\ Rev.\  D {\bf 31}, 273 (1985).
  
\bibitem{chaotic}
A. Linde,
  ``Chaotic Inflation,''
  Phys.\ Lett.\  B {\bf 129}, 177 (1983).
  
\bibitem{Feldman}
H. Feldman and R. Brandenberger,
  ``Chaotic Inflation With Metric And Matter Perturbations,''
  Phys.\ Lett.\  B {\bf 227}, 359 (1989).

\bibitem{hybrid}
A. Linde,
  ``Hybrid inflation,''
  Phys.\ Rev.\  D {\bf 49}, 748 (1994).

\bibitem{ShellVil}
A. Vilenkin and E.P.S. Shellard, \textit{Cosmic Strings and other
Topological Defects} (Cambridge Univ. Press, Cambridge, 1994).

\bibitem{HK}
M.~B.~Hindmarsh and T.~W.~B.~Kibble,
  ``Cosmic strings,''
  Rept.\ Prog.\ Phys.\  {\bf 58}, 477 (1995)
  [arXiv:hep-ph/9411342].
  
\bibitem{RHBtoprev}
R.~H.~Brandenberger,
  ``Topological defects and structure formation,''
  Int.\ J.\ Mod.\ Phys.\ A {\bf 9}, 2117 (1994)
  [arXiv:astro-ph/9310041].

\bibitem{Allahverdi}
R.~Allahverdi, R.~Brandenberger, F.~Y.~Cyr-Racine and A.~Mazumdar,
  ``Reheating in Inflationary Cosmology: Theory and Applications,''
  Ann. Rev. Nucl. Part. Science {\bf 60}, 27 (2010)
  [arXiv:1001.2600 [hep-th]].
  
 \bibitem{initial}
 L. F. Abbott,  E. Farhi and M. Wise, 
  ``Particle Production In The New Inflationary Cosmology,''
  Phys.\ Lett.\  B {\bf 117}, 29 (1982);\\
  A. Dolgov and A. Linde,
  ``Baryon Asymmetry In Inflationary Universe,''
  Phys.\ Lett.\  B {\bf 116}, 329 (1982);\\
  A. Albrecht, P. J. Steinhardt,  M. S. Turner and  F. Wilczek,
  ``Reheating An Inflationary Universe,''
  Phys.\ Rev.\ Lett.\  {\bf 48}, 1437 (1982).

\bibitem{TB}
J. Traschen and R. Brandenberger,
  ``Particle Production During Out-Of-Equilibrium Phase Transitions,''
  Phys.\ Rev.\  D {\bf 42}, 2491 (1990).
  
 \bibitem{KLS}
L. Kofman, A. Linde and A. Starobinsky,
  ``Reheating after inflation,''
  Phys.\ Rev.\ Lett.\  {\bf 73}, 3195 (1994).

\bibitem{STB}
Y. Shtanov, J. Traschen and R. Brandenberger,
  ``Universe reheating after inflation,''
  Phys.\ Rev.\  D {\bf 51}, 5438 (1995);\\
Shtanov, Y., Ukr. Fiz. Zh. {\bf 38}, 1425 (1993).

\bibitem{KLS2}
L. Kofman, A. Linde and A. Starobinsky,
  ``Towards the theory of reheating after inflation,''
  Phys.\ Rev.\  D {\bf 56}, 3258 (1997).
  
\bibitem{Ramos}
M. Gleiser and R. Ramos,
  ``Microphysical approach to nonequilibrium dynamics of quantum fields,''
  Phys.\ Rev.\  D {\bf 50}, 2441 (1994).

\bibitem{DK}
A. Dolgov and D. Kirilova,
  ``Production of particles by a variable scalar field,''
  Sov.\ J.\ Nucl.\ Phys.\  {\bf 51}, 172 (1990)
  [Yad.\ Fiz.\  {\bf 51}, 273 (1990)].
 
 \bibitem{Mathieu}
N. W. McLachlan, ``Theory and Applications of Mathieu Functions" (Oxford Univ.
Press, Clarendon, 1947).
 
 \bibitem{BKM}
B. Bassett, D. Kaiser and R. Maartens,
  ``General relativistic preheating after inflation,''
  Phys.\ Lett.\  B {\bf 455}, 84 (1999).

\bibitem{Gordon}
C.~Gordon, D.~Wands, B.~A.~Bassett and R.~Maartens,
  ``Adiabatic and entropy perturbations from inflation,''
  Phys.\ Rev.\  D {\bf 63}, 023506 (2001)
  [arXiv:astro-ph/0009131].
  
\bibitem{BFL}
R. Brandenberger, A. Frey and L. Lorenz,
  ``Entropy Fluctuations in Brane Inflation Models,''
  Int.\ J.\ Mod.\ Phys.\  A {\bf 24}, 4327 (2009).

\bibitem{ABD}
R. Brandenberger, K. Dasgupta and A.-C. Davis,
  ``A Study of Structure Formation and Reheating in the D3/D7 Brane Inflation
  Model,''
  Phys.\ Rev.\  D {\bf 78}, 083502 (2008).

\bibitem{Laurence}
L.~P.~Levasseur, G.~Laporte and R.~Brandenberger,
  ``Analytical Study of Mode Coupling in Hybrid Inflation,''
  Phys.\ Rev.\  D {\bf 82}, 123524 (2010).
  arXiv:1004.1425 [hep-th].
  
\bibitem{Francis}
F.-Y. Cyr-Racine and R. Brandenberger,
  ``Study of the Growth of Entropy Modes in MSSM Flat Directions Decay:
   Constraints on the Parameter Space,''
  JCAP {\bf 0902}, 022 (2009).
  
\bibitem{Shaposh}
 F.~L.~Bezrukov and M.~Shaposhnikov,
  ``The Standard Model Higgs boson as the inflaton,''
  Phys.\ Lett.\  B {\bf 659}, 703 (2008)
  [arXiv:0710.3755 [hep-th]].
  
\bibitem{Adams}
 F.~C.~Adams, K.~Freese and A.~H.~Guth,
  ``Constraints on the scalar field potential in inflationary models,''
  Phys.\ Rev.\  D {\bf 43}, 965 (1991).
   
\bibitem{RHBrev3}
 R.~H.~Brandenberger,
  ``Inflationary cosmology: Progress and problems,''
  arXiv:hep-ph/9910410.
   
\bibitem{Jerome1}
R.~H.~Brandenberger and J.~Martin, 
``The Robustness of inflation to changes in superPlanck scale physics,''
Mod.~Phys.~Lett.~A~{\bf 16}, 999
(2001), [arXiv:astro-ph/0005432];\\
J.~Martin and R.~H.~Brandenberger,
``The TransPlanckian problem of inflationary cosmology,''
Phys.~Rev.~D~{\bf 63}, 123501 (2001), [arXiv:hep-th/0005209].
  
\bibitem{Niemeyer}
 J.~C.~Niemeyer,
``Inflation with a high frequency cutoff,''
Phys.~Rev.~D~{\bf 63}, 123502 (2001),
[arXiv:astro-ph/0005533];\\
J.~C.~Niemeyer and R.~Parentani,
``Minimal modifications of the primordial power spectrum from an  adiabatic
  short distance cutoff,''
  Phys.~Rev.~D~{\bf 64}, 101301 (2001),
[arXiv:astro-ph/0101451];\\
S.~Shankaranarayanan,
  ``Is there an imprint of Planck scale physics on inflationary cosmology?,''
  Class.\ Quant.\ Grav.\  {\bf 20}, 75 (2003)
  [arXiv:gr-qc/0203060].
   
\bibitem{Unruh}
 W.~G.~Unruh,
  ``Sonic Analog Of Black Holes And The Effects Of High Frequencies On Black
  Hole Evaporation,''
  Phys.\ Rev.\  D {\bf 51}, 2827 (1995).
  
\bibitem{CJ}
 S.~Corley and T.~Jacobson,
  ``Hawking Spectrum and High Frequency Dispersion,''
  Phys.\ Rev.\  D {\bf 54}, 1568 (1996)
  [arXiv:hep-th/9601073].
  
\bibitem{Jerome2}
 R.~H.~Brandenberger and J.~Martin,
  ``Back-reaction and the trans-Planckian problem of inflation revisited,''
  Phys.\ Rev.\ D {\bf 71}, 023504 (2005)
  [arXiv:hep-th/0410223].
   
\bibitem{Tanaka}
   T.~Tanaka,
``A comment on trans-Planckian physics in inflationary universe,''
 arXiv:astro-ph/0012431.
  
\bibitem{Starob4}
A.~A.~Starobinsky,
``Robustness of the inflationary perturbation spectrum to trans-Planckian
  physics,''
  Pisma Zh.~Eksp.~Teor.~Fiz.~{\bf 73}, 415 (2001),
[JETP Lett.\ {\bf 73}, 371 (2001)], [arXiv:astro-ph/0104043].
  
\bibitem{HE}
S.~W.~Hawking and G.~F.~R.~Ellis,
  ``The Large scale structure of space-time,''
{\it  Cambridge University Press, Cambridge, 1973}
  
\bibitem{Borde}
 A.~Borde and A.~Vilenkin,
  ``Eternal inflation and the initial singularity,''
  Phys.\ Rev.\ Lett.\  {\bf 72}, 3305 (1994)
  [arXiv:gr-qc/9312022].
   
\bibitem{RHBrev4}
R.~H.~Brandenberger,
  ``Back reaction of cosmological perturbations and the cosmological constant
  problem,''
  arXiv:hep-th/0210165.
  
\bibitem{PBB}
M.~Gasperini and G.~Veneziano,
  ``Pre - big bang in string cosmology,''
  Astropart.\ Phys.\  {\bf 1}, 317 (1993)
  [arXiv:hep-th/9211021].
  
\bibitem{Ekp}
J.~Khoury, B.~A.~Ovrut, P.~J.~Steinhardt and N.~Turok,
  ``The ekpyrotic universe: Colliding branes and the origin of the hot big
  bang,''
  Phys.\ Rev.\  D {\bf 64}, 123522 (2001)
  [arXiv:hep-th/0103239].

\bibitem{ghost}
J.~M.~Cline, S.~Jeon and G.~D.~Moore,
  ``The phantom menaced: Constraints on low-energy effective ghosts,''
  Phys.\ Rev.\  D {\bf 70}, 043543 (2004)
  [arXiv:hep-ph/0311312].

\bibitem{quintom}
B.~Feng, X.~L.~Wang and X.~M.~Zhang,
  ``Dark Energy Constraints from the Cosmic Age and Supernova,''
  Phys.\ Lett.\  B {\bf 607}, 35 (2005)
  [arXiv:astro-ph/0404224];\\
B.~Feng, M.~Li, Y.~S.~Piao and X.~Zhang,
  ``Oscillating quintom and the recurrent universe,''
  Phys.\ Lett.\  B {\bf 634}, 101 (2006)
  [arXiv:astro-ph/0407432].
  
\bibitem{LW}
B.~Grinstein, D.~O'Connell and M.~B.~Wise,
  ``The Lee-Wick standard model,''
  Phys.\ Rev.\  D {\bf 77}, 025012 (2008)
  [arXiv:0704.1845 [hep-ph]].
 
 \bibitem{Karouby}
 J.~Karouby and R.~Brandenberger,
  ``A Radiation Bounce from the Lee-Wick Construction?,''
  Phys.\ Rev.\  D {\bf 82}, 063532 (2010)
  [arXiv:1004.4947 [hep-th]].
  
\bibitem{Chunshan}
 C.~Lin, R.~H.~Brandenberger and L.~P.~Levasseur,
  ``A Matter Bounce By Means of Ghost Condensation,''
  arXiv:1007.2654 [hep-th].
  
 \bibitem{Arkani}
   N.~Arkani-Hamed, H.~C.~Cheng, M.~A.~Luty and S.~Mukohyama,
  ``Ghost condensation and a consistent infrared modification of gravity,''
  JHEP {\bf 0405}, 074 (2004)
  [arXiv:hep-th/0312099].
  
\bibitem{mirage}
A.~Kehagias and E.~Kiritsis,
  ``Mirage cosmology,''
  JHEP {\bf 9911}, 022 (1999)
  [arXiv:hep-th/9910174].

\bibitem{HV}
J.~c.~Hwang and E.~T.~Vishniac,
  ``Gauge-invariant joining conditions for cosmological perturbations,''
  Astrophys.\ J.\  {\bf 382}, 363 (1991).
  
\bibitem{DM}
N.~Deruelle and V.~F.~Mukhanov,
  ``On matching conditions for cosmological perturbations,''
  Phys.\ Rev.\  D {\bf 52}, 5549 (1995)
  [arXiv:gr-qc/9503050].

\bibitem{Durrer2}
R.~Durrer and F.~Vernizzi,
  ``Adiabatic perturbations in pre big bang models: Matching conditions and
  scale invariance,''
  Phys.\ Rev.\  D {\bf 66}, 083503 (2002)
  [arXiv:hep-ph/0203275].

\bibitem{Yifu} Y. Cai, R. Brandenberger and X. Zhang, 
``The Matter Bounce Curvaton Scenario",
arXiv:1101.0822 [hep-th].

\bibitem{Cerioni1}
A.~Cerioni and R.~H.~Brandenberger,
  ``Cosmological Perturbations in the Projectable Version of Horava-Lifshitz
  Gravity,''
  arXiv:1007.1006 [hep-th].
  
\bibitem{Cerioni2}
A.~Cerioni and R.~H.~Brandenberger,
  ``Cosmological Perturbations in the 'Healthy Extension'' of Horava-Lifshitz
  gravity,''
  arXiv:1008.3589 [hep-th].
  
\bibitem{Xue}
 Y.~F.~Cai, W.~Xue, R.~Brandenberger and X.~Zhang,
  ``Non-Gaussianity in a Matter Bounce,''
  JCAP {\bf 0905}, 011 (2009)
  [arXiv:0903.0631 [astro-ph.CO]].
  
\bibitem{Xingang}
X.~Chen,
  ``Primordial Non-Gaussianities from Inflation Models,''
  arXiv:1002.1416 [astro-ph.CO].
  
\bibitem{Malda}
J.~M.~Maldacena,
  ``Non-Gaussian features of primordial fluctuations in single field
  inflationary models,''
  JHEP {\bf 0305}, 013 (2003)
  [arXiv:astro-ph/0210603].

\bibitem{LiHong} 
H.~Li, J.~Q.~Xia, R.~Brandenberger and X.~Zhang,
  ``Constraints on Models with a Break in the Primordial Power Spectrum,''
  arXiv:0903.3725 [astro-ph.CO].
  
\bibitem{Lindecrit}
N.~Kaloper, A.~D.~Linde and R.~Bousso,
  ``Pre-big-bang requires the universe to be exponentially large from the  very
  beginning,''
  Phys.\ Rev.\  D {\bf 59}, 043508 (1999)
  [arXiv:hep-th/9801073].
  
\bibitem{Ellis2}
G.~F.~R.~Ellis and R.~Maartens,
  ``The emergent universe: Inflationary cosmology with no singularity,''
  Class.\ Quant.\ Grav.\  {\bf 21}, 223 (2004)
  [arXiv:gr-qc/0211082].
  
\bibitem{Malda2}
J.~M.~Maldacena,
  ``The large N limit of superconformal field theories and supergravity,''
  Adv.\ Theor.\ Math.\ Phys.\  {\bf 2}, 231 (1998)
  [Int.\ J.\ Theor.\ Phys.\  {\bf 38}, 1113 (1999)]
  [arXiv:hep-th/9711200].
  
\bibitem{RHBrev6}
R.~H.~Brandenberger,
  ``Challenges for string gas cosmology,''
  arXiv:hep-th/0509099.

\bibitem{BattWatrev}
T.~Battefeld and S.~Watson,
  ``String gas cosmology,''
  Rev.\ Mod.\ Phys.\  {\bf 78}, 435 (2006)
  [arXiv:hep-th/0510022].
 
\bibitem{ABE}
 S.~Alexander, R.~H.~Brandenberger and D.~Easson,
  ``Brane gases in the early universe,''
  Phys.\ Rev.\ D {\bf 62}, 103509 (2000)
  [arXiv:hep-th/0005212].
 
\bibitem{Pol}
J. Polchinski, \textit{String Theory, Vols. 1 and 2},
(Cambridge Univ. Press, Cambridge, 1998).

\bibitem{Boehm}
T.~Boehm and R.~Brandenberger,
  ``On T-duality in brane gas cosmology,''
  JCAP {\bf 0306}, 008 (2003)
  [arXiv:hep-th/0208188].

\bibitem{Hotta}
K.~Hotta, K.~Kikkawa and H.~Kunitomo,
  ``Correlation between momentum modes and winding modes in
  Brandenberger-Vafa's string cosmological model,''
  Prog.\ Theor.\ Phys.\  {\bf 98}, 687 (1997)
  [arXiv:hep-th/9705099].
  
\bibitem{Osorio}
 M.~A.~R.~Osorio and M.~A.~Vazquez-Mozo,
  ``A Cosmological Interpretation Of Duality,''
  Phys.\ Lett.\  B {\bf 320}, 259 (1994)
  [arXiv:hep-th/9311080].

\bibitem{TV}
A.~A.~Tseytlin and C.~Vafa,
  ``Elements of string cosmology,''
  Nucl.\ Phys.\ B {\bf 372}, 443 (1992)
  [arXiv:hep-th/9109048].
  
\bibitem{Ven}
G.~Veneziano,
  ``Scale factor duality for classical and quantum strings,''
  Phys.\ Lett.\ B {\bf 265}, 287 (1991).

\bibitem{Tseytlin}
 A.~A.~Tseytlin,
  ``Dilaton, winding modes and cosmological solutions,''
  Class.\ Quant.\ Grav.\  {\bf 9}, 979 (1992)
  [arXiv:hep-th/9112004].
  
\bibitem{Kanno1}
R.~H.~Brandenberger, A.~R.~Frey and S.~Kanno,
  ``Towards A Nonsingular Tachyonic Big Crunch,''
  Phys.\ Rev.\  D {\bf 76}, 063502 (2007)
  [arXiv:0705.3265 [hep-th]].
  
\bibitem{Sduality}
S.~Arapoglu, A.~Karakci and A.~Kaya,
  ``S-duality in string gas cosmology,''
  Phys.\ Lett.\  B {\bf 645}, 255 (2007)
  [arXiv:hep-th/0611193].

\bibitem{Mairi}
 M.~Sakellariadou,
  ``Numerical Experiments in String Cosmology,''
  Nucl.\ Phys.\  B {\bf 468}, 319 (1996)
  [arXiv:hep-th/9511075].

\bibitem{Cleaver}
G.~B.~Cleaver and P.~J.~Rosenthal,
  ``String cosmology and the dimension of space-time,''
  Nucl.\ Phys.\  B {\bf 457}, 621 (1995)
  [arXiv:hep-th/9402088].

\bibitem{BEK}
R.~Brandenberger, D.~A.~Easson and D.~Kimberly,
  ``Loitering phase in brane gas cosmology,''
  Nucl.\ Phys.\ B {\bf 623}, 421 (2002)
  [arXiv:hep-th/0109165].

\bibitem{Col2}
R.~Easther, B.~R.~Greene, M.~G.~Jackson and D.~Kabat,
  ``String windings in the early universe,''
  JCAP {\bf 0502}, 009 (2005)
  [arXiv:hep-th/0409121].
  
\bibitem{Danos}
R.~Danos, A.~R.~Frey and A.~Mazumdar,
  ``Interaction rates in string gas cosmology,''
  Phys.\ Rev.\ D {\bf 70}, 106010 (2004)
  [arXiv:hep-th/0409162].

\bibitem{Kabat}
B.~Greene, D.~Kabat and S.~Marnerides,
  ``Dynamical Decompactification and Three Large Dimensions,''
  arXiv:0908.0955 [hep-th].
  
\bibitem{Watson1}
S.~Watson and R.~H.~Brandenberger,
  ``Isotropization in brane gas cosmology,''
  Phys.\ Rev.\ D {\bf 67}, 043510 (2003)
  [arXiv:hep-th/0207168].

\bibitem{Watson2}
S.~Watson and R.~Brandenberger,
  ``Stabilization of extra dimensions at tree level,''
  JCAP {\bf 0311}, 008 (2003)
  [arXiv:hep-th/0307044].
  
\bibitem{Subodh1}
S.~P.~Patil and R.~Brandenberger,
  ``Radion stabilization by stringy effects in general relativity and  dilaton
  gravity,''
  Phys.\ Rev.\ D {\bf 71}, 103522 (2005)
  [arXiv:hep-th/0401037].
  
\bibitem{Subodh2}
S.~P.~Patil and R.~H.~Brandenberger,
  ``The cosmology of massless string modes,''
  JCAP {\bf 0601}, 005 (2006)
  [arXiv:hep-th/0502069].
  
\bibitem{Watson3} 
S.~Watson,
  ``Moduli stabilization with the string Higgs effect,''
  Phys.\ Rev.\ D {\bf 70}, 066005 (2004)
  [arXiv:hep-th/0404177].

\bibitem{Eva2} 
L.~Kofman, A.~Linde, X.~Liu, A.~Maloney, L.~McAllister and E.~Silverstein,
  ``Beauty is attractive: Moduli trapping at enhanced symmetry points,''
  JHEP {\bf 0405}, 030 (2004)
  [arXiv:hep-th/0403001].

\bibitem{RHBrev5}
R.~H.~Brandenberger,
  ``Moduli stabilization in string gas cosmology,''
  Prog.\ Theor.\ Phys.\ Suppl.\  {\bf 163}, 358 (2006)
  [arXiv:hep-th/0509159].
  
\bibitem{Edna}
R.~Brandenberger, Y.~K.~Cheung and S.~Watson,
  ``Moduli stabilization with string gases and fluxes,''
  JHEP {\bf 0605}, 025 (2006)
  [arXiv:hep-th/0501032].

\bibitem{Danos2}
R.~J.~Danos, A.~R.~Frey and R.~H.~Brandenberger,
  ``Stabilizing moduli with thermal matter and nonperturbative effects,''
  Phys.\ Rev.\  D {\bf 77}, 126009 (2008)
  [arXiv:0802.1557 [hep-th]].

\bibitem{BNPV2}
R.~H.~Brandenberger, A.~Nayeri, S.~P.~Patil and C.~Vafa,
  ``String gas cosmology and structure formation,''
  Int.\ J.\ Mod.\ Phys.\  A {\bf 22}, 3621 (2007)
  [arXiv:hep-th/0608121].

\bibitem{BNPV1}
R.~H.~Brandenberger, A.~Nayeri, S.~P.~Patil and C.~Vafa,
  ``Tensor modes from a primordial Hagedorn phase of string cosmology,''
  Phys.\ Rev.\ Lett.\  {\bf 98}, 231302 (2007)
  [arXiv:hep-th/0604126].

\bibitem{Betal}
R.~H.~Brandenberger {\it et al.},
  ``More on the spectrum of perturbations in string gas cosmology,''
  JCAP {\bf 0611}, 009 (2006)
  [arXiv:hep-th/0608186].

 \bibitem{KKLM}
 N.~Kaloper, L.~Kofman, A.~Linde and V.~Mukhanov,
  ``On the new string theory inspired mechanism of generation of  cosmological
  perturbations,''
  JCAP {\bf 0610}, 006 (2006)
  [arXiv:hep-th/0608200].

\bibitem{Deo}
N.~Deo, S.~Jain, O.~Narayan and C.~I.~Tan,
  ``The Effect of topology on the thermodynamic limit for a string gas,''
  Phys.\ Rev.\  D {\bf 45}, 3641 (1992).

\bibitem{Ali}
A.~Nayeri,
   ``Inflation free, stringy generation of scale-invariant cosmological
  fluctuations in D = 3 + 1 dimensions,''
  arXiv:hep-th/0607073.
 
 \bibitem{SGNG}
 B.~Chen, Y.~Wang, W.~Xue and R.~Brandenberger,
  ``String Gas Cosmology and Non-Gaussianities,''
  arXiv:0712.2477 [hep-th].
  
 \bibitem{Witten}
 E.~Witten,
  ``Cosmic Superstrings,''
  Phys.\ Lett.\  B {\bf 153}, 243 (1985).
  
 \bibitem{KS}
 N.~Kaiser and A.~Stebbins,
  ``Microwave Anisotropy Due To Cosmic Strings,''
  Nature {\bf 310}, 391 (1984).

 \bibitem{Amsel}
 S.~Amsel, J.~Berger and R.~H.~Brandenberger,
  ``Detecting Cosmic Strings in the CMB with the Canny Algorithm,''
  JCAP {\bf 0804}, 015 (2008)
  [arXiv:0709.0982 [astro-ph]];\\
A.~Stewart and R.~Brandenberger,
  ``Edge Detection, Cosmic Strings and the South Pole Telescope,''
  arXiv:0809.0865 [astro-ph];\\
R.~J.~Danos and R.~H.~Brandenberger,
  ``Canny Algorithm, Cosmic Strings and the Cosmic Microwave Background,''
  arXiv:0811.2004 [astro-ph];\\
R.~J.~Danos and R.~H.~Brandenberger,
  ``Searching for Signatures of Cosmic Superstrings in the CMB,''
  arXiv:0910.5722 [astro-ph.CO].
 
\bibitem{Kaloper}
  N.~Kaloper and S.~Watson,
  ``Geometric Precipices in String Cosmology,''
  Phys.\ Rev.\  D {\bf 77}, 066002 (2008)
  [arXiv:0712.1820 [hep-th]].
  
\bibitem{Biswas2}
T.~Biswas, R.~Brandenberger, A.~Mazumdar and W.~Siegel,
  ``Non-perturbative gravity, Hagedorn bounce and CMB,''
  JCAP {\bf 0712}, 011 (2007)
  [arXiv:hep-th/0610274].
 
 \bibitem{SZ}
 R.~A.~Sunyaev, Y.~.B.~Zeldovich,
  ``Small scale fluctuations of relic radiation,''
  Astrophys.\ Space Sci.\  {\bf 7}, 3-19 (1970).
  
 \bibitem{Peebles}
  P.~J.~E.~Peebles, J.~T.~Yu,
  ``Primeval adiabatic perturbation in an expanding universe,''
  Astrophys.\ J.\  {\bf 162}, 815-836 (1970).
  
\end{thebibliography}
\end{document}